\documentclass{article}
\usepackage[active]{srcltx}
\usepackage{amsmath}
\usepackage{amsfonts}
\usepackage{amssymb}
\usepackage{amscd}
\usepackage{a4wide}
\usepackage{bm}
\usepackage{amssymb}
\usepackage{latexsym}
\usepackage{amsmath}
\usepackage{amsfonts}
\usepackage{color}
\usepackage{dsfont}
\usepackage{indentfirst}
\usepackage[latin1]{inputenc}
\usepackage{mathrsfs}
\usepackage{bbm}
\usepackage{yfonts}

\def\qed{\hfill $\square$}

\newtheorem{definition}{\underline{Definition}}[section]
\newtheorem{proposition}[definition]{Proposition}
\newtheorem{theorem}[definition]{Theorem}

\newtheorem{remark}[definition]{Remark}
\newtheorem{lema}[definition]{Lemma}
\newtheorem{corollary}[definition]{Corollary}

\numberwithin{equation}{section}

\DeclareMathAlphabet{\mathpzc}{OT1}{pzc}{m}{it}

\begin{document}
\begin{center}
{\Large{ \textbf{On the atomic orbital magnetism: A rigorous derivation of \\ the Larmor and
Van Vleck  contributions.}}}

\medskip

\today
\end{center}

\begin{center}
\small{Baptiste Savoie\footnote{Department of Mathematical Sciences, University of Aarhus, Ny Munkegade, Building 1530, DK-8000 Aarhus C, Denmark; e-mail: baptiste.savoie@gmail.com .}}

\end{center}
\vspace{1.0cm}

\begin{abstract}
The purpose of this paper is to rigorously investigate the orbital magnetism of core electrons in 3-dimensional crystalline ordered solids and in the zero-temperature regime. To achieve that, we consider a non-interacting Fermi gas subjected to an external periodic potential modeling the crystalline field within the tight-binding approximation (i.e. when the distance between two consecutive ions is large). For a fixed number of particles in the Wigner-Seitz cell and in the zero-temperature limit, we derive an asymptotic expansion for the bulk zero-field orbital susceptibility. We prove that the leading term is the superposition of the Larmor diamagnetic contribution, generated by the quadratic part of the Zeeman Hamiltonian, together with the 'complete' orbital Van Vleck paramagnetic contribution, generated by the linear part of the Zeeman Hamiltonian, and related to field-induced electronic transitions.
\end{abstract}
\vspace{1.0cm}

\noindent
\textbf{PACS-2010 number}: 75.20.-g, 51.60.+a, 75.20.Ck, 75.30.Cr.

\medskip

\noindent
\textbf{MSC-2010 number}: 81Q10, 82B10, 82B21, 82D05, 82D20, 82D40.

\medskip

\medskip

\tableofcontents
\medskip
\vspace{1.0cm}

\section{Introduction \& the main result.}

\subsection{An historical review.}
\label{intr}

The first important contribution to the understanding of diamagnetism of ions (we assimilate an atom to an ion of charge zero) and molecules (including the polar ones) goes back at least to 1905 with the three papers \cite{Lan0,Lan1,Lan2} of P. Langevin. We mention all the same that, in all likelihood, W. Weber brought between 1852 and 1871 the pioneer ideas through his molecular theory of magnetism, in which he already introduced the idea that the magnetic effects are due to orbiting motion of electric charges around fixed charges of opposite sign, see \cite{AWW}. The Langevin's microscopic theory essentially leans on the classical Maxwell equations of electromagnetism. Putting things back into context (the nuclear structure of atoms was experimentally discovered in 1911), Langevin considered that matter is formed by electrons in stable periodic motion, the mutual actions between electrons assuring the mechanical stability. In particular, the molecules contain at least one closed electron orbit with a fixed
magnetic moment out of any external field (electrons are assimilated with particulate Amp\`ere's currents), and the different orbits in each molecule have such a moment and such orientations, that their resultant moment may vanish, or not. With these assumptions, Langevin calculated the mean variation of the magnetic moment (orthogonal to the orbit) of electron moving in intramolecular closed orbits under the influence of an external constant magnetic field. This led to the so-called \textit{Langevin formula} for the diamagnetic susceptibility per unit volume of $n$ electrons, see \cite[pp. 89]{Lan2}:
\begin{equation}
\label{Langevin}
\chi_{\mathrm{La}}^{\mathrm{dia}} = - n \frac{e^{2}}{4mc^2}\left\langle r^{2} \right\rangle,
\end{equation}
where $r$ is the distance from the molecule's centre of mass to the electron (considering its motion in the projected orbit on the plane perpendicular to the magnetic field) and $\langle r^2 \rangle$ is the average $r^2$ over all the molecular electrons. Here and hereafter, $e,m,c$ are universal constants denoting respectively the elementary electric charge, the electron rest mass and the speed of light in vacuum. Note that \eqref{Langevin} is independent of the temperature (provided that the orbits retain the same size) and independent of whether or not the initial resultant magnetic moment of the molecule is null. For completeness' sake, we mention that Langevin derived also in \cite{Lan2} the analytic expression of the Curie's empirical law for molecules of paramagnetic substances. The paramagnetic susceptibility per unit volume of $N$ identical molecules with a non-zero resultant magnetic moment with magnitude $M$ reads as, see \cite[pp. 119]{Lan2}:
\begin{equation}
\label{Langevin2}
\chi_{\mathrm{La}}^{\mathrm{para}}(\beta) = \frac{N}{3} \beta M^{2},\quad \beta:= \frac{1}{k_{B}T},
\end{equation}
where $T$ is the absolute temperature and $k_{B}$ denotes the Boltzmann constant. Note that, in the derivation of \eqref{Langevin2}, Langevin used the Boltzmann factor to determine the spatial distribution of the permanent magnetic moment. \\
\indent On the 1910's, a series of works came up in response to the Langevin's theory. For a review, see \cite[Sec. X]{VL}. Essentially, the purpose consisted in computing, within the framework of classical statistical mechanics, the thermal average magnetic moment of various sorts of molecules taking into account specific kinds of collisions. Among these works, we mention the paper \cite{VL} of J.H. Van Leeuwen which played a crucial role in the understanding of the origin of magnetism. In \cite[Sec. VIII]{VL}, she investigated the magnetic response of the free electrons in metals within the Drude model, in which the electrons are assimilated to solid spheres and the velocity distribution is given at thermal equilibrium by the Maxwell-Boltzmann distribution. In the presence of a constant magnetic field, and assuming that the electrons collide elastically with the molecules (forming the matter) assumed to be fixed, then she showed that, whenever a state of equilibrium exists, the thermal average magnetic moment of the electron gas in an arbitrary volume element is identically zero. In other words, within the framework of classical statistical mechanics, the free electron gas displays no diamagnetic effect. This result is referred in Physics literature to as the \textit{Bohr-Van Leeuwen theorem}. This apparent contradiction with the Langevin results (also stated within the classical theory) will be removed by the development of quantum mechanics: the assumptions made by Langevin (stationary of the electron orbits and permanence of the magnetic moment) are actually of a quantum nature.\\
\indent In 1920, W. Pauli was interested in diamagnetism of monoatomic gases in \cite{Pauli}. Still within the framework of the classical theory, his approach slightly differs from the one of Langevin, in the sense that it is based on the Larmor precession theorem. This latter states that, when placed in an external constant magnetic field with intensity $B$, an atom with any number of electron orbits precesses around an axis through the nucleus and parallel to the magnetic field. It causes the angular velocity of an electron in a periodic orbit to be increased by $-\frac{Be}{2mc}$ without the orbit undergoing any change. This led to the formula for the diamagnetic susceptibility per unit volume of $N$ identical atoms assumed to have random orientation in space, see \cite[pp. 203]{Pauli}:
\begin{equation}
\label{Pauli}
\chi_{\mathrm{La}^{*}}^{\mathrm{dia}} = - N \frac{e^{2}}{6mc^2} \sum_{i} \overline{r_{i}^{2}},
\end{equation}
where the sum is over all the electrons of a single atom, $r^{2}_{i}$ is the square distance from the nucleus to the $i$-th electron, and $\overline{r_{i}^{2}}$ has to be understood as the time average of $r^{2}_{i}$. Note that the additional $\frac{2}{3}$-factor (see \eqref{Langevin}) comes from the statistical mean (the orientations of the atoms are random). \eqref{Pauli} is sometimes referred to as the \textit{Langevin formula in the form given by Pauli}.\\

\indent In 1927--1928, J.H. Van Vleck revisited in a series of three papers \cite{VV4,VV3,VV2} the Debye theory on dielectric constant and Langevin theory on magnetic susceptibility of ions/molecules within the framework of quantum mechanics. In particular, he derived a general formula in \cite[Eq. (13)]{VV2} for the total magnetic susceptibility per unit volume, at any 'temperature' $\beta>0$ and in the zero-field limit, of $N$ non-interacting randomly oriented identical ions/molecules with a non-zero (time-averaged) resultant magnetic moment. Only the contribution of electrons is considered (the contribution of nuclei is assumed to be negligible), and the interactions between electrons are disregarded. The 'proof' given by Van Vleck requires the assumption that the energy intervals between the various component levels of the low-energy states of the ions/molecules are small compared to $\beta^{-1}$, see also \cite[Sec. 2]{VV4} and below pp. \pageref{GenVVgd}. This assumption implies that the Van Vleck's results do
not cover the zero-temperature regime. The formula
\cite[Eq. (13)]{VV2} is a generalization of the complete classical Langevin formula (obtained by adding \eqref{Pauli} and \eqref{Langevin2}) but including quantum effects, and consists of the sum of two contributions:
\begin{itemize}
\item A $\beta$-dependent contribution (second term in the r.h.s. of \cite[Eq. (13)]{VV2}).
\end{itemize}
It is purely paramagnetic, depends linearly on $\beta$ and arises from the presence of the non-zero resultant magnetic moment. Van Vleck points out that this term vanishes when the total electronic orbital and spin angular momentum of ions both are null. Moreover, its classical equivalent is \eqref{Langevin2}.
\begin{itemize}
\item A $\beta$-independent contribution (first term in the r.h.s. of \cite[Eq. (13)]{VV2}).
\end{itemize}
It is assumed to represent the diamagnetic effect, since experimentally, diamagnetism in gases is known to be  temperature-independent (at constant density). Denoted by $N\alpha$, it is the sum of two terms,
see \cite[Eq. (15)]{VV2}:
\begin{equation}
\label{Nalp}
N\alpha = N\left(\mathpzc{X}_{\mathrm{vV}} + \mathpzc{X}_{\mathrm{La}}\right).
\end{equation}
Here, $\mathpzc{X}_{\mathrm{La}}$ is the so-called \textit{Larmor contribution} which is purely diamagnetic, and $N\mathpzc{X}_{\mathrm{La}}$ reduces to the Langevin formula in \eqref{Pauli} in the classical limit.  As for $\mathpzc{X}_{\mathrm{vV}}$, it is purely paramagnetic and has no equivalent in the classical theory. It is often referred to as \textit{(orbital) Van Vleck paramagnetic susceptibility}. Besides, Van Vleck analyzed the origin of each one of these two contributions, see e.g. \cite[Sec. VIII.49]{VV1}. Restricting to the case of a single ion, he showed that $\mathpzc{X}_{\mathrm{vV}}$ and $\mathpzc{X}_{\mathrm{La}}$ are generated respectively by the the linear part and the quadratic part of the Zeeman Hamiltonian of the electron gas which are defined by:
\begin{equation}
\label{Zeeman}
H_{\mathrm{Z},\mathrm{lin}} := \mu_{B}\left(\mathbf{L} + g_{0}\mathbf{S}\right)\cdot \mathbf{B},\quad  H_{\mathrm{Z},\mathrm{qua}} :=  \frac{e^{2}}{2m c^{2}}  \sum_{i}  \left(\frac{\mathbf{B} \times \mathbf{r}_{i}}{2}\right)^2,
\end{equation}
in the case where the magnetic vector potential is chosen to be in the symmetric gauge.
Here and hereafter, $\mathbf{L}$ and $\mathbf{S}$ stand for the total electronic orbital and spin angular momentum respectively, $g_{0}$ and $\mu_{B}$ for the electronic $g$-factor and Bohr magneton respectively, $\mathbf{B}$ for the external constant magnetic field and $\mathbf{r}_{i}$ for the position vector of the $i$-th electron. 
From the foregoing, Van Vleck concluded that $\mathpzc{X}_{\mathrm{La}}$ always exists and is in competition with the paramagnetic contribution $\mathpzc{X}_{\mathrm{vV}}$ when the ion has either its total electronic spin angular momentum or orbital angular momentum different from zero in its low-energy states. Moreover, he claimed that $\mathpzc{X}_{\mathrm{vV}}$ does not vanish in great generality in the case of molecules, see
\cite[Sec. X.69]{VV1}. Hence $N\mathpzc{X}_{\mathrm{La}}$ is an upper bound limit to the diamagnetism of electrons in all cases (ions/molecules).\\
\indent Let us outline the 'proof' of \cite[Eq. (13)]{VV2}, we refer to \cite[pp. 594--598]{VV2} and also \cite[Sec. 3]{VV4}. Van Vleck took into account many degrees of freedom of ions/molecules (such that the temperature rotation, the orientations relative to the magnetic field, etc...). We simplify the model and restrict to the case of a single ion. Assume for simplicity that the magnetic field $\mathbf{B}$ is constant and parallel to the $k$-th ($k\in \{1,2,3\}$) direction, i.e. $\mathbf{B} = B \mathbf{e}_{k}$. The derivation consists of three steps. \textit{Step 1: Expressing the average induced magnetic moment with the Boltzmann distribution.} Let $E_{j}(B)$, $j \geq 0$ be the energy levels of the ion in the magnetic field. Let $M_{k}(j,B) := - \frac{d E_{j}}{d B}(B)$ be the induced magnetic moment per unit volume of the state of energy $E_{j}(B)$. The average value of the induced magnetic moment per unit volume $\langle M_{k}(\beta,B) \rangle$ at  'temperature' $\beta>0$ is defined as the thermal equilibrium average (by the Boltzmann distribution) of the $M_{k}(j,B)$'s:
\begin{equation}
\label{MaxBol}
\langle M_{k}(\beta,B) \rangle := - \frac{1}{\sum_{j \geq 0} \mathrm{e}^{- \beta E_{j}(B)}} \sum_{j \geq 0} \frac{d E_{j}(B)}{dB} \mathrm{e}^{- \beta E_{j}(B)}.
\end{equation}
Here, the energy levels are assumed to be non-degenerate. \textit{Step 2: Expanding the $E_{j}(B)$'s in powers of $B$}. Provided that $B$ is small enough, the asymptotic perturbation theory allows to compute the changes in the energy levels by treating the magnetic field as a perturbation. Denoting by $\vert j \rangle$ the $j$-th state, one has, up to the second order correction (in the absence of degeneracy):
\begin{equation*}
E_{j}(B) = E_{j}(0) + \left\langle j \left\vert H_{\mathrm{Z},\mathrm{lin}} + H_{\mathrm{Z},\mathrm{qua}} \right\vert j\right\rangle + \sum_{l\neq j} \frac{\left\vert \left\langle j \left\vert H_{\mathrm{Z},\mathrm{lin}} + H_{\mathrm{Z},\mathrm{qua}} \right\vert l\right\rangle\right\vert^{2}}{E_{j}(0) - E_{l}(0)} + \dotsb,
\end{equation*}
where $H_{\mathrm{Z},\mathrm{lin}}$ and $H_{\mathrm{Z},\mathrm{qua}}$ are defined in \eqref{Zeeman}. This leads to the expansion:
\begin{gather}
\label{EjBser}
E_{j}(B) = E_{j}(0) + B E_{j}^{(1)}(0) + B^{2} E_{j}^{(2)}(0)+ \dotsb,\quad \textrm{with}:\\
\label{correct1}
E_{j}^{(1)}(0) := \mu_{B} \left\langle j \left\vert  \left(\mathbf{L} + g_{0} \mathbf{S}\right)\cdot \mathbf{e}_{k} \right\vert j \right\rangle, \\
\label{correct2}
E_{j}^{(2)}(0) := \frac{e^2}{8 m c^{2}} \sum_{i} \left\langle j \left\vert \left(\mathbf{r}_{i} \times \mathbf{e}_{k}\right)^{2} \right\vert j \right\rangle +  \mu_{B}^{2} \sum_{l \neq j} \frac{\left\vert \left\langle j\left\vert  \left(\mathbf{L} + g_{0} \mathbf{S}\right)\cdot \mathbf{e}_{k} \right\vert l \right\rangle\right\vert^{2}}{E_{j}(0) - E_{l}(0)}.
\end{gather}
\textit{Step 3: Deriving the zero-field magnetic susceptibility}. The rest consists in substituting \eqref{EjBser} into \eqref{MaxBol}, then expanding the Boltzmann factor $\mathrm{e}^{-\beta E_{j}(B)}$ (in the numerator and denominator) in Taylor series in $B$. Assuming that the field-dependent corrections to the energy
in \eqref{EjBser} are smaller than $\beta^{-1}$, an expansion in powers of $B$ of $\langle M_{k}(\beta,B) \rangle$ is derived. Within the framework of the linear response theory, the zero-field magnetic susceptibility per unit volume $\mathpzc{X}(\beta,0)$ corresponds to the coefficient of the linear term in $B$. Since the induced magnetic moment is null in vanishing field, $\mathpzc{X}(\beta,0)$ reduces to:
\begin{equation}
\label{GenVVgd}
\mathpzc{X}(\beta,0)=\frac{1}{\sum_{j \geq 0} \mathrm{e}^{- \beta E_{j}(0)}} \sum_{j \geq 0} \left\{\beta \left(E_{j}^{(1)}(0)\right)^{2} - 2 E_{j}^{(2)}(0)\right\} \mathrm{e}^{- \beta E_{j}(0)}.
\end{equation}
In the derivation of \cite[Eq. (13)]{VV2}, the excited states (i.e. $j \geq 1$) are assumed to be unoccupied. Its counterpart for the particular case we consider here is:
\begin{equation}
\label{mathpzX}
\mathpzc{X}^{(0)}(\beta,0) = \beta \left(E_{0}^{(1)}(0)\right)^{2}- 2 E_{0}^{(2)}(0).
\end{equation}
The $\beta$-independent part in \eqref{mathpzX} can be identified from \eqref{correct2} and reads as:
\begin{equation}
\label{ES0}
- \frac{e^2}{4 m c^{2}} \sum_{i} \left\langle 0 \left\vert \left(\mathbf{r}_{i} \times \mathbf{e}_{k}\right)^{2} \right\vert 0 \right\rangle + 2 \mu_{B}^{2} \sum_{l \neq 0} \frac{\left\vert \left\langle 0\left\vert  \left(\mathbf{L} + g_{0} \mathbf{S}\right)\cdot \mathbf{e}_{k} \right\vert l \right\rangle\right\vert^{2}}{E_{l}(0) - E_{0}(0)}.
\end{equation}
The first and second term are the Larmor and Van Vleck contributions respectively. If the ion has its total electronic orbital and spin angular momentum both equal to zero, \eqref{ES0} reduces to the Larmor contribution.
In accordance with Hund's rules, this is the case of an ion having all its electron shells filled.

In regard to the magnetism of ions/molecules in the zero-temperature regime, another approach is commonly encountered in Physics literature, see e.g. \cite[Chap. 31]{AM}. Let us give the main idea. Consider a system of $N$ non-interacting identical ions subjected to a constant magnetic field at thermal equilibrium. For simplicity, assume that the magnetic field is given by $\mathbf{B} = B \mathbf{e}_{k}$, $k \in \{1,2,3\}$. Let $E_{j}(B)$, $j\geq 0$ be the energy levels of a single ion in the magnetic field. In the canonical ensemble of statistical mechanics, the zero-field magnetic susceptibility per unit volume is defined by:
\begin{equation}
\label{freeEn}
\mathpzc{X}(\beta,N,B=0) := - \frac{\partial^{2} \mathcal{F}}{\partial B^{2}}(\beta,N,B=0),\quad \beta>0,
\end{equation}
where $\mathcal{F}(\beta,N,B)$ stands for the Helmholtz free energy of the system:
\begin{equation}
\label{Helmfree}
\mathcal{F}(\beta,N,B) := - \frac{1}{\beta} \ln\left(\sum_{j \geq 0} \mathrm{e}^{- N \beta E_{j}(B)}\right).
\end{equation}
Based on the principle that only the ground-state of the system is occupied in the zero-temperature regime for $B$ small enough, the Helmholtz free energy reduces to the ground-state energy: $N E_{0}(B)$. From \eqref{freeEn} along with the expansion \eqref{EjBser} obtained within the asymptotic perturbation theory:
\begin{equation*}
\lim_{\beta \uparrow \infty} \mathpzc{X}(\beta,0) =  - 2 N E_{0}^{(2)}(0),
\end{equation*}
which holds in the absence of degeneracy. If $N=1$, this is nothing but the $\beta$-independent part of the magnetic susceptibility derived in \eqref{ES0}.

\subsection{What are the motivations of this paper?}

In the light of the works mentioned in Sec. \ref{intr}, a formula for the zero-field magnetic susceptibility of ions in the low/high-temperature regime can be derived (at least, at the formal level) knowing the energy levels of the ion likely to be occupied, and the associated stationary states.\\
\indent The Van Vleck's approach and the one in \cite[Chap. 31]{AM} have many similarities. Firstly, an ion is modeled as a (semi)classical object in its own right, characterized by a discrete set of energy levels. In the wake of this model, the Maxwell-Boltzmann distribution is used to derive an expression for the thermal average induced magnetic moment/magnetic susceptibility, see \eqref{MaxBol} and \eqref{Helmfree}-\eqref{freeEn}. The use of the Maxwell-Boltzmann distribution turns out to be essential to write down 'easy-handled' formulae for the quantities of interest. However, such formulae hold only for temperatures large enough or in the semiclassical limit. Quantum mechanics is used only at the second stage in order to compute the changes in the energy levels by treating the magnetic field as a perturbation, see \eqref{EjBser}. Subsequently, when computing the zero-field magnetic susceptibility, only the contribution involving the corrections to the ground-state energy (i.e. which couple the ground-state energy with the excited-states energies) are considered, whereas the contribution involving the corrections to the excited-states energies (i.e. which couple the energy levels between them) are discarded, see the transition from \eqref{GenVVgd} to \eqref{mathpzX}. Such a simplification is justified by the fact that the excited-states energies of an ion are not occupied in a regime of weak magnetic field or low temperature, and actually dodges the questions of convergence of the series.\\
\indent Treating an ion as a whole, and retaining only the energy levels which may be occupied dodge the difficulty to compute directly the magnetic susceptibility of the core electrons within the framework of quantum statistical mechanics, whose an expression is more delicate to derive. Indeed, for a wide class of potentials (including Coulomb) used to model the interaction electron-nucleus, the counterpart of the quantity \eqref{MaxBol} for the core electrons is divergent. As a result, some natural question arise: do the results mentioned in Sec. \ref{intr} still hold true if one computes the magnetic susceptibility of core electrons within the framework of quantum statistical mechanics? Does the susceptibility representing the orbital effects still consist of two contributions? If so, how they differ from the 'usual' Larmor and Van Vleck contributions? This paper takes place in this direction.\\
\indent The aim of this paper is to rigorously revisit the atomic orbital magnetism (i.e. we focus on the magnetic effects which do not arise from spin effects) in the zero-temperature regime, the most complicated situation. Our approach is substantially different from the ones mentioned in Sec. \ref{intr}. To model the core electrons of an ion, we consider a non-interacting Fermi gas subjected to an external periodic potential (modeling an ideal lattice of fixed nuclei) within the tight-binding approximation. Under this approximation, we suppose that the distance $R>0$ between two consecutive ions is sufficiently large so that the Fermi gas 'feels' mainly the potential energy generated by one single nucleus. To investigate the atomic orbital magnetism, our starting-point is the expression derived in \cite[Thm. 1.2]{BS} for the bulk zero-field orbital susceptibility of the Fermi gas valid for any 'temperature' $\beta>0$ and $R>0$. We emphasize that this expression is derived from the usual rules of quantum statistical mechanics. Our main result is Theorem \ref{rewLTyp}, and Remark \ref{lienVV} makes the connection with the works mentioned in Sec. \ref{intr}.\\
\indent In the framework of Mathematical-Physics, the rigorous study of orbital magnetism and more generally of diamagnetism, have been the subject of numerous works. Let us list the main ones among them. The first rigorous proof of the Landau susceptibility formula for free electron gases came as late as 1975, due to Angelescu \textit{et al.} in \cite{ABN}. Then in 1990, Helffer \textit{et al.} developed for the first time in \cite{HeSj} a rigorous theory based on the Peierls substitution and considered the connection with the diamagnetism of Bloch electrons and the de Haas-Van Alphen effect. These and many more results were reviewed in 1991 by Nenciu in \cite{N3}. In 2001, Combescure \textit{et al.} recovered in \cite{CR} the Landau susceptibility formula in the semiclassical limit. In 2012, Briet \textit{et al.} gave for the first time in \cite{BCS2} a rigorous justification of the Landau-Peierls approximation for the susceptibility of Bloch electron gases. Finally we mention the following papers \cite{F1,F2,F3} in connection with atomic magnetism.

\subsection{The setting and the main result.}

Consider a three-dimensional quantum gas composed of a large number of non-relativistic identical and indistinguishable particles, with charge $q \neq 0$ and mass $m=1$, obeying the Fermi-Dirac statistics, and subjected to an external constant magnetic field. Furthermore, each particle interacts with an external periodic electric potential modeling the ideal lattice of fixed ions in crystalline ordered solids. The spin of particles is not taken into consideration since we are only interested in orbital effects arising from the 'motion' of particles in the medium (the spin-orbit coupling is disregarded). Moreover, the interactions between particles are neglected (strongly diluted gas assumption) and the gas is at equilibrium with a thermal and particles bath.\\
\indent Let us make our assumptions more precise. The gas is confined in a cubic box $\Lambda_{L}:= (-\frac{L}{2},\frac{L}{2})^{3}$ with $L>0$, centered at the origin of coordinates. We denote its Lebesgue-measure by $\vert \Lambda_{L}\vert$. We consider a uniform magnetic field $\mathbf{B} := B \mathbf{e}_{3}$, $\mathbf{e}_{3}:=(0,0,1)$ parallel to the third direction of the canonical basis of $\mathbb{R}^{3}$. We choose the symmetric gauge, i.e. the magnetic vector potential is defined  by $\mathbf{A}(\mathbf{x}) := \frac{1}{2} \mathbf{B} \times \mathbf{x} = B\mathbf{a}(\mathbf{x})$, $\mathbf{a}(\mathbf{x}) := \frac{1}{2}(-x_{2},x_{1},0)$ so that $\mathbf{B} = \nabla \times \bold{A}(\mathbf{x})$ and $\nabla \cdot \mathbf{A}(\mathbf{x}) = 0$. Hereafter, we denote by $b := \frac{q}{c}B \in \mathbb{R}$ the cyclotron frequency. The potential energy modeling the interaction between each particle and the ideal lattice of fixed nuclei is given by:
\begin{equation}
\label{VR}
V_{R} := \sum_{\boldsymbol{\upsilon} \in \mathbb{Z}^{3}} u(\cdot\, - R \boldsymbol{\upsilon}), \qquad R >0,
\end{equation}
where the single-site potential $u$ satisfies the following assumption:
\begin{itemize}
\item[($\mathscr{A}_{\mathrm{r}}$)] $u \in \mathcal{C}^{1}(\mathbb{R}^{3};\mathbb{R})$ is \textit{compactly supported}.
\end{itemize}
We denote by $\Omega_{R}$, $R>0$  the Wigner-Seitz cell of the $R\mathbb{Z}^{3}$-lattice centered at the origin of coordinates. Its Lebesgue-measure is denoted by $\vert \Omega_{R}\vert$.\\
\indent Introduce the one-particle Hamiltonian. On $\mathcal{C}_0^\infty (\Lambda_L)$, define $\forall R >0$:
\begin{equation}
\label{HL}
H_{R,L}(b) := \frac{1}{2} (-i \nabla - b\mathbf{a})^{2} + V_{R}, \qquad b  \in \mathbb{R}.
\end{equation}
By standard arguments, $\forall R>0$ and $\forall b\in\mathbb{R}$, \eqref{HL} extends to a family of self-adjoint and semi-bounded operators for any $L\in(0,\infty)$, denoted again by $H_{R,L}(b)$, with domain $\mathcal{H}_{0}^{1}(\Lambda_{L}) \cap \mathcal{H}^{2}(\Lambda_{L})$. This definition corresponds to choose Dirichlet boundary conditions on $\partial\Lambda_{L}$. Moreover, since the inclusion $\mathcal{H}_{0}^{1}(\Lambda_{L}) \hookrightarrow L^{2}(\Lambda_{L})$ is compact, then $H_{R,L}(b)$ has purely discrete spectrum with an accumulation point at infinity. We denote by $\{\lambda_{R,L}^{(j)}(b)\}_{j \geq 1}$ the set of eigenvalues of $H_{R,L}(b)$ counting multiplicities and in increasing order. As well, $\forall R>0$ and $\forall b \in \mathbb{R}$, denote by $N_{R,L}(E)$, $E \in \mathbb{R}$ the number of eigenvalues (counting multiplicities) of $H_{R,L}(b)$ in the interval $(-\infty,E)$.\\
\indent When $\Lambda_{L}$ fills the whole space, on $\mathcal{C}_{0}^{\infty}(\mathbb{R}^{3})$ define  $\forall R>0$:
\begin{equation}
\label{Hinfini}
H_{R}(b):=  \frac{1}{2}(-i \nabla - b\mathbf{a})^{2} + V_{R}, \qquad b \in \mathbb{R}.
\end{equation}
By \cite[Thm. X.22]{RS2}, $\forall R>0$ and $\forall b \in \mathbb{R}$, \eqref{Hinfini} is essentially self-adjoint and its self-adjoint extension, denoted again by $H_{R}(b)$, is bounded from below. Sometimes, we will use the shorthand notation $H_{R}= H_{R}(b=0)$. Moreover, the operator $H_{R}(b)$ has only essential spectrum since it commutes with the magnetic translations (of the lattice). Further, $\forall R>0$ and $\forall b \in \mathbb{R}$ the integrated density of states of the operator $H_{R}(b)$ exists, see e.g. \cite[Thm. 3.1]{HLMW}. Denoting by $\mathrm{P}_{I}(H_{R}(b))$ the spectral projection of $H_{R}(b)$ corresponding to the interval $I \subset \mathbb{R}$, it is given for any $E \in \mathbb{R}$ by the limit:
\begin{equation}
\label{IDS}
N_{R}(E) := \lim_{L \uparrow \infty} \frac{N_{R,L}(E)}{\vert \Lambda_{L}\vert} = \lim_{L \uparrow \infty} \frac{\mathrm{Tr}_{L^{2}(\mathbb{R}^{3})}\left\{\chi_{\Lambda_{L}}\mathrm{P}_{(-\infty,E)}(H_{R}(b)) \chi_{\Lambda_{L}}\right\}}{\vert \Lambda_{L}\vert},
\end{equation}
where $\chi_{\Lambda_{L}}$ is the multiplication operator by the characteristic function
of $\Lambda_{L}$.\\

Let us now define some quantities related to the  confined Fermi gas
introduced above within the framework  of quantum statistical mechanics. In the grand-canonical situation, let $(\beta,z,\vert \Lambda_{L}\vert)$ be the fixed external parameters. Here $\beta := (k_{B}T)^{-1} > 0$ ($k_{B}$ stands for the Boltzmann constant) is the 'inverse' temperature and $z:= \mathrm{e}^{\beta \mu} >0$ ($\mu \in \mathbb{R}$ stands for the chemical potential) is the fugacity. For any $\beta>0$, $z >0$ and $b\in \mathbb{R}$, the finite-volume pressure and density are respectively defined $\forall R >0$ by, see e.g. \cite{H,AC,ABN}:
\begin{gather}
\label{pressurefv}
P_{R,L}(\beta,z,b) := \frac{1}{\beta \vert \Lambda_{L}\vert} \mathrm{Tr}_{L^{2}(\Lambda_{L})} \left\{\ln\left(\mathbbm{1} + z \mathrm{e}^{-\beta H_{R,L}(b)}\right)\right\}  = \frac{1}{\beta \vert \Lambda_{L}\vert} \sum_{j=1}^{\infty} \ln\left(1 + z \mathrm{e}^{-\beta \lambda_{R,L}^{(j)}(b)}\right),\\
\label{densityfv}
\rho_{R,L}(\beta,z,b) := \beta z \frac{\partial P_{R,L}}{\partial z}(\beta,z,b) = \frac{1}{\vert \Lambda_{L}\vert}
\sum_{j=1}^{\infty} \frac{z \mathrm{e}^{-\beta \lambda_{R,L}^{(j)}(b)}}{1 + z\mathrm{e}^{-\beta \lambda_{R,L}^{(j)}(b)}}.
\end{gather}
Note that the series in \eqref{pressurefv}-\eqref{densityfv} are absolutely convergent since $\forall b \in \mathbb{R}$ and $\forall R >0$, the semigroup $\{\mathrm{e}^{-\beta H_{R,L}(b)},\, \beta >0\}$ is trace-class on $L^{2}(\Lambda_{L})$, see e.g.
\cite[Eq. (2.12)]{BCS1}. Moreover, from \cite[Thm. 1.1]{BCS1}, then $\forall \beta>0$ and $\forall R>0$, $P_{R,L}(\beta,\cdot\,,\cdot\,)$ is jointly real analytic in $(z,b) \in (0,\infty)\times \mathbb{R}$. This allows to define the finite-volume orbital susceptibility as the second derivative of the pressure w.r.t. the intensity $B$ of the magnetic field, see e.g. \cite[Prop. 2]{ABN}:
\begin{equation*}
\mathcal{X}_{R,L}(\beta,z,b) :=  \left(\frac{q}{c}\right)^{2}  \frac{\partial^{2} P_{R,L}}{\partial b^{2}}(\beta, z,b),\qquad \beta>0,\, z>0,\, b \in \mathbb{R},\, R>0.
\end{equation*}
When $\Lambda_{L}$ fills the whole space (i.e. in the limit $L \uparrow \infty$), the thermodynamic limits of the three grand-canonical quantities defined above generically exist, see e.g. \cite[Thms. 1.1 \& 1.2]{BS} and \cite[Sec. 3.1]{BCS2}. Denoting $\forall \beta>0$, $\forall z>0$, $\forall b \in \mathbb{R}$ and $\forall R>0$ the bulk pressure by $P_{R}(\beta,z,b) := \lim_{L \uparrow \infty} P_{R,L}(\beta,z,b)$, then under our conditions, one has the following pointwise convergences:
\begin{gather}
\label{densitylim}
\rho_{R}(\beta,z,b) := \beta z \frac{\partial P_{R}}{\partial z}(\beta,z,b) = \lim_{L \uparrow \infty} \beta z  \frac{\partial P_{R,L}}{\partial z}(\beta,z,b), \\
\label{susceptilim}
\mathcal{X}_{R}(\beta,z,b) := \left(\frac{q}{c}\right)^{2} \frac{\partial^{2} P_{R}}{\partial b^{2}}(\beta,z,b) = \lim_{L \uparrow \infty} \left(\frac{q}{c}\right)^{2} \frac{\partial^{2} P_{R,L}}{\partial b^{2}}(\beta,z,b),
\end{gather}
and the convergences are compact w.r.t. $(\beta,z,b)$. The limit commutes with the first derivative (resp. the second derivative) of the pressure w.r.t. the fugacity $z$ (resp. the cyclotron frequency $b$). Moreover, $\forall \beta >0$ and $\forall R>0$, $P_{R}(\beta,\cdot\,,\cdot\,)$ is jointly smooth in $(z,b) \in (0,\infty) \times \mathbb{R}$, see e.g. \cite{Sa}.\\

Next, we switch to the canonical conditions and assume that the density of particles $\rho_{0}>0$ becomes, in addition with the 'inverse' temperature, a fixed external parameter.
Seeing the bulk density in \eqref{densitylim} as a function of the $\mu$-variable, denote by $\mu_{R}^{(0)}(\beta,\rho_{0},b) \in \mathbb{R}$, the unique solution of the equation:
\begin{equation*}
\rho_{R}(\beta,\mu,b) = \rho_{0}, \qquad \beta>0,\, b \in \mathbb{R},\, R>0.
\end{equation*}
The inversion of the relation between the bulk density and the chemical potential is ensured by the fact that $\forall \beta>0$, $\forall b \in \mathbb{R}$ and $\forall R>0$, $\mu \mapsto \rho_{R}(\beta,\mu,b)$ is a strictly increasing function on $\mathbb{R}$, and actually it defines a $\mathcal{C}^{\infty}$-diffeomorphism of $\mathbb{R}$ into $(0,\infty)$, see e.g. \cite{Sa,BCS2}. In the following, we consider the situation in which the density of particles is given by:
\begin{equation}
\label{fisdeni}
\rho_{0}(R) = \frac{n_{0}}{\vert \Omega_{R} \vert}, \qquad n_{0} \in \mathbb{N}^{*},\, R>0,
\end{equation}
where $n_{0}$ plays the role of the number of particles in the Wigner-Seitz cell. From \eqref{susceptilim} and under the above conditions, the bulk zero-field orbital susceptibility at fixed $\beta>0$ and density $\rho_{0}(R)$, $R>0$ is defined by:
\begin{equation}
\label{ZFOSrho}
\mathcal{X}_{R}\left(\beta,\rho_{0}(R),b=0\right):= \mathcal{X}_{R}\left(\beta, \mu_{R}^{(0)}(\beta,\rho_{0}(R),b=0),b=0\right).
\end{equation}
\vspace{0.1cm}

\indent Before giving our main theorem related to the quantity in \eqref{ZFOSrho}, let us introduce some reference operators which will be involved in its statements.\\
\indent On $\mathcal{C}_{0}^{\infty}(\mathbb{R}^{3})$, define the 'single atom' Schr\"odinger operator:
\begin{equation}
\label{HamiW}
H_{P} := \frac{1}{2}(-i\nabla)^{2} + u,
\end{equation}
where $u$ is the function appearing in \eqref{VR} and obeying assumption ($\mathscr{A}_{r}$). By standard arguments, $H_{P}$ is essentially self-adjoint and its self-adjoint extension, denoted again by $H_{P}$, is semi-bounded with domain $\mathcal{H}^{2}(\mathbb{R}^{3})$. Moreover, $\sigma_{\mathrm{ess}}(H_{P}) = [0,\infty)$ is absolutely continuous, and $H_{P}$ has finitely many eigenvalues in $(-\infty,0)$ if any, see e.g.
\cite[Thm. XIII.15]{RS4}. Throughout, we suppose:
\begin{itemize}
\item[($\mathscr{A}_{\mathrm{m}}$)] $H_{P}$ has at least one eigenvalue in $(-\infty,0)$,
\end{itemize}
together with the non-degeneracy assumption:
\begin{itemize}
\item[($\mathscr{A}_{\mathrm{nd}}$)] All the eigenvalues of $H_{P}$ in $(-\infty,0)$ are \textit{non-degenerate}.
\end{itemize}
Hereafter, we denote by $\{\lambda_{l}\}_{l=1}^{\tau}$, $\tau \in \mathbb{N}^{*}$ the set of eigenvalues of $H_{P}$ in $(-\infty,0)$ counting in increasing order, and by $\{\Phi_{l}\}_{l=1}^{\tau}$ the set of corresponding normalized eigenfunctions. Also, we denote by $\Pi_{l}$ the orthogonal projection onto the eigenvector $\Phi_{l}$, and define $\Pi_{l}^{\perp} := \mathbbm{1}-\Pi_{l}$.  Since $\Pi_{l}$ commutes with $H_{P}$, one has the decomposition in direct sum: $L^{2}(\mathbb{R}^{3}) = \Pi_{l} L^{2}(\mathbb{R}^{3}) \oplus \Pi_{l}^{\perp} L^{2}(\mathbb{R}^{3})$.\\
\indent In the presence of the uniform magnetic field (as in \eqref{Hinfini}), define on $\mathcal{C}_{0}^{\infty}(\mathbb{R}^{3})$, the 'single atom' magnetic Schr\"odinger operator:
\begin{equation}
\label{HamiWb}
H_{P}(b) := \frac{1}{2}(-i\nabla - b \mathbf{a})^{2} + u, \qquad b \in \mathbb{R}.
\end{equation}
By \cite[Thm. B.13.4]{Si1}, $\forall b \in \mathbb{R}$ \eqref{HamiWb} is essentially self-adjoint and its self-adjoint extension, denoted again by $H_{P}(b)$, is bounded from below. The nature of the spectrum of $H_{P}(b)$ is not known in general, except for $b$ small enough. Indeed from \cite[Thm. 6.1]{AHS}, the eigenvalues of $H_{P}$ in $(-\infty,0)$ are stable under the perturbation by the magnetic field $H_{P}(b)-H_{P} = -b \mathbf{a}\cdot (-i\nabla) + \frac{b^{2}}{2} \mathbf{a}^{2}$ provided that it is weak. From the asymptotic perturbation theory in \cite[Sec. VIII]{K} and due to the assumption ($\mathscr{A}_{\mathrm{nd}}$), then there exists $\mathfrak{b}>0$ s.t. $\forall \vert b\vert \leq \mathfrak{b}$ and for any $l \in \{1,\ldots,\tau\}$, $H_{P}(b)$ has exactly one and only one eigenvalue $\lambda_{l}(b)$ near $\lambda_{l}$ which reduces to $\lambda_{l}$ in the limit $b \rightarrow 0$. In particular, for any $l \in \{1,\ldots,\tau\}$, $\lambda_{l}(\cdot\,)$ can be written in terms of an asymptotic power series in $b$, see e.g. \cite[Thm. 1.2]{BC}. Hereafter we denote by $\{\lambda_{l}(b)\}_{l=1}^{\tau}
$, $\tau \in \mathbb{N}^{*}$ and $\vert b\vert \leq \mathfrak{b}$, the set of these eigenvalues counting in increasing order.\\
\indent Getting back to the 'single atom' operator in \eqref{HamiW}, by \cite[Thm. B.7.2]{Si1}, for any $\xi \in \mathbb{C}\setminus [\inf \sigma(H_{P}),\infty)$, the resolvent operator $(H_{P}-\xi)^{-1}$ is an integral operator with integral kernel $(H_{P}-\xi)^{-1}(\cdot\,,\cdot\,)$ jointly continuous on $\mathbb{R}^{6}\setminus D$, $D:= \{(\mathbf{x},\mathbf{y}) \in \mathbb{R}^{6}: \mathbf{x}=\mathbf{y}\}$. Introduce on $L^{2}(\mathbb{R}^{3})$ the operators $T_{P,j}(\xi)$, $j=1,2$  generated via their kernel respectively defined on $\mathbb{R}^{6}\setminus D$ as:
\begin{gather}
\label{WP1}
T_{P,1}(\mathbf{x},\mathbf{y};\xi) := \mathbf{a}(\mathbf{x}-\mathbf{y}) \cdot (i\nabla_{\mathbf{x}}) (H_{P} - \xi)^{-1}(\mathbf{x},\mathbf{y}),\\
\label{WP2}
T_{P,2}(\mathbf{x},\mathbf{y};\xi) := \frac{1}{2} \mathbf{a}^{2}(\mathbf{x}-\mathbf{y})(H_{P} - \xi)^{-1}(\mathbf{x},\mathbf{y}).
\end{gather}

Here is our main result related with the quantity defined in \eqref{ZFOSrho}:

\begin{theorem}
\label{rewLTyp}
Suppose that the assumptions $\mathrm{(\mathscr{A}_{r})}$, $\mathrm{(\mathscr{A}_{m})}$ and $\mathrm{(\mathscr{A}_{nd})}$ hold.\\ Assume that the number of particles $n_{0} \in \mathbb{N}^{*}$ in the Wigner-Seitz cell is fixed and satisfies $n_{0} \leq \tau$, while the density is given by \eqref{fisdeni}. Then:\\
$\mathrm{(i)}$. For any $0< \alpha < 1$, there exists a $R$-independent constant $c>0$ s.t.
\begin{equation}
\label{thmfond}
\mathcal{X}_{R}\left(\rho_{0}(R)\right) := \lim_{\beta \uparrow \infty} \mathcal{X}_{R}\left(\beta,\rho_{0}(R),b=0\right) = \frac{1}{\vert \Omega_{R} \vert} \mathscr{X}_{P}(n_{0}) + \mathcal{O}\left(\mathrm{e}^{- c R^{\alpha}}\right),
\end{equation}
with:
\begin{equation}
\label{defcalLp1}
\mathscr{X}_{P}(n_{0}) := - \left(\frac{q}{c}\right)^{2} \frac{i}{\pi} \mathrm{Tr}_{L^{2}(\mathbb{R}^{3})} \left\{\int_{\Gamma_{n_{0}}} \mathrm{d}\xi\, \xi (H_{P}-\xi)^{-1}\left[T_{P,1}(\xi)T_{P,1}(\xi) - T_{P,2}(\xi)\right]\right\},
\end{equation}
where $\Gamma_{n_{0}}$ is any positively oriented simple closed contour enclosing the $n_{0}$ smallest eigenvalues of $H_{P}$, while letting the rest of the spectrum outside.\\
$\mathrm{(ii)}$. The $R$-independent quantity in \eqref{defcalLp1} can be identified with:
\begin{equation}
\label{defcalLp2}
\mathscr{X}_{P}(n_{0}) = -\left(\frac{q}{c}\right)^{2} \sum_{l=1}^{n_{0}} \frac{\mathrm{d}^{2} \lambda_{l}}{\mathrm{d} b^{2}}(b=0).
\end{equation}
$\mathrm{(iii)}$. \eqref{defcalLp1} can be rewritten as a sum of two contributions:
\begin{equation}
\label{deride}
\frac{1}{\vert \Omega_{R}\vert} \mathscr{X}_{P}(n_{0}) = \frac{1}{\vert \Omega_{R}\vert} \mathscr{X}_{La}(n_{0}) + \frac{1}{\vert \Omega_{R}\vert} \mathscr{X}_{vV}(n_{0}),
\end{equation}
with:
\begin{gather}
\label{chiLW}
\frac{1}{\vert \Omega_{R}\vert} \mathscr{X}_{La}(n_{0}) := - \left(\frac{q}{c}\right)^{2} \frac{1}{4 \vert \Omega_{R}\vert}  \sum_{l=1}^{n_{0}} \left\langle \Phi_{l}, \left(X_{1}^{2} + X_{2}^{2}\right) \Phi_{l}\right\rangle_{L^{2}(\mathbb{R}^{3})},\\
\label{chiVV}
\frac{1}{\vert \Omega_{R}\vert} \mathscr{X}_{vV}(n_{0}) := \left(\frac{q}{c}\right)^{2} \frac{1}{2 \vert \Omega_{R}\vert} \sum_{l=1}^{n_{0}} \left\langle L_{3} \Phi_{l}, \left(\Pi_{l}^{\perp}\left(H_{P} - \lambda_{l}\right) \Pi_{l}^{\perp}\right)^{-1} L_{3} \Phi_{l} \right\rangle_{L^{2}(\mathbb{R}^{3})}.
\end{gather}
Here, $X_{k}:= \mathbf{X} \cdot \mathbf{e}_{k}$, $k\in \{1,2,3\}$ stands for the position operator projected in the $k$-th direction, and $L_{k} := \mathbf{L} \cdot \mathbf{e}_{k}$ the $k$-th component of the orbital angular momentum operator $\mathbf{L} := \mathbf{X} \times (-i\nabla)$.
\end{theorem}

\begin{remark}[$R$-behavior in the asymptotic \eqref{thmfond}]
The leading term decreases as $R^{-3}$, the remainder (explicitly identified in the proof) decreases exponentially in $R^{\alpha}$, $0<\alpha <1$.
\end{remark}

\begin{remark}[Rewriting of \eqref{chiVV} in a sum of two contributions]
\label{decose}
Denote by $\Pi_{\mathrm{ac}}$ the orthogonal projection onto the absolutely continuous spectrum of the 'single atom' operator $H_{P}$. Since $\Pi_{l}$, $l \in \{1,\ldots,\tau\}$ and $\Pi_{\mathrm{ac}}$ commute with $H_{P}$, then one has the decomposition in direct sum:
\begin{equation*}
\Pi_{l}^{\perp}L^{2}(\mathbb{R}^{3}) = \bigoplus_{\substack{m=1 \\ m\neq l}}^{\tau} \Pi_{m} L^{2}(\mathbb{R}^{3}) \oplus  \Pi_{\mathrm{ac}}L^{2}(\mathbb{R}^{3}).
\end{equation*}
Due to this decomposition, the reduced resolvent operator appearing in \eqref{chiVV} can be rewritten (below, we use the bra-ket notation) as:
\begin{equation*}
\left(\Pi_{l}^{\perp} \left(H_{P} - \lambda_{l}\right) \Pi_{l}^{\perp}\right)^{-1} = \sum_{\substack{m=1 \\ m\neq l}}^{\tau} \frac{\left\vert \Phi_{m}\right\rangle \left\langle \Phi_{m} \right\vert}{\lambda_{m} - \lambda_{l}} + \left(\Pi_{\mathrm{ac}} \left(H_{P} - \lambda_{l}\right) \Pi_{\mathrm{ac}}\right)^{-1}.
\end{equation*}
In view of \eqref{chiVV}, one has under the same conditions the rewriting:
\begin{equation}
\label{redecompo}
\frac{1}{\vert \Omega_{R}\vert} \mathscr{X}_{vV}(n_{0}) = \frac{1}{\vert \Omega_{R}\vert} \mathscr{X}_{vV}^{\mathrm{d}}(n_{0}) + \frac{1}{\vert \Omega_{R}\vert} \mathscr{X}_{vV}^{\mathrm{ac}}(n_{0}),
\end{equation}
with:
\begin{equation}
\label{chiVVd}
\begin{split}
\mathscr{X}_{vV}^{\mathrm{d}}(n_{0}) :&= \left(\frac{q}{c}\right)^{2} \frac{1}{2}  \sum_{l=1}^{n_{0}} \sum_{\substack{m=1 \\ m\neq l}}^{\tau} \frac{\left\vert \left\langle  \Phi_{m}, L_{3} \Phi_{l} \right\rangle\right\vert^{2}}{\lambda_{m} - \lambda_{l}} \\
&= \left\{\begin{array}{ll} \displaystyle{\left(\frac{q}{c}\right)^{2} \frac{1}{2} \sum_{l=1}^{n_{0}} \sum_{m=n_{0}+1}^{\tau} \frac{\left\vert \left\langle  \Phi_{m}, L_{3} \Phi_{l} \right\rangle\right\vert^{2}}{\lambda_{m} - \lambda_{l}}},\, &\textrm{if $n_{0} < \tau$},\\
0, &\textrm{if $n_{0}=\tau$}
\end{array}\right.;
\end{split}
\end{equation}
\begin{equation}
\label{chiVVac}
\mathscr{X}_{vV}^{\mathrm{ac}}(n_{0}) := \left(\frac{q}{c}\right)^{2} \frac{1}{2} \sum_{l=1}^{n_{0}} \left\langle L_{3} \Phi_{l}, \left(\Pi_{\mathrm{ac}}\left(H_{P} - \lambda_{l}\right) \Pi_{\mathrm{ac}}\right)^{-1} L_{3} \Phi_{l} \right\rangle_{L^{2}(\mathbb{R}^{3})},
\end{equation}
and the quantity in \eqref{chiVVac} can be expressed in terms of the generalized eigenfunctions of $H_{P}$ by:
\begin{equation*}
\left(\Pi_{\mathrm{ac}} \left(H_{P}  - \lambda_{l}\right)\Pi_{\mathrm{ac}}\right)^{-1} = \int_{[0,\infty)} \mathrm{d}E \int_{\mathbb{S}^{2}} \mathrm{d}\omega \, \frac{\left\vert\Psi_{E,\omega}\right\rangle \left\langle \Psi_{E,\omega} \right\vert}{E - \lambda_{l}}.
\end{equation*}
Hence, \eqref{chiVV} is the superposition of two contributions: \eqref{chiVVd} involving the corrections to the isolated eigenvalues which couple the isolated eigenvalues between them, and \eqref{chiVVac} involving the corrections to the isolated eigenvalues coupling with the absolutely continuous part of the spectrum.
\end{remark}

\begin{remark}[Connection with the Van Vleck results]
\label{lienVV}
Under the condition of large separation of ions and in the zero-temperature limit, the leading term of the bulk zero-field orbital susceptibility consists of the superposition of two contributions, see \eqref{deride}.
\begin{itemize}
\item Contribution in \eqref{chiLW}.
\end{itemize}
It is purely diamagnetic and generated by the quadratic part of the Zeeman Hamiltonian. Still assuming one single ion in the Wigner-Seitz cell, it reduces in the classical limit to the Langevin formula, see \eqref{Langevin}:
\begin{equation}
\label{LanW}
- \left(\frac{q}{c}\right)^{2} \frac{1}{4 \vert \Omega_{R}\vert}  \sum_{l=1}^{n_{0}} r_{l}^{2},
\end{equation}
where $r_{l}$ is the distance from the origin to the $l$-th particle in the plane orthogonal to $\mathbf{e}_{3}$ (as a rule, the centre of mass of the ion nucleus is at the origin, and the fixed axis passing through it, is taken parallel to the magnetic field).
\begin{itemize}
\item Contribution in \eqref{chiVV}.
\end{itemize}
In view of the rewriting \eqref{redecompo} with \eqref{chiVVd}-\eqref{chiVVac}, it is purely paramagnetic and generated by the linear part of the Zeeman Hamiltonian. Arising from the quantum description, it has no classical equivalent. It is related to field-induced electronic transitions. \\
\indent Let us turn to the comparison with the Van Vleck results in Sect. \ref{intr}.
The contribution in \eqref{chiLW} exhibits the same features than $\mathpzc{X}_{La}$ in \eqref{Nalp}, and its counterpart within the Van Vleck formulation corresponds to the first term in \eqref{ES0}. Then, it identifies with the diamagnetic Larmor susceptibility. The contribution in \eqref{chiVV} exhibits the features of $\mathpzc{X}_{vV}$ in \eqref{Nalp}, but to identify which quantity from the decomposition \eqref{redecompo} corresponds to the second term in \eqref{ES0} within the Van Vleck formulation, we need the following. As pointed out below \eqref{ES0}, when dealing with ions, the orbital Van Vleck paramagnetic susceptibility vanishes when the ion has all its electron shells filled, in accordance with Hund's rules. In a way, this condition (i.e. the electron shells filled) corresponds to have $n_{0}=\tau$ within our model, i.e. the number of negative isolated eigenvalues coincides with the number of electrons in the Wigner-Seitz cell. Since \eqref{chiVVd} vanishes when $n_{0}=\tau$, then its counterpart within the Van Vleck formulation corresponds to the second term in \eqref{ES0}. From the foregoing, this means that, within our model, the quantity in \eqref{chiVVac} is an additional contribution to the 'usual' orbital Van Vleck susceptibility. Involving the corrections to the isolated negative eigenvalues coupling with the absolutely continuous part of the spectrum, it has no counterpart within the Van Vleck formulation. Due to this feature, \eqref{chiVV} represents the 'complete' orbital Van Vleck paramagnetic susceptibility. In that sense, Theorem \ref{rewLTyp} is a generalization of the Van Vleck results.
\end{remark}

\subsection{Outline of the proof of Theorem \ref{rewLTyp}.}
\label{outline}

The starting-point is the expression \eqref{ZFOSR} for the bulk zero-field orbital susceptibility (under the grand-canonical conditions) which holds for any $\beta>0$ and $R>0$. The derivation of such an expression is the main subject of \cite{BS}. Its proof relies on the \textit{gauge invariant magnetic perturbation theory} applied on the magnetic resolvent operator. It allows to keep a good control on the linear growth of the potential vector, see below for further details.\\
\indent The difficulties coming up when proving Theorem \ref{rewLTyp} are threefold:\\
(1). Isolating the main $R$-dependent contribution (which decays polynomially in $R$) and a remainder term (which decays exponentially in $R^{\alpha}$) from \eqref{ZFOSR} when switching to the canonical conditions (the density of particles is fixed, and given by \eqref{fisdeni}). (2). Performing the zero-temperature limit, while taking into account the location of the Fermi energy defined in \eqref{EnFerm}. (3). Identifying the leading term of the asymptotic expansion in the zero-temperature regime with the corresponding relevant physical quantities.

\begin{itemize}
\item Outline of the proof of $\mathrm{(i)}$.
\end{itemize}

The proof of the asymptotic expansion in \eqref{thmfond} requires three steps.\\
\textit{Step 1: Isolating the main $R$-dependent contribution - Part 1}. The formula in \eqref{ZFOSR} involves the trace per unit volume:
\begin{equation}
\label{tracdes1}
\vert \Omega_{R} \vert^{-1} \mathrm{Tr}_{L^{2}(\mathbb{R}^{3})}\left\{\chi_{\Omega_{R}} (H_{R}-\xi)^{-1} \left[T_{R,1}(\xi) T_{R,1}(\xi) - T_{R,2}(\xi)\right]\chi_{\Omega_{R}}\right\},
\end{equation}
where $\chi_{\Omega_{R}}$ denotes the operator multiplication by the indicator function of the Wigner-Seitz cell $\Omega_{R}$, and $T_{R,j}(\xi)$, with $j=1,2$ involve the resolvent operator $(H_{R} - \xi)^{-1}$, see \eqref{WR1}-\eqref{WR2}. In the tight-binding situation, we naturally expect the error made in replacing each resolvent $(H_{R}-\xi)^{-1}$ with $(H_{P}-\xi)^{-1}$ ($H_{P}$ is the 'single atom' operator in \eqref{HamiW}) inside \eqref{tracdes1} to be 'small' for large values of the $R$-parameter. Arising from this approximation, the 'corrective term' \eqref{calGRbm} of \eqref{ZFOSP} involves the trace per unit volume:
\begin{multline}
\label{traqu}
\vert \Omega_{R} \vert^{-1} \mathrm{Tr}_{L^{2}(\mathbb{R}^{3})}\left\{\chi_{\Omega_{R}}\left((H_{R}-\xi)^{-1} \left[T_{R,1}(\xi) T_{R,1}(\xi) - T_{R,2}(\xi)\right]\right. \right. \\
-\left. \left. (H_{P}-\xi)^{-1} \left[T_{P,1}(\xi) T_{P,1}(\xi) - T_{P,2}(\xi)\right]\right)\chi_{\Omega_{R}}\right\},
\end{multline}
where $T_{P,j}(\xi)$, $j=1,2$ involve the resolvent $(H_{P} - \xi)^{-1}$, see \eqref{WP1}-\eqref{WP2}. In view of the operators entering in the trace, to control the $R$-behavior of \eqref{traqu}, it is necessary to control for large values of $R$ the error made in approximating $\chi_{\Omega_{R}}(H_{R} - \xi)^{-1}$  by $\chi_{\Omega_{R}}(H_{P}-\xi)^{-1}$ (or with $\chi_{\Omega_{R}}$ from the right) in the Hilbert-Schmidt sense. To that purpose, we use a geometric perturbation theory in Sec. \ref{GPT}. The key-idea consists in isolating in the Wigner-Seitz cell a region close to the boundary: $\mathpzc{B}_{R}(\kappa) := \{\mathbf{x} \in \overline{\Omega}_{R}: \mathrm{dist}(\mathbf{x}, \partial \Omega_{R}) \leq \kappa R^{\alpha}\}$ for some $\kappa>0$ and $0<\alpha<1$, from the bulk where only the 'single atom' operator $H_{P}$ will act. It comes down to approximate $(H_{R}-\xi)^{-1}$ with:
\begin{equation*}
\hat{g}_{R} (H_{P} - \xi)^{-1} g_{R} + \hat{\hat{g}}_{R} (H_{R} - \xi)^{-1} (1 - g_{R}),
\end{equation*}
where $g_{R}$, $\hat{g}_{R}$ and $\hat{\hat{g}}_{R}$ are smooth cutoff functions; $g_{R}, \hat{g}_{R}$ are supported in the bulk and satisfy \eqref{supot1}-\eqref{disjsup1} and $\hat{\hat{g}}_{R}$ is supported outside and satisfies \eqref{supot2}-\eqref{disjsup2}. Starting with this approximation, we arrive at \eqref{complr}. From the exponential decay of the resolvent's kernel along with the properties \eqref{disjsup1}-\eqref{disjsup2} on the cutoff functions, then provided that the single-site potential $u$ is compactly supported, one gets that the Hilbert-Schmidt norm of:
\begin{equation*}
\chi_{\Omega_{R}}\left\{(H_{R}-\xi)^{-1} - (H_{P}-\xi)^{-1}\right\},\quad \left\{(H_{R}-\xi)^{-1} - (H_{P}-\xi)^{-1}\right\}\chi_{\Omega_{R}},
\end{equation*}
are exponentially small in $R^{\alpha}$, $0<\alpha < 1$. We emphasize that the assumption $u$ is compactly supported in $(\mathscr{A}_{\mathrm{r}})$ is crucial to get the exponential decay in $R^{\alpha}$ of the norms. Indeed, \eqref{complr} involves the operator defined in \eqref{tild2TR}, and the support of $\chi_{\Omega_{R}} \hat{\hat{g}}_{R}$ (or $(1-g_{R}) \chi_{\Omega_{R}}$) is disjoint from the one of $\breve{V}_{R}$ only if $u$ is compactly supported. Based on the foregoing, we prove that the trace in \eqref{traqu} is exponentially small in $R^{\alpha}$, $0<\alpha<1$, see Lemmas \ref{lem1ma}-\ref{lem4ma}.\\

\textit{Step 2: Switching to the canonical conditions and performing the zero-temperature limit}. Remind that, under the assumptions $(\mathscr{A}_{\mathrm{m}})$-$(\mathscr{A}_{\mathrm{nd}})$, $\{\lambda_{l}\}_{l=1}^{\tau}$, $\tau \in \mathbb{N}^{*}$ denotes the set of eigenvalues of $H_{P}$ in $(-\infty,0)$ counting in increasing order. Now, we switch to the canonical conditions. We assume that the number of particles $n_{0} \in \mathbb{N}^{*}$ in the Wigner-Seitz cell  is fixed and obeys $n_{0} \leq \tau$, while the density is given by \eqref{fisdeni}. (The reason for this will become clear below). The starting-point is \eqref{aprlimb} corresponding to the sum of \eqref{ZFOSP} with \eqref{calGRbm}, but with $\mu_{R}^{(0)}(\beta,\rho_{0}(R),b=0)$ instead of $\mu$, see \eqref{ZFOSrho}. With the aim of performing the zero-temperature limit in \eqref{aprlimb}, we need beforehand to turn to the location for large values of $R$ of the Fermi energy defined as the limit:
\begin{equation*}
\mathscr{E}_{R}(\rho_{0}(R)) := \lim_{\beta \uparrow \infty} \mu_{R}^{(0)}(\beta,\rho_{0}(R),b=0).
\end{equation*}
Recall that, since $H_{R}$ commutes with the translations of the $R\mathbb{Z}^{3}$-lattice, then $\sigma(H_{R})$ is absolutely continuous and consists of the union of compact intervals $\mathcal{E}_{R,l}$ (the Bloch bands): $\sigma(H_{R})= \cup_{l=1}^{\infty} \mathcal{E}_{R,l}$, with $\min \mathcal{E}_{R,l} \leq \min \mathcal{E}_{R,l+1}$ and $\max \mathcal{E}_{R,l} \leq \max \mathcal{E}_{R,l+1}$ (we refer to Sec. \ref{FeEn} for the definition of the $\mathcal{E}_{R,l}$'s within the Bloch-Floquet theory). If $\max \mathcal{E}_{R,l} < \min \mathcal{E}_{R,l+1}$ for some $l\geq 1$ then a spectral gap occur. Under the assumption $(\mathscr{A}_{\mathrm{r}})$, we know that the Fermi energy always exists, see \cite[Thm. 1.1]{BCS2}.  Moreover, only two situations can occur: either the Fermi energy lies in the middle of a spectral gap (this corresponds to the semiconducting/isolating situation if $\max \mathcal{E}_{R,s} < \min \mathcal{E}_{R,s+1}$, the semimetal situation if $\max \mathcal{E}_{R,s}= \min \mathcal{E}_{R,s+1}$), or in the interior of a Bloch band (this is the metallic situation). In Proposition \ref{locferm}, we state that for large values of $R$, our assumptions $(\mathscr{A}_{\mathrm{m}})$-$(\mathscr{A}_{\mathrm{nd}})$ along with the fact that $\rho_{0}(R)$ obeys \eqref{fisdeni}, \textit{automatically lead to the isolating situation}:
\begin{equation}
\label{semcondsitu}
\mathscr{E}_{R}(\rho_{0}(R)) = \frac{\max \mathcal{E}_{R,n_{0}} + \min \mathcal{E}_{R,n_{0}+1}}{2} < 0.
\end{equation}
The proof leans on two ingredients. The first one concerns the location of the negative Bloch bands of $H_{R}$ at negative eigenvalues of $H_{P}$ for large $R$, see Lemma \ref{lemRsuf}. Since the $\lambda_{l}$'s ($l=1,\ldots,\tau$) are simple, then for $R$ sufficiently large, the Bloch bands $\mathcal{E}_{R,l}$ ($l=1,\ldots,\tau$) are simple, isolated from each other and from the rest of the spectrum. Moreover, $\forall l \in \{1,\ldots,\tau\}$ $\mathcal{E}_{R,l}$ is localized in a neighborhood of $\lambda_{l}$, and reduces to $\lambda_{l}$ when $R \uparrow \infty$. The second one is a relation between the I.D.S. of $H_{R}$ and the zero-temperature limit of the grand-canonical density expressed in terms of the eigenvalues of the Bloch Hamiltonian, see Lemma \ref{densityBF}. Then, it remains to use the criterion in \cite[Thm 1.1]{BCS2}.
Let us get back to the zero-temperature limit. Performing it needs special attention. In \eqref{ZFOSP} and \eqref{calGRbm}, the dependence in $\beta$ is contained in
\begin{equation*}
\beta^{-1} \mathfrak{f}\left(\beta,\mu_{R}^{(0)}(\beta,\rho_{0}(R),b=0);\xi\right),\quad \mathfrak{f}(\beta,\mu;\xi) := \ln(1 + \mathrm{e}^{\beta(\mu - \xi)}),
\end{equation*}
and $\forall \mu \in \mathbb{R}$, $\mathfrak{f}(\beta,\mu;\cdot\,)$ is holomorphic on $\{\zeta \in \mathbb{C}: \Im \zeta \in (- \frac{\pi}{\beta}, \frac{\pi}{\beta})\}$. The difficulty encountered is as follows. The contours involved in the integration w.r.t. $\xi$ in \eqref{ZFOSP}-\eqref{calGRbm} depend on $\beta$, see \eqref{cbetaP}-\eqref{cbeta}. When $\beta \uparrow \infty$, they reduce to a half-line on the real axis. Here the isolating situation in \eqref{semcondsitu} plays a crucial role allowing to dodge this difficulty. In fact, our assumptions have been chosen to force this situation. Indeed, for $R$ sufficiently large, $\mathscr{E}_{R}(\rho_{0}(R))$ lies in the interior of an open set $I_{n_{0}}$ containing $\frac{\lambda_{n_{0}} + \lambda_{n_{0}+1}}{2}$ (we set $\lambda_{\tau+1}:=0$) and s.t. $I_{n_{0}} \cap \sigma(H_{R}) = \emptyset$ and $I_{n_{0}} \cap \sigma(H_{P}) = \emptyset$. In this way, we can decompose the contour \eqref{cbetaP} into three parts: the first one enclosing the first $n_{0}$ eigenvalues of $H_{P}$,
the second one enclosing $I_{n_{0}}$ but no eigenvalue, and the third one enclosing the rest of $\sigma(H_{P})$, see \eqref{decomCbeta}. Note that, for $R$ large enough, these contours do not intersect $\sigma(H_{R})$. Since the second contour lies in the holomorphic domain of the integrand, the integral is null. Since for a given $\mu_{0} \in \mathbb{R}$, $\mathfrak{f}(\beta,\mu_{0};\cdot\,)$ is holomorphic on $\{\zeta \in \mathbb{C}: \Re \zeta \neq \mu_{0}\}$, then the first and third contours can be deformed in some $\beta$-independent contours. It remains to use the Lebesgue's dominated theorem knowing \eqref{pointwis}. From \eqref{aprlimb}, Propositions \ref{PRO1}-\ref{PRO2} together lead to \eqref{expandZFOS}. The behavior for large $R$ of the remainder in \eqref{expandZFOS} follows from the one of \eqref{traqu} discussed in Step 1.\\ 
We mention that, if $(\mathscr{A}_{\mathrm{nd}})$ is disregarded, then the insulating situation can still occur if one sets some restrictions on the $n_{0}$, see Sec. \ref{discus} and Sec. \ref{oppb}.\\

\textit{Step 3: Isolating the main $R$-dependent contribution - Part 2}. We start by removing the $R$-dependence arising from the indicator functions $\chi_{\Omega_{R}}$ inside the trace of \eqref{tildZFOSP}. Getting the trace out the integration w.r.t. the $\xi$-variable, it remains to control the $R$-behavior of the error made in extending the trace to the whole space. In Proposition \ref{PRO3}, we state that it is exponentially small in $R$. To do so, the crucial ingredient is the exponential localization of the eigenfunctions associated with the eigenvalues of $H_{P}$ in $(-\infty,0)$. Next, the Fermi energy can be removed from \eqref{vvvv} without changing the value of the trace, see Proposition \ref{PRO4}. Thus, the main $R$-dependent term behaves like $R^{-3}$.

\begin{itemize}
\item Outline of the proof of $\mathrm{(ii)}$ and $\mathrm{(iii)}$.
\end{itemize}

The proofs heavily rely on the stability of the eigenvalues $\{\lambda_{l}\}_{l=1}^{\tau}$, $\tau \in \mathbb{N}^{*}$ of $H_{P}$ in $(-\infty,0)$ under the perturbation $W(b):= H_{P}(b)-H_{P}$.\\
\indent \textit{Outline of the proof of $\mathrm{(ii)}$}. For $b$ small enough, the key-idea consists in expressing the sum over the $n_{0}$ eigenvalues $\lambda_{l}(b)$ by means of the Riesz integral formula for orthogonal projections, see \eqref{avanpr}. In the rest of the proof, we show that the second derivative w.r.t. $b$ at $b=0$ of the r.h.s. of \eqref{avanpr} is nothing but \eqref{defcalLp1}, see Proposition \ref{dernPr}. From the rewriting of the trace in \eqref{avanpr} as the integral of the diagonal kernel \eqref{diagke}, it comes down to prove that the integral kernel of $(H_{P}(b) - \xi)^{-1}$, far away the diagonal, is twice differentiable in a neighborhood of $b=0$, and to write down a formula for its second derivative, see Lemma \ref{oubli}. To that purpose, the crucial ingredient is the so-called gauge invariant magnetic perturbation theory, see e.g.
\cite{N2, BC, CN1, BCS2, BS, Sa2}. The idea behind is to isolate the singularity of the magnetic perturbation (arising from the linear growth of the vector potential) via an exponential factor involving the magnetic phase in \eqref{phase}. Then $(H_{P}(b)-\xi)^{-1}$ can be approximated by $\tilde{R}_{P}(b,0,\xi)$ generated by the kernel in \eqref{tildR}, while the 'corrective term' behaves like $\mathcal{O}(\vert b \vert)$, see \eqref{startequi}. Iterating twice \eqref{startequi} in the kernels sense,  then expanding the exponential factor in Taylor series, one obtains the beginning of an expansion in powers of $b$ for the kernel of $(H_{P}(b)-\xi)^{-1}$.  We mention that the exponential localization of the eigenfunctions associated with the $\lambda_{j}(b)$'s for $b$ small enough (see Lemma \ref{stabb}) plays a crucial role to control the involved quantities, see Corollary \ref{coro1}.\\
\indent  \textit{Outline of the proof of $\mathrm{(iii)}$}. From \eqref{defcalLp2}, the aim is to derive an expression for the second derivative w.r.t. $b$ at $b=0$ of the perturbed eigenvalue $\lambda_{l}(b)$, $l \in \{1,\ldots,n_{0}\}$. To do that, we use the Feshbach formula to write down 'the corrections' to the unperturbed eigenvalue $\lambda_{l}$ under the  perturbation $W(b) = -b \mathbf{a}\cdot(-i\nabla) + \frac{b^{2}}{2} \mathbf{a}^{2}$ for small values of $b$, see \eqref{comre}. Here, the exponential localization of the eigenfunction associated with $\lambda_{l}$ allows to control the linear growth of $\mathbf{a}$. Iterating \eqref{comre}, we obtain the beginning of an expansion in powers of $b$ for $\lambda_{l}(\cdot\,)$, see \eqref{Fesch2}. To conclude, it remains to use that $\lambda_{l}(\cdot\,)$ has an asymptotic expansion series, see e.g. \cite[Thm. 1.2]{BC}. Note that the asymptotic perturbation theory and the gauge-invariant magnetic perturbation theory both are used to control the $b$-behavior of the
remainder in \eqref{Fesch2}.

\subsection{Discussion on the assumptions.}
\label{discus}
Let us first discuss the assumption ($\mathscr{A}_{\mathrm{r}}$) on the single-site potential $u$. The physical modeling requires that the 'single atom' operator $H_{P}$ possesses \textit{finitely} many eigenvalues (at least one) below the essential spectrum. Then $u$ has to be chosen accordingly. As emphasized in Sec. \ref{outline}, choosing $u$ compactly supported plays a crucial role in our analysis since it gives rise to the exponentially decreasing behavior in $R^{\alpha}$, $0<\alpha<1$ of the remainder in \eqref{thmfond}, see Proposition \ref{PRO2}. However, we believe that the optimal $\alpha$  is $\alpha=1$, i.e. the remainder should behave like $\mathcal{O}(\mathrm{e}^{-cR})$, $c>0$. We think that it could be obtained from our analysis by using a more refined geometric perturbation theory to approximate the resolvent $(H_{R}-\xi)^{-1}$. Furthermore, we point out that the leading term in the asymptotic expansion \eqref{thmfond} is unchanged if the single-site potential is not compactly supported.
Consider the assumption:
\begin{itemize}
\item[($\mathscr{A}_{\mathrm{r}^{*}}$)] $u \in \mathcal{C}^{1}(\mathbb{R}^{3};\mathbb{R})$ with $u = \mathcal{O}(\vert \mathbf{x}\vert^{-(3+\epsilon)})$ for $\vert \mathbf{x}\vert$ sufficiently large.
\end{itemize}
From \cite[Thm. XIII.6]{RS4}, $\sigma_{\mathrm{ess}}(H_{P}) = [0,\infty)$ and $H_{P}$ has a finite number of bound states  in $(-\infty,0)$. Replacing ($\mathscr{A}_{\mathrm{r}}$) with ($\mathscr{A}_{\mathrm{r}^{*}}$) in Theorem \ref{rewLTyp}, then under the same conditions, one may expect the behavior of the remainder to be only polynomially decreasing with $R$ owing to the 'tail' of the potential. Finally, let us mention that $u$ is chosen continuously differentiable to ensure some regularities for the eigenvectors of $H_{P}$, see \eqref{fpexpo} and Lemma \ref{stabb}.\\
\indent Let us discuss the non-degeneracy assumption ($\mathscr{A}_{\mathrm{nd}}$). Our analysis is based on the insulating situation which occurs when the Fermi energy lies in the middle of a spectral gap of  $H_{R}$, see \cite[Thm 1.1]{BCS2}. When the number of particles $n_{0}$ in the Wigner-Seitz cell is any integer lesser than the number of negative eigenvalues of $H_{P}$, while the density is given by \eqref{fisdeni}, this together with  ($\mathscr{A}_{\mathrm{nd}}$) automatically lead to the insulating situation for $R$ sufficiently large, see Proposition \ref{locferm}. Nonetheless, we stress the point that, when disregarding the assumption ($\mathscr{A}_{\mathrm{nd}}$),
then the insulating condition can still occur for $R$ sufficiently large provided that one sets some restrictions on the $n_{0}$'s in \eqref{fisdeni}. It has to obey (see proof of Proposition \ref{locferm} and Remark \ref{rem1}):
\begin{equation}
\label{cond}
n_{0} \leq \tau \quad \textrm{and} \quad \exists \varkappa\in\{1,\ldots,\nu\}\,\,\,\textrm{s.t.}\,\,\, n_{0} = \mathrm{dim}\, \mathpzc{E}_{1} + \dotsb + \mathrm{dim}\, \mathpzc{E}_{\varkappa}.
\end{equation}
Here, $\tau$ is the number of eigenvalues of $H_{P}$ in $(-\infty,0)$ counting multiplicities, $\nu \leq \tau$ is the number of distinct eigenvalues and $\mathpzc{E}_{l}$, $l\in\{1,\ldots, \nu\}$ stands for the eigenspace associated with the (possibly degenerate) eigenvalue $\lambda_{l}$ of $H_{P}$, $\lambda_{l} = \{\lambda_{l}^{(m)}\}_{m=1}^{\mathrm{dim} \mathpzc{E}_{l}}$. Supposing that the assumptions ($\mathscr{A}_{\mathrm{r}}$)-($\mathscr{A}_{\mathrm{m}}$) hold, and assuming that the number of particles $n_{0} \in \mathbb{N}^{*}$ in the Wigner-Seitz cell is fixed and obeys \eqref{cond}, while the density of particles is given by \eqref{fisdeni}, then the formulae in Theorem \ref{rewLTyp} $\mathrm{(ii)}$-$\mathrm{(iii)}$ have to be modified accordingly (the statement in $\mathrm{(i)}$ is unchanged). Indeed, \eqref{defcalLp2} becomes:
\begin{equation*}
\mathscr{X}_{P}(n_{0}) = -\left(\frac{q}{c}\right)^{2} \sum_{l=1}^{\varkappa} \sum_{m=1}^{\mathrm{dim} \mathpzc{E}_{l}} \frac{\mathrm{d}^{2} \lambda_{l}^{(m)}}{\mathrm{d} b^{2}}(b=0);
\end{equation*}
and \eqref{chiLW}-\eqref{chiVV} respectively become (with intuitive notations):
\begin{equation*}
\frac{1}{\vert \Omega_{R}\vert} \mathscr{X}_{La}(n_{0}) := - \left(\frac{q}{c}\right)^{2} \frac{1}{4 \vert \Omega_{R}\vert}  \sum_{l=1}^{\varkappa}  \sum_{m=1}^{\mathrm{dim} \mathpzc{E}_{l}} \left\langle \Phi_{l}^{(m)}, \left(X_{1}^{2} + X_{2}^{2}\right) \Phi_{l}^{(m)}\right\rangle_{L^{2}(\mathbb{R}^{3})},
\end{equation*}
\begin{equation*}
\frac{1}{\vert \Omega_{R}\vert} \mathscr{X}_{vV}(n_{0}) :=
\left(\frac{q}{c}\right)^{2} \frac{1}{2 \vert \Omega_{R}\vert}  \sum_{l=1}^{\varkappa}  \sum_{m=1}^{\mathrm{dim} \mathpzc{E}_{l}} \left\langle L_{3} \Phi_{l}^{(m)}, \left(\Pi_{l}^{(m),\perp}\left(H_{P} - \lambda_{l}^{(m)}\right) \Pi_{l}^{(m),\perp}\right)^{-1} L_{3} \Phi_{l}^{(m)} \right\rangle_{L^{2}(\mathbb{R}^{3})}.
\end{equation*}
Note that the contribution $\mathscr{X}_{vV}(n_{0})$ can be decompose into two contributions as in Remark \ref{decose}.

\subsection{An open problem.}
\label{oppb}

From the foregoing, a natural question arises: does the insulating situation occur when $n_{0}$ is any integer less than the number of negative eigenvalues of $H_{P}$ counting multiplicities if some of the eigenvalues are degenerate, while the density is given by \eqref{fisdeni}? Such a problem comes up when the single-site potential is chosen spherically symmetric. To tackle it, we need to know precisely the behavior of the negative spectral bands of $H_{R}$ near the degenerate eigenvalues of $H_{P}$ in $(-\infty,0)$ for large values of the $R$-parameter. For instance, suppose that one of the negative eigenvalue of $H_{P}$, say $\lambda_{c}$, is two-fold degenerate. For $R$ sufficiently large, it is well-known that the spectrum of $H_{R}$ in a neighborhood of $\lambda_{c}$ consists of the union of two Bloch bands, see Lemma \ref{lemRsuf} and also \cite[Thm. 2.1]{Dau}. But we need to know much more; in particular we need to control how the Bloch bands behave the one relative to the other. Especially, do they always overlap
for $R$ sufficiently large or do they overlap only in the limit $R\uparrow \infty$? How fast each of the Bloch bands reduces to $\lambda_{c}$? This remains a challenging spectral problem.

\subsection{The content of the paper.}

Our current paper is organized as follows. In Sec. \ref{GPT}, we use a geometric perturbation theory to approximate the resolvent operator $(H_{R}-\xi)^{-1}$ in the tight-binding situation. Sec. \ref{Bproof} is devoted to the proof of Theorem \ref{rewLTyp}. In Sec. \ref{FeEn}, we show that under our assumptions, only the insulating situation can occur in the tight-binding situation. Subsequently, in Sec. \ref{Fgh12}-\ref{Fgh13} we prove the asymptotic expansion in \eqref{thmfond}. In Sec. \ref{defcal}, we prove the identity \eqref{defcalLp2}. In Sec. \ref{Fesh} we prove the formula \eqref{deride}. In Sec. \ref{append}, we have collected all the proofs of the technical intermediary results needed in Sec. \ref{Bproof}.

\section{An approximation of the resolvent via a geometric perturbation theory.}
\label{GPT}

The method we use below is borrowed from \cite{CN1,CFFH}.\\
\indent For any $0 < \alpha <1$, $0 < \kappa \leq 10$ and $R>0$ define:
\begin{equation*}
\mathpzc{B}_{R}(\kappa) := \left\{\mathbf{x} \in \overline{\Omega_{R}}: \mathrm{dist}\left(\mathbf{x}, \partial \Omega_{R}\right) \leq \kappa R^{\alpha}\right\}.
\end{equation*}
Hence, for $R$ sufficiently large, $\mathpzc{B}_{R}(\kappa)$ models a 'thin' compact subset of $\Omega_{R}$ near the boundary with Lebesgue-measure $\vert \mathpzc{B}_{R}(\kappa)\vert$ of order $\mathcal{O}(R^{2+\alpha})$.\\
Let $0 < \alpha < 1$ be fixed. Below, by $R$ sufficiently large, we mean:
\begin{equation}
\label{defR0}
R \geq R_{0}\quad \textrm{with}\quad R_{0}=R_{0}(\alpha) \geq 1 \quad \textrm{s.t.} \quad \mathpzc{B}_{R_{0}}(8) \subsetneq \Omega_{R_{0}}.
\end{equation}
\indent Let us introduce some well-chosen families of smooth cutoff functions. \\
Let $\{g_{R}\}$ and $\{\hat{g}_{R}\}$ satisfying for $R \geq R_{0}$:
\begin{gather*}
\mathrm{Supp}(g_{R}) \subset \left(\Omega_{R} \setminus \mathpzc{B}_{R}(3)\right),\quad g_{R} = 1\,\, \textrm{if}\,\, \mathbf{x} \in \left(\Omega_{R} \setminus \mathpzc{B}_{R}(4)\right),\quad
0 \leq g_{R} \leq 1;\\
\mathrm{Supp}(\hat{g}_{R}) \subset \left(\Omega_{R} \setminus \mathpzc{B}_{R}(1)\right),\quad
\hat{g}_{R} = 1\,\, \textrm{if}\,\, \mathbf{x} \in \left(\Omega_{R} \setminus \mathpzc{B}_{R}(2)\right),\quad 0 \leq \hat{g}_{R} \leq 1.
\end{gather*}
Moreover, there exists a constant $C>0$ s.t.
\begin{equation*}
\forall R \geq R_{0},\quad  \max\left\{\left\Vert D^{s} g_{R} \right\Vert_{\infty}, \left\Vert D^{s} \hat{g}_{R} \right\Vert_{\infty}\right\} \leq C  R^{- \vert s\vert \alpha},\quad \forall \vert s \vert \leq 2,\,s \in \mathbb{N}^{3}.
\end{equation*}
Also, let $\{\hat{\hat{g}}_{R}\}$ satisfying for $R \geq R_{0}$:
\begin{equation*}
\mathrm{Supp}\left(\hat{\hat{g}}_{R}\right) \subset \left(\overline{\Omega_{R} \setminus \mathpzc{B}_{R}(6)}\right)^\complement,\,\,\,
\hat{\hat{g}}_{R} = 1\,\, \textrm{if}\,\, \mathbf{x} \in \left(\overline{\Omega_{R} \setminus \mathpzc{B}_{R}(5)}\right)^\complement, \,\,\,
0 \leq \hat{\hat{g}}_{R} \leq 1.
\end{equation*}
Moreover, there exists a constant $C>0$ s.t.
\begin{equation*}
\forall R \geq R_{0},\quad  \left\Vert D^{s} \hat{\hat{g}}_{R} \right\Vert_{\infty} \leq C  R^{- \vert s\vert \alpha},\quad \forall \vert s \vert \leq 2,\,s \in \mathbb{N}^{3}.
\end{equation*}
By gathering all these properties together, one straightforwardly gets:
\begin{gather}
\label{supot1}
\hat{g}_{R} g_{R} = g_{R},\\
\label{disjsup1}
\mathrm{dist}\left(\mathrm{Supp}(D^{s} \hat{g}_{R}), \mathrm{Supp}(g_{R})\right) \geq C R^{\alpha}, \quad \forall 1\leq \vert s\vert \leq 2,\\
\label{supot2}
\hat{\hat{g}}_{R}(1 - g_{R}) = (1 - g_{R}),\\
\label{disjsup2}
\mathrm{dist}\left(\mathrm{Supp}\left(D^{s} \hat{\hat{g}}_{R}\right), \mathrm{Supp}\left(1-g_{R}\right)\right) \geq C R^{\alpha}, \quad \forall 1\leq \vert s\vert \leq 2,
\end{gather}
for some $R$-independent constant $C>0$.\\

\indent Let us now define a series of operators. At first, introduce $\forall R\geq R_{0}$ and $\forall  \xi \in \varrho(H_{R}) \cap \varrho(H_{P})$ (here $\varrho(\cdot\,)$ denotes the resolvent set) on $L^{2}(\mathbb{R}^{3})$:
\begin{equation}
\label{tildSR}
\mathcal{R}_{R}(\xi) := \hat{g}_{R} (H_{P} - \xi)^{-1} g_{R} + \hat{\hat{g}}_{R}(H_{R} - \xi)^{-1}(1 - g_{R}).
\end{equation}
Due to the support of the cutoff functions, one has:
\begin{equation*}
(H_{R} - \xi) \hat{g}_{R} = (H_{P} - \xi) \hat{g}_{R}.
\end{equation*}
Using that $\mathrm{Ran}(\mathcal{R}_{R}(\xi)) \subset \mathrm{Dom}(H_{R})$, then from \eqref{supot1} along with \eqref{supot2}:
\begin{equation*}
(H_{R} - \xi) \mathcal{R}_{R}(\xi) = \mathbbm{1} + \mathcal{W}_{R}(\xi),
\end{equation*}
where, $\forall R\geq R_{0}$ and $\forall \xi \in \varrho(H_{R})\cap \varrho(H_{P})$:
\begin{equation*}
\begin{split}
\mathcal{W}_{R}(\xi) := &\left\{-\frac{1}{2} \left(\Delta \hat{g}_{R}\right) - \left(\nabla \hat{g}_{R}\right) \cdot \nabla \right\} (H_{P} - \xi)^{-1} g_{R}\\
& + \left\{-\frac{1}{2} \left(\Delta \hat{\hat{g}}_{R}\right) - \left(\nabla \hat{\hat{g}}_{R}\right) \cdot \nabla\right\} (H_{R} - \xi)^{-1} \left(1-g_{R}\right).
\end{split}
\end{equation*}
Since $\mathcal{W}_{R}(\xi)$ is bounded on $L^{2}(\mathbb{R}^{3})$, see e.g. \cite[Lem. 5.1]{BS}, this means that:
\begin{equation}
\label{eqsecresQ}
(H_{R} - \xi)^{-1} = \mathcal{R}_{R}(\xi) - (H_{R}-\xi)^{-1} \mathcal{W}_{R}(\xi).
\end{equation}
Next, $\mathcal{R}_{R}(\xi)$ in \eqref{tildSR} can be rewritten by the second resolvent equation as:
\begin{equation}
\label{rewtildSR}
\mathcal{R}_{R}(\xi) = \mathscr{R}_{R}(\xi)  - \mathscr{W}_{R}(\xi),
\end{equation}
where, $\forall R\geq R_{0}$ and $\forall \xi \in \varrho(H_{R})\cap \varrho(H_{P})$:
\begin{gather}
\label{tild2SR}
\mathscr{R}_{R}(\xi) := \hat{g}_{R}(H_{P} - \xi)^{-1}g_{R} + \hat{\hat{g}}_{R}(H_{P} - \xi)^{-1}(1 - g_{R}),\\
\label{tild2TR}
\mathscr{W}_{R}(\xi) := \hat{\hat{g}}_{R}(H_{R} - \xi)^{-1}\breve{V}_{R}(H_{P} - \xi)^{-1}(1 - g_{R}),
\end{gather}
with:
\begin{equation}
\label{hatVR}
\breve{V}_{R} := \sum_{\boldsymbol{\upsilon} \in \mathbb{Z}^{3}\setminus \{\mathbf{0}\}} u(\cdot\, - R \boldsymbol{\upsilon}).
\end{equation}
Finally, $\mathscr{R}_{R}(\xi)$ in \eqref{tild2SR} can be rewritten as:
\begin{equation}
\label{rewtildSR2}
\mathscr{R}_{R}(\xi) = (H_{P}-\xi)^{-1}  + \mathfrak{W}_{R}(\xi),
\end{equation}
where, $\forall R \geq R_{0}$ and $\forall \xi \in \varrho(H_{R})\cap \varrho(H_{P})$:
\begin{equation*}
\begin{split}
\mathfrak{W}_{R}(\xi) :=   &(H_{P} - \xi)^{-1} \left\{-\frac{1}{2} \left(\Delta \hat{g}_{R}\right) - \left(\nabla \hat{g}_{R}\right) \cdot \nabla\right\} (H_{P} - \xi)^{-1} g_{R}\\
&+ (H_{P} - \xi)^{-1}\left\{-\frac{1}{2} \left(\Delta \hat{\hat{g}}_{R}\right) - \left(\nabla \hat{\hat{g}}_{R}\right) \cdot \nabla\right\} (H_{P} - \xi)^{-1} \left(1-g_{R}\right).
\end{split}
\end{equation*}
Gathering \eqref{eqsecresQ}, \eqref{rewtildSR} and \eqref{rewtildSR2} together, $\forall R \geq R_{0}$ and $\forall \xi \in \varrho(H_{R})\cap \varrho(H_{P})$:
\begin{equation}
\label{complr}
(H_{R}-\xi)^{-1} = (H_{P}-\xi)^{-1} + \mathfrak{W}_{R}(\xi) - \mathscr{W}_{R}(\xi) - (H_{R}-\xi)^{-1} \mathcal{W}_{R}(\xi),
\end{equation}
and \eqref{complr} holds in the bounded operators sense on $L^{2}(\mathbb{R}^{3})$.\\
\indent From \eqref{complr} and taking into account the assumption $(\mathscr{A}_{\mathrm{r}})$, let us now prove that for $R \geq R_{0}$ sufficiently large, the error made in approximating $\chi_{\Omega_{R}} (H_{R} - \xi)^{-1}$ with $\chi_{\Omega_{R}}(H_{P} - \xi)^{-1}$ (or with the indicator function $\chi_{\Omega_{R}}$ from the right) is exponentially small in $R^{\alpha}$, $0<\alpha < 1$ in the Hilbert-Schmidt norm sense. To do that, define $R_{1}\geq1$ so that:
\begin{equation}
\label{defR1}
\mathrm{Supp}(u) \subset (\Omega_{R_{1}} \setminus \mathpzc{B}_{R_{1}}(7)).
\end{equation}
Such $R_{1}$ exists since the support of $u$ is compact, see assumption ($\mathscr{A}_{\mathrm{r}}$). Now, look at the r.h.s. of \eqref{complr}. For any $R \geq R_{0}$, $(H_{R}-\xi)^{-1} \mathcal{W}_{R}(\xi)$ and $\mathfrak{W}_{R}(\xi)$ have their operator norm exponentially small in $R^{\alpha}$, see \eqref{estgeN2}-\eqref{estgeN} below. However, this is not the case for the operator norm of $\mathscr{W}_{R}(\xi)$ even for $R\geq R_{0}$ and large enough, see \eqref{estgeN}. This comes from the fact that the support of $\hat{\hat{g}}_{R}$ (or $(1-g_{R})$) and the one of $\breve{V}_{R}$ in \eqref{hatVR} are not disjoint. But for any $R \geq \max\{R_{0},R_{1}\}$, the Hilbert-Schmidt (H-S) norm of $\mathscr{W}_{R}(\xi)$ when multiplied by the indicator function $\chi_{\Omega_{R}}$ from the left (or from the right) is exponentially small in $R^{\alpha}$, see \eqref{normI}. The same holds true for the H-S norms of $(H_{R}-\xi)^{-1} \mathcal{W}_{R}(\xi)$ and $\mathfrak{W}_{R}(\xi)$ when multiplied by the indicator function $\chi_{\Omega_{R}}$, see \eqref{normI0}. This last feature, which results from the fact that $u$ is compactly supported, will turn out to be decisive to get the exponential decay of the remainder term in the asymptotic expansion \eqref{thmfond}.\\

We end this section by giving a series of estimates we will use throughout. For any $\xi \in \mathbb{C}$ and real number $\ell>0$, we use the shorthand notation:
\begin{equation}
\label{shorthand}
\ell_{\xi} := \ell (1 + \vert \xi \vert)^{-1}.
\end{equation}

\begin{lema}
\label{lemhgfP}
Let $\Xi = P$ or $R$. For every $\eta>0$ (and $\forall R >0$ if $\Xi =R$), there exists a constant $\vartheta=\vartheta(\eta)>0$ and a polynomial $p(\cdot\,)$ s.t. $\forall \xi \in \mathbb{C}$ satisfying $\mathrm{dist}(\xi,\sigma(H_{\Xi})) \geq \eta$ and $\forall(\mathbf{x},\mathbf{y}) \in \mathbb{R}^{6}\setminus D$:
\begin{gather}
\label{kertildR}
\left\vert (H_{\Xi}-\xi)^{-1}(\mathbf{x},\mathbf{y}) \right\vert \leq p(\vert \xi \vert) \frac{\mathrm{e}^{- \vartheta_{\xi}\vert \mathbf{x} - \mathbf{y}\vert}}{\vert \mathbf{x} - \mathbf{y}\vert},\\
\label{kertildRder}
\left\vert \nabla_{\mathbf{x}} (H_{\Xi}-\xi)^{-1}(\mathbf{x},\mathbf{y}) \right\vert \leq p(\vert \xi \vert) \frac{\mathrm{e}^{- \vartheta_{\xi}\vert \mathbf{x} - \mathbf{y}\vert}}{\vert \mathbf{x} - \mathbf{y}\vert^{2}}.
\end{gather}
\end{lema}

\noindent \textit{\textbf{Proof}}. See \cite[Thm. B.7.2]{Si1} and \cite[Lem. 2.4]{BS} respectively.  \qed

\begin{lema}
\label{lemhgf}
Let $0<\alpha < 1$ and $R_{0}=R_{0}(\alpha) \geq 1$ as in \eqref{defR0}. For every $\eta>0$, there exists a constant $\vartheta=\vartheta(\eta)>0$ and a polynomial $p(\cdot\,)$ s.t. \\
$\mathrm{(i)}$. $\forall R \geq R_{0}$, $\forall \xi \in \mathbb{C}$ obeying $\mathrm{dist}(\xi,\sigma(H_{R})\cap \sigma(H_{P})) \geq \eta$, $\forall(\mathbf{x},\mathbf{y})\in \mathbb{R}^{6}\setminus D$:
\begin{gather}
\label{kertildS}
\max\left\{ \left\vert (\mathcal{R}_{R}(\xi))(\mathbf{x},\mathbf{y})\right\vert,\left\vert(\mathscr{R}_{R}(\xi))(\mathbf{x},\mathbf{y})
\right\vert\right\} \leq p(\vert \xi \vert) \frac{\mathrm{e}^{- \vartheta_{\xi}\vert \mathbf{x} - \mathbf{y}\vert}}{\vert \mathbf{x} - \mathbf{y}\vert},\\
\label{kertildSder}
\max\left\{\left\vert \nabla_{\mathbf{x}}(\mathcal{R}_{R}(\xi))(\mathbf{x},\mathbf{y})\right\vert, \left\vert \nabla_{\mathbf{x}} (\mathscr{R}_{R}(\xi))(\mathbf{x},\mathbf{y})\right\vert\right\} \leq p(\vert \xi \vert)\frac{\mathrm{e}^{- \vartheta_{\xi}\vert \mathbf{x} - \mathbf{y}\vert}}{\vert \mathbf{x}-\mathbf{y}\vert^{2}}.
\end{gather}
$\mathrm{(ii)}$. $\forall R \geq R_{0}$, $\forall \xi \in \mathbb{C}$ obeying $\mathrm{dist}(\xi,\sigma(H_{R})\cap \sigma(H_{P})) \geq \eta$ and $\forall(\mathbf{x},\mathbf{y}) \in \mathbb{R}^{6}$:
\begin{gather}
\label{kerhatT}
\left\vert (\mathscr{W}_{R}(\xi))(\mathbf{x},\mathbf{y})\right\vert \leq p(\vert \xi \vert) \mathrm{e}^{- \vartheta_{\xi}\vert \mathbf{x} - \mathbf{y}\vert}, \\
\label{ledernier}
\left\vert \nabla_{\bold{x}}(\mathscr{W}_{R}(\xi))(\mathbf{x},\mathbf{y})\right\vert \leq p(\vert \xi \vert) \frac{\mathrm{e}^{- \vartheta_{\xi}\vert \mathbf{x} - \mathbf{y}\vert}}{\vert \bold{x} - \bold{y}\vert},\quad \bold{x} \neq \bold{y},\\
\label{kertildT}
\max\left\{\left\vert (\mathcal{W}_{R}(\xi))(\mathbf{x},\mathbf{y})\right\vert,\left\vert (\mathfrak{W}_{R}(\xi))(\mathbf{x},\mathbf{y})\right\vert,
\left\vert \nabla_{\mathbf{x}}(\mathfrak{W}_{R}(\xi))(\mathbf{x},\mathbf{y})\right\vert\right\}
\leq p(\vert \xi \vert) \mathrm{e}^{- \vartheta_{\xi} R^{\alpha}} \mathrm{e}^{- \vartheta_{\xi}\vert \mathbf{x} - \mathbf{y}\vert}.
\end{gather}
$\mathrm{(iii)}$. $\forall R \geq \max\{R_{0},R_{1}\}$ ($R_{1}$ is defined through \eqref{defR1}), $\forall \xi \in \mathbb{C}$ obeying $\mathrm{dist}(\xi,\sigma(H_{R})\cap \sigma(H_{P})) \geq \eta$ and $\forall(\mathbf{x},\mathbf{y}) \in \mathbb{R}^{6}$:
\begin{multline*}
\max\left\{\left\vert \left(\chi_{\Omega_{R}}\hat{\hat{g}}_{R}\right)(\mathbf{x}) (H_{R} - \xi)^{-1}(\mathbf{x},\mathbf{y}) \breve{V}_{R}(\mathbf{y}) \right\vert,
 \left\vert \breve{V}_{R}(\mathbf{x}) (H_{P} - \xi)^{-1}(\mathbf{x},\mathbf{y}) \left(\chi_{\Omega_{R}}(1-g_{R})\right)(\mathbf{y}) \right\vert\right\} \\
\leq p(\vert \xi \vert) \mathrm{e}^{- \vartheta_{\xi} R^{\alpha}} \mathrm{e}^{- \vartheta_{\xi}\vert \mathbf{x} - \mathbf{y}\vert}.
\end{multline*}
\end{lema}

\noindent \textit{\textbf{Proof}}. We start by $\mathrm{(i)}$. In view of \eqref{tildSR} and \eqref{tild2SR}, \eqref{kertildS} directly follows from \eqref{kertildR} knowing that $0 \leq g_{R},\hat{g}_{R},\hat{\hat{g}}_{R} \leq 1$. Next, one has  $\forall (\mathbf{x},\mathbf{y})\in \mathbb{R}^{6}\setminus D$:
\begin{equation*}
\begin{split}
\nabla_{\mathbf{x}}(\mathcal{R}_{R}(\xi))(\mathbf{x},\mathbf{y})
= &\left\{\left(\nabla \hat{g}_{R}\right)(\mathbf{x}) (H_{P} - \xi)^{-1}(\mathbf{x},\mathbf{y}) +
\hat{g}_{R}(\mathbf{x}) \nabla_{\mathbf{x}}(H_{P} - \xi)^{-1} (\mathbf{x},\mathbf{y})\right\}g_{R}(\mathbf{y})
  \\ &+
\left\{\left(\nabla \hat{\hat{g}}_{R}\right)(\mathbf{x}) (H_{R} - \xi)^{-1}(\mathbf{x},\mathbf{y})+ \hat{\hat{g}}_{R}(\mathbf{x}) \nabla_{\mathbf{x}}(H_{R} - \xi)^{-1} (\mathbf{x},\mathbf{y})\right\}(1 - g_{R}(\mathbf{y})),
\end{split}
\end{equation*}
and by replacing $(H_{R} - \xi)^{-1}$ with $(H_{P} - \xi)^{-1}$ above, we get $\nabla_{\mathbf{x}}(\mathscr{R}_{R}(\xi))(\cdot\,,\cdot\,)$.
Then \eqref{kertildSder} follows from \eqref{kertildR}-\eqref{kertildRder}. Let us turn to $\mathrm{(ii)}$. From \eqref{tild2TR}, \eqref{kerhatT} results from \eqref{kertildR}, our assumptions on $u$ along with
\cite[Lem. A.2]{BS}. Moreover:
\begin{equation*}
\begin{split}
\nabla_{\bold{x}}(\mathscr{W}_{R}(\xi))(\bold{x},\bold{y}) =
&\left(\nabla \hat{\hat{g}}_{R}\right)(\bold{x}) \int_{\mathbb{R}^{3}} \mathrm{d}\bold{z}\, (H_{R}-\xi)^{-1}(\bold{x},\bold{z}) \breve{V}_{R}(\bold{z}) (H_{P}-\xi)^{-1}(\bold{z},\bold{y})(1-g_{R})(\bold{y}) \\
&+ \hat{\hat{g}}_{R}(\bold{x}) \int_{\mathbb{R}^{3}} \mathrm{d}\bold{z}\, \nabla_{\bold{x}} (H_{R}-\xi)^{-1}(\bold{x},\bold{z}) \breve{V}_{R}(\bold{z}) (H_{P}-\xi)^{-1}(\bold{z},\bold{y})(1-g_{R})(\bold{y}).
\end{split}
\end{equation*}
Then, \eqref{ledernier} follows from \eqref{kertildR}-\eqref{kertildRder} combined with \cite[Lem. A.2]{BS}. Next, we prove \eqref{kertildT}. The properties \eqref{disjsup1}-\eqref{disjsup2} are the key-ingredient to make appear the exponential decay in $R^{\alpha}$. Under the conditions of the lemma:
\begin{multline}
\label{autilis}
\max \left\{\left\vert \left(\nabla \hat{\hat{g}}_{R}\right)(\mathbf{x}) \nabla_{\mathbf{x}}(H_{\Xi} - \xi)^{-1}(\mathbf{x},\mathbf{y}) (1 - g_{R})(\mathbf{y})\right\vert,
\left\vert \left(\nabla \hat{g}_{R}\right)(\mathbf{x}) \nabla_{\mathbf{x}}(H_{\Xi} - \xi)^{-1}(\mathbf{x},\mathbf{y}) g_{R}(\mathbf{y})\right\vert, \right.\\
 \left. \left\vert \left(\Delta \hat{g}_{R}\right)(\mathbf{x}) (H_{\Xi} - \xi)^{-1}(\mathbf{x},\mathbf{y}) g_{R}(\mathbf{y})\right\vert, \left\vert \left(\Delta \hat{\hat{g}}_{R}\right)(\mathbf{x}) (H_{\Xi} - \xi)^{-1}(\mathbf{x},\mathbf{y}) (1 - g_{R})(\mathbf{y})\right\vert\right\} \\
\leq p(\vert \xi\vert) \mathrm{e}^{- \vartheta_{\xi} R^{\alpha}} \mathrm{e}^{- \vartheta_{\xi}\vert \mathbf{x} - \mathbf{y}\vert},
\end{multline}
for another $R$-independent $\vartheta>0$ and polynomial $p(\cdot\,)$. This leads to \eqref{kertildT} for the kernel of $\mathcal{W}_{R}(\xi)$ and $\mathfrak{W}_{R}(\xi)$. As for $\nabla_{\mathbf{x}}(\mathfrak{W}_{R}(\xi))(\cdot\,,\cdot\,)$, it is enough to use \eqref{autilis}, \eqref{kertildRder} along with
\cite[Eq. (A.12)]{BS}. Finally $\mathrm{(iii)}$ follows from \eqref{kertildR} with \eqref{defR1} assuring that $\mathrm{dist}(\mathrm{Supp}(u), \mathrm{Supp}(\chi_{\Omega_{R}}\hat{\hat{g}}_{R})) \geq R^{\alpha}$. \qed

\begin{remark}
From Lemma \ref{lemhgfP} and Lemma \ref{lemhgf} $\mathrm{(i)}$-$\mathrm{(ii)}$, together with the Shur-Holmgren criterion, one has $\forall R\geq R_{0}$:
\begin{gather}
\label{estgeN}
\max\left\{\Vert (H_{R} - \xi)^{-1} \Vert, \Vert \mathcal{R}_{R}(\xi)\Vert, \Vert \mathscr{R}_{R}(\xi) \Vert, \Vert \mathscr{W}_{R}(\xi)\Vert,
\Vert \nabla \mathcal{R}_{R}(\xi)\Vert,
\Vert \nabla \mathscr{R}_{R}(\xi)\Vert, \Vert \nabla\mathscr{W}_{R}(\xi)\Vert\right\} \leq p(\vert \xi \vert),\\
\label{estgeN2}
\max\left\{\Vert \mathcal{W}_{R}(\xi)\Vert, \Vert \mathfrak{W}_{R}(\xi)\Vert\, \Vert \nabla \mathfrak{W}_{R}(\xi)\Vert\right\} \leq p(\vert \xi\vert) \mathrm{e}^{- \vartheta_{\xi} R^{\alpha}},
\end{gather}
for another $R$-independent constant $\vartheta>0$ and polynomial $p(\cdot\,)$.
\end{remark}
\begin{remark}
Let $(\mathfrak{I}_{2}(L^{2}(\mathbb{R}^{3})), \Vert \cdot\, \Vert_{\mathfrak{I}_{2}})$ be the Banach space of Hilbert-Schmidt operators. From Lemma \ref{lemhgf} $\mathrm{(ii)}$ and the $*$-ideal property of $\mathfrak{I}_{2}(L^{2}(\mathbb{R}^{3}))$:
\begin{multline}
\label{normI0}
\forall R \geq R_{0},\quad \max\left\{\Vert \chi_{\Omega_{R}} \mathfrak{W}_{R}(\xi) \Vert_{\mathfrak{I}_{2}}, \Vert \chi_{\Omega_{R}} (H_{R} - \xi)^{-1} \mathcal{W}_{R}(\xi) \Vert_{\mathfrak{I}_{2}},\right.\\
\left. \Vert \mathfrak{W}_{R}(\xi) \chi_{\Omega_{R}} \Vert_{\mathfrak{I}_{2}}, \Vert  (H_{R} - \xi)^{-1} \mathcal{W}_{R}(\xi) \chi_{\Omega_{R}} \Vert_{\mathfrak{I}_{2}}\right\} \leq p(\vert \xi\vert) \mathrm{e}^{- \vartheta_{\xi} R^{\alpha}},
\end{multline}
and from Lemma \ref{lemhgf} $\mathrm{(iii)}$ one has $\forall R \geq \max\{R_{0},R_{1}\}$:
\begin{equation}
\label{normI}
\max\left\{\Vert \chi_{\Omega_{R}} \mathscr{W}_{R}(\xi) \Vert_{\mathfrak{I}_{2}}, \Vert \mathscr{W}_{R}(\xi)\chi_{\Omega_{R}} \Vert_{\mathfrak{I}_{2}}\right\} \leq p(\vert \xi\vert) \mathrm{e}^{- \vartheta_{\xi} R^{\alpha}},
\end{equation}
for another $R$-independent constant $\vartheta>0$ and polynomial $p(\cdot\,)$.
\end{remark}


\section{Proof of Theorem \ref{rewLTyp}.}
\label{Bproof}

This section is organized as follows. The first part is devoted to the proof of the asymptotic expansion \eqref{thmfond} in the tight-binding situation and in the zero-temperature limit. The second and third part are respectively concerned with the proof of \eqref{defcalLp2} and \eqref{deride}. For reader's convenience, the proofs of technical intermediary results are collected in Appendix, see Sec. \ref{append}.

\subsection{Proof of $\mathrm{(i)}$.}
\label{i}

\subsubsection{The location of the Fermi energy.}
\label{FeEn}

Here, we are interested in the location of the Fermi energy in the tight-binding situation when the number of particles $n_{0} \in \mathbb{N}^{*}$ in the Wigner-Seitz cell is fixed, while the density is given by \eqref{fisdeni}. We recall that under our conditions, the Fermi energy defined as:
\begin{equation}
\label{EnFerm}
\mathscr{E}_{R}(\rho_{0}(R)) := \lim_{\beta \uparrow \infty} \mu^{(0)}_{R}(\beta,\rho_{0}(R),b=0),\quad R>0,
\end{equation}
always exists, see \cite[Thm 1.1]{BCS2}.\\

Before stating the main result, let us introduce some basic elements of Bloch-Floquet theory. We refer to \cite[Sec. 3.5]{BeSh} and also \cite{Wil}. We mention that the results we give below hold true for any $R>0$. Denote by $\Omega_{R}^{*}$ the unit cell of the dual lattice $(2\pi/R) \mathbb{Z}^{3}$ (the so-called first Brillouin zone) of the Bravais lattice $R \mathbb{Z}^{3}$. Denoting by $\mathscr{S}(\mathbb{R}^{3})$ the Schwartz space of rapidly decreasing functions on $\mathbb{R}^{3}$, the Bloch-Floquet(-Zak) transformation is defined by:
\begin{equation*}
\begin{split}
\mathcal{U}: \mathscr{S}(\mathbb{R}^{3}) &\mapsto L^{2}\left(\Omega_{R}^{*}, L^{2}(\Omega_{R})\right) \cong \int_{\Omega_{R}^{*}}^{\oplus} \mathrm{d}\mathbf{k}\, L^{2}(\Omega_{R}) \\
(\mathcal{U} \phi)(\underline{\mathbf{x}};\mathbf{k}) &= \frac{1}{\sqrt{\vert \Omega_{R}^{*} \vert}} \sum_{\boldsymbol{\upsilon} \in R \mathbb{Z}^{3}} \mathrm{e}^{-i \mathbf{k} \cdot (\underline{\mathbf{x}} + \boldsymbol{\upsilon})} \phi(\underline{\mathbf{x}} + \boldsymbol{\upsilon}),\quad \mathbf{k} \in \Omega_{R}^{*},\, \underline{\mathbf{x}} \in \Omega_{R},
\end{split}
\end{equation*}
which can be continued in a unitary operator on $L^{2}(\mathbb{R}^{3})$. The unitary transformation of $H_{R}$ is decomposable into a direct integral:
\begin{equation*}
\mathcal{U} H_{R} \mathcal{U}^{*} = \int_{\Omega_{R}^{*}}^{\oplus} \mathrm{d}\mathbf{k}\, h_{R}(\mathbf{k}), \quad h_{R}(\mathbf{k}) := \frac{1}{2}(-i\nabla + \mathbf{k})^{2} + V_{R},
\end{equation*}
and the Bloch Hamiltonian $h_{R}$ lives in $L^{2}(\mathbb{R}^{3}/R\mathbb{Z}^{3})$. By standard arguments, $h_{R}$ is essentially self-adjoint on $\mathcal{C}^{\infty}(\mathbb{R}^{3}/R\mathbb{Z}^{3})$ and the domain of its closure is $\mathcal{H}^{2}(\mathbb{R}^{3}/R \mathbb{Z}^{3})$. For each $\mathbf{k} \in \Omega_{R}^{*}$, $h_{R}(\mathbf{k})$ has purely discrete spectrum with an accumulation point at infinity. We denote by $\{E_{R,l}(\mathbf{k})\}_{l\geq 1}$ the set of eigenvalues counting multiplicities and in increasing order. Due to this choice of labeling, the $E_{R,l}$'s are periodic and Lipschitz continuous on $\Omega_{R}^{*}$. Indeed, when crossing-points occur, the $E_{R,l}$'s are not differentiable on a zero Lebesgue-measure subset of $\Omega_{R}^{*}$ corresponding to crossing-points. The spectrum of $H_{R}$ is absolutely continuous and given (as a set of points) by $\sigma(H_{R}) = \bigcup_{l=1}^{\infty} \mathcal{E}_{R,l}$ where $\forall l \geq 1$, $\mathcal{E}_{R,l} :=
[\min_{\mathbf{k} \in \Omega_{R}^{*}}E_{R,l}(\mathbf{k}),\max_{\mathbf{k} \in \Omega_{R}^{*}} E_{R,l}(\mathbf{k})]$ stands for the $l$-th Bloch band function. Note that the sets $\mathcal{E}_{R,l}$ can overlap each other in many ways, and some of them can coincide. The energy bands are disjoint unions of $\mathcal{E}_{R,l}$'s. If $\max \mathcal{E}_{R,l} < \min\mathcal{E}_{R,l+1} $ for some $l\geq 1$, then a spectral gap occurs. Since the Bethe-Sommerfeld conjecture holds true under our conditions, see e.g.
\cite[Coro. 2.3]{HM}, then the number of spectral gaps is finite, if not zero.\\

Our main result below states that, under our conditions and in the tight-binding situation, the Fermi energy always lies in the middle of a spectral gap of $H_{R}$ (in other words, only the insulating situation can occur, see \cite[Thm. 1.1]{BCS2}), and moreover, it provides an asymptotic expansion:

\begin{proposition}
\label{locferm}
Suppose that assumptions $\mathrm{(\mathscr{A}_{r})}$, $\mathrm{(\mathscr{A}_{m})}$ and $\mathrm{(\mathscr{A}_{nd})}$ hold, and assume that the number of particles $n_{0} \in \mathbb{N}^{*}$ in the Wigner-Seitz cell is fixed and satisfies $n_{0} \leq \tau$, while the density is given by \eqref{fisdeni}.\\
$\mathrm{(i)}$. For any $\beta >0$, let $\mu_{R}^{(0)}(\beta,\rho_{0}(R),b=0) \in \mathbb{R}$ be the unique solution of the equation $\rho_{R}(\beta,\mu,b=0)=\rho_{0}(R)$. Then for $R$ sufficiently large,
\begin{equation}
\label{EnFer}
\mathscr{E}_{R}(\rho_{0}(R)) := \lim_{\beta \uparrow \infty} \mu_{R}^{(0)}(\beta,\rho_{0}(R),b=0) = \frac{\max \mathcal{E}_{R,n_{0}} + \min \mathcal{E}_{R,n_{0}+1}}{2}<0.
\end{equation}
$\mathrm{(ii)}$. Define:
\begin{equation}
\label{defEF}
\mathscr{E}_{P}(n_{0}) :=  \frac{1}{2} \times \left\{\begin{array}{ll} (\lambda_{n_{0}} + \lambda_{n_{0}+1})\,\,\,&\textrm{if $n_{0} < \tau$},\\
\lambda_{\tau}\,\,\, &\textrm{if $n_{0}=\tau$.}\end{array}\right.
\end{equation}
Then, under the additional assumption that $n_{0} < \tau$, one has:
\begin{equation*}
\label{compasyEf}
\mathscr{E}_{R}(\rho_{0}(R)) =  \mathscr{E}_{P}(n_{0}) + \mathcal{O}\left(\mathrm{e}^{- \sqrt{\vert \lambda_{n_{0}+1}\vert}R }\right).
\end{equation*}
\end{proposition}

\begin{remark} $\mathrm{(i)}$. We stress the point that the non-degeneracy assumption ($\mathscr{A}_{\mathrm{nd}}$) along with the fact that $\rho_{0}(R)$ is given as in \eqref{fisdeni}, together imply the insulating situation for $R$ sufficiently large.\\
$\mathrm{(ii)}$. We will see in Proposition \ref{PRO4} that the Fermi energy plays in fact no role in the statements of Theorem \ref{rewLTyp} since it can be removed from the quantities of interest without changing their values.
\end{remark}

The rest of this paragraph is devoted to the proof of Proposition \ref{locferm}.\\
Let us start by writing down an expression for the bulk density of particles. Under the grand-canonical conditions, let $\beta>0$ and $\mu \in \mathbb{R}$. $\forall R>0$, let $\mathscr{C}_{\beta}^{(R)}$ be the positively oriented simple contour around $[\inf \sigma(H_{R}),\infty)$ defined as:
\begin{gather}
\label{cbeta}
\mathscr{C}_{\beta}^{(R)} := \left\{ \Re \xi \in [\delta_{R},\infty),\Im\xi = \pm \frac{\pi}{2\beta}\right\} \cup \left\{ \Re \xi = \delta_{R},\Im\xi \in \left[-\frac{\pi}{2\beta},\frac{\pi}{2\beta}\right]\right\}, \\
\delta_{R} := \inf \sigma(H_{R}) - 1. \nonumber
\end{gather}
Let us mention that, for any $R>0$, the closed subset surrounding by $\mathscr{C}_{\beta}^{(R)}$ is a strict subset of the holomorphic domain $\mathfrak{D}:= \{ \zeta \in \mathbb{C}: \Im \zeta \in (-\frac{\pi}{\beta},\frac{\pi}{\beta})\}$ of the Fermi-Dirac distribution function $\mathfrak{f}_{FD}(\beta,\mu;\xi) := \mathrm{e}^{\beta(\mu - \xi)}(1 + \mathrm{e}^{\beta(\mu - \xi)})^{-1}$. From \eqref{densitylim}, and seen as a function of the $\mu$-variable, the bulk zero-field density of particles  reads $\forall \beta>0$, $\forall \mu \in \mathbb{R}$, $\forall R>0$ as, see e.g.
\cite[Eq. (6.3)]{BS}:
\begin{equation}
\label{bulkdens}
\rho_{R}(\beta,\mu,b=0)\\
:= \frac{1}{\vert \Omega_{R}\vert} \frac{i}{2\pi} \mathrm{Tr}_{L^{2}(\mathbb{R}^{3})} \left\{\chi_{\Omega_{R}}\bigg(\int_{\mathscr{C}_{\beta}^{(R)}} \mathrm{d}\xi\, \mathfrak{f}_{FD}(\beta,\mu;\xi) (H_{R} - \xi)^{-1}\bigg)\chi_{\Omega_{R}}\right\}.
\end{equation}
Another way to express the bulk zero-field density consists in bringing into play the integrated density of states of $H_{R}$. Under the conditions of \eqref{bulkdens},
\begin{equation*}
\rho_{R}(\beta,\mu,b=0) = - \int_{-\infty}^{\infty} \mathrm{d}t\, \frac{\partial \mathfrak{f}_{FD}}{\partial t}(\beta,\mu;t) N_{R}(t),
\end{equation*}
where $N_{R}$ is the integrated density of states of the operator $H_{R}=H_{R}(b=0)$ defined in \eqref{IDS}. We recall that, when the magnetic field vanishes, $N_{R}$ is a positive, continuous and non-decreasing function, and $N_{R}$ is piecewise constant when the energy parameter belongs to a spectral gap.\\
Next, in order to write down a convenient expression for the bulk zero-field density of particles in the zero-temperature limit, we need to rewrite \eqref{bulkdens} by the use of the Bloch-Floquet decomposition, see
\cite[Sec. 2]{BCS2}. Using the notation introduced previously, all the needed results are collected in:

\begin{lema}
\label{densityBF}
$\mathrm{(i)}$. Let $\beta > 0$ and $\mu \in \mathbb{R}$. Then for any $R>0$:
\begin{equation*}
\rho_{R}(\beta,\mu,b=0) = \frac{1}{\vert \Omega_{R} \vert \vert \Omega_{R}^{*}\vert} \sum_{j=1}^{\infty} \int_{\Omega_{R}^{*}} \mathrm{d}\mathbf{k}\, \mathfrak{f}_{FD}(\beta,\mu; E_{R,j}(\mathbf{k})).
\end{equation*}
$\mathrm{(ii)}$. For any $R>0$, let $\mu \geq \inf\sigma(H_{R})$ be fixed. One has the identity:
\begin{equation}
\label{betaroinf}
\lim_{\beta \uparrow \infty} \rho_{R}(\beta,\mu,b=0) = \frac{1}{\vert \Omega_{R}\vert \vert \Omega_{R}^{*}\vert} \sum_{j=1}^{\infty} \int_{\Omega_{R}^{*}} \mathrm{d}\mathbf{k}\, \chi_{[\inf\sigma(H_{R}),\mu]}(E_{R,j}(\mathbf{k})) = N_{R}(\mu),
\end{equation}
where $\chi_{[\inf\sigma(H_{R}),\mu]}$ denotes the indicator function of the compact interval $[\inf\sigma(H_{R}),\mu]$, and $N_{R}$ the integrated density of states of the operator $H_{R}$.
\end{lema}

Now, let us get back to the location of the Fermi energy in the tight-binding situation when the density is given by \eqref{fisdeni}. We need first to know how the negative spectral bands of the operator $H_{R}$ are localized at negative eigenvalues of the operator $H_{P}$ for large values of the $R$-parameter. For completeness' sake, in the below lemma, we disregard assumption ($\mathscr{A}_{\mathrm{nd}}$) and we allow the negative eigenvalues of $H_{P}$ to be (possibly) finitely degenerate:

\begin{lema}
\label{lemRsuf}
Let $\tau \in \mathbb{N}^{*}$ be the number of the eigenvalues of $H_{P}$ in $(-\infty,0)$ counting multiplicities. Denote by $\{\lambda_{l}\}_{l=1}^{\nu}$, $\nu \leq \tau$ the set of distinct eigenvalues counting in increasing order. Then, for $R$ sufficiently large, there exist $2\nu + 1$ real numbers: $c_{R,l}, d_{R,l}<0$ with $l =1,\ldots, \nu$ and $C_{R,\nu+1}$ satisfying:
\begin{equation*}
- \lambda_{1} - \frac{1}{2} < c_{R,1}<d_{R,1} < \dotsb < c_{R,\nu} < d_{R,\nu} < C_{R,\nu+1}, \quad \textrm{s.t.}
\end{equation*}
\begin{itemize}
\item [$\mathrm{(i)}$] $\sigma(H_{R})_{\upharpoonright(-\infty,0)} \subset \bigcup_{l=1}^{\nu} [c_{R,l},d_{R,l}]$,
\item [$\mathrm{(ii)}$] $[c_{R,l},d_{R,l}] \cap [c_{R,m},d_{R,m}] = \emptyset$ if $l \neq m$,
\item [$\mathrm{(iii)}$] $\lambda_{l} \in (c_{R,l},d_{R,l})$,
\item [$\mathrm{(iv)}$] $d_{R,\nu} + C_{R,\nu+1} < 0$;
\end{itemize}
together with the properties that $c_{R,l}, d_{R,l} \rightarrow \lambda_{l}$ and $C_{R,\nu+1} \rightarrow 0$ when $R \uparrow \infty$.\\
Furthermore, one has the following relations with the Bloch bands of $H_{R}$:
\begin{gather*}
[c_{R,l},d_{R,l}] \cap \sigma(H_{R}) = \bigcup_{m=1}^{\mathrm{dim} \mathpzc{E}_{l}} \mathcal{E}_{R,m}, \quad l=1,\ldots,\nu,\\
C_{R,\nu+1} = \min \mathcal{E}_{R,\tau+1}.
\end{gather*}
Here, $\mathpzc{E}_{l}$ denotes the eigenspace associated with the degenerate eigenvalue $\lambda_{l}$.
\end{lema}

\noindent \textit{\textbf{Proof}}. The lemma  follows from \cite[Thm. 2.1]{Dau} taking into account the labeling in increasing order for the $E_{R,l}$'s. Note that \cite[Thm. 2.1]{Dau} is stated under the assumptions that $V_{R}$ is smooth and sufficiently fast decaying at infinity. But the statements still hold true under our conditions on the potential $V_{R}$, see \cite[Thm. 2]{GOP}. \qed \\

Now we are ready for:\\
\textit{Proof of Proposition \ref{locferm}.} Let us first prove $\mathrm{(i)}$. Consider the equation:
\begin{equation*}
\frac{1}{\vert \Omega_{R}^{*}\vert} \sum_{l \geq 1} \int_{\Omega_{R}^{*}} \mathrm{d}\mathbf{k}\, \chi_{[\inf \sigma (H_{R}),E]}(E_{R,l}(\mathbf{k})) = n_{0}.
\end{equation*}
Due to the assumption ($\mathscr{A}_{\mathrm{nd}}$), Lemma \ref{lemRsuf} ensures that the Bloch bands $\mathcal{E}_{R,l}$, $l=1,\ldots,\tau$ are simple, isolated from each other and from the rest of the spectrum for $R$ sufficiently large. Hence, if $n_{0} \in \mathbb{N}^{*}$ satisfies $n_{0} \leq \tau$, then $E$ must belong to $[\max_{\mathbf{k} \in \Omega_{R}^{*}} E_{R,n_{0}}(\mathbf{k}), \min_{\mathbf{k} \in \Omega_{R}^{*}} E_{R,n_{0}+1}(\mathbf{k})]$. This comes from the fact that the Lebesgue-measure of the set $\{\mathbf{k} \in \Omega_{R}^{*}: E_{R,l}(\mathbf{k}) \leq E\}$ equals $\vert \Omega_{R}^{*}\vert$ if and only if $E\geq \max_{\mathbf{k} \in \Omega_{R}^{*}} E_{R,l}(\mathbf{k})$. Getting back to \eqref{betaroinf}, this means that for $R$ sufficiently large:
\begin{equation}
\label{eqfongl}
\forall E \in [\max \mathcal{E}_{R,n_{0}}, \min \mathcal{E}_{R,n_{0}+1}], \quad N_{R}(E) = \rho_{0}(R).
\end{equation}
In view of \eqref{eqfongl}, it remains to use \cite[Thm. 1.1]{BCS2} what leads to \eqref{EnFer}.\\
Let us turn to $\mathrm{(ii)}$. Assume that $n_{0} < \tau$. It is enough to use that:
\begin{equation*}
\mathscr{E}_{R}(\rho_{0}(R)) - \mathscr{E}_{P}(n_{0}) = \frac{\max \mathcal{E}_{R,n_{0}} - \lambda_{n_{0}}}{2} + \frac{\min \mathcal{E}_{R,n_{0}+1} - \lambda_{n_{0}+1}}{2},
\end{equation*}
along with the estimate which holds uniformly in $\mathbf{k}$, see e.g.
\cite[Thm. 2]{GOP}:
\begin{equation*}
\left\vert \sqrt{\vert E_{R,l}(\mathbf{k})\vert} - \sqrt{\vert \lambda_{l}\vert} \right\vert = \mathcal{O}\left(\frac{\mathrm{e}^{- \sqrt{\vert \lambda_{l}\vert} R}}{R}\right),\quad l=1,\ldots,\tau. \tag*{\qed}
\end{equation*}

\begin{remark}
\label{rem1}
From the foregoing, one can infer a sufficient condition which leads to the insulating situation when allowing the eigenvalues of $H_{P}$ in $(-\infty,0)$ to be finitely degenerate. From Lemma \ref{lemRsuf}, then for $R$ sufficiently large, \eqref{eqfongl} holds $\forall E$ belonging to a spectral gap of $H_{R}$ provided that:
\begin{equation*}
\exists \varkappa \in \{1,\ldots,\nu\}\quad s.t.\quad n_{0} = \sum_{l=1}^{\varkappa} \mathrm{dim}\,\mathpzc{E}_{l}.
\end{equation*}
\end{remark}

\subsubsection{Isolating the main $R$-dependent contribution at zero-temperature.}
\label{Fgh12}

We start by writing down an expression for the bulk zero-field orbital susceptibility. In the grand-canonical situation, let $\beta>0$ and $\mu \in \mathbb{R}$. For any $R>0$, let $\mathscr{C}_{\beta}^{(R)}$ be the positively oriented simple contour around $[\inf \sigma(H_{R}),\infty)$ defined in \eqref{cbeta}. The closed subset surrounding by $\mathscr{C}_{\beta}^{(R)}$ is a strict subset of the holomorphic domain $\mathfrak{D} := \{ \zeta \in \mathbb{C}: \Im \zeta \in (-\frac{\pi}{\beta},\frac{\pi}{\beta})\}$ of $\xi \mapsto \mathfrak{f}(\beta,\mu;\xi) := \ln(1 + \mathrm{e}^{\beta(\mu -  \xi)})$ satisfying $-\frac{1}{\beta} (\partial_{\xi}\mathfrak{f})(\beta,\mu;\xi) = \mathfrak{f}_{FD} (\beta,\mu;\xi)$. Note that $\mathfrak{f}(\beta,\mu;\cdot\,)$ admits an exponential decay on $\mathscr{C}_{\beta}^{(R)}$, i.e. there exists a constant $c>0$ s.t.
\begin{equation}
\label{expdeC}
\forall \beta>0,\,\forall \xi \in \mathscr{C}_{\beta}^{(R)},\quad \vert \mathfrak{f}(\beta,\mu;\xi)\vert \leq c \mathrm{e}^{\beta \mu} \mathrm{e}^{- \beta \Re \xi}.
\end{equation}
From \eqref{susceptilim}, and seen as a function of $\mu$, the bulk zero-field orbital susceptibility reads $\forall \beta>0$, $\forall \mu \in \mathbb{R}$ and $\forall R >0$ as, see e.g.
\cite[Eq. (1.21)]{BS}:
\begin{equation}
\label{ZFOSR}
\begin{split}
\mathcal{X}_{R} (\beta,\mu,b=0)  := &\left(\frac{q}{c}\right)^{2}\frac{2}{\beta \vert \Omega_{R}\vert} \frac{i}{2\pi} \int_{\mathscr{C}_{\beta}^{(R)}} \mathrm{d}\xi\, \mathfrak{f}(\beta,\mu;\xi) \\
&\times \mathrm{Tr}_{L^{2}(\mathbb{R}^{3})}\left\{\chi_{\Omega_{R}}(H_{R} - \xi)^{-1} \left[T_{R,1}(\xi)T_{R,1}(\xi)- T_{R,2}(\xi)\right]\chi_{\Omega_{R}}\right\},
\end{split}
\end{equation}
where $T_{R,j}(\xi)$, $j=1,2$ are bounded operators on $L^{2}(\mathbb{R}^{3})$ generated via their kernel respectively defined on $\mathbb{R}^{6}\setminus D$ ($D:=\{(\bold{x},\bold{y}) \in \mathbb{R}^{6}: \bold{x}=\bold{y}\}$) as:
\begin{gather}
\label{WR1}
T_{R,1}(\mathbf{x},\mathbf{y};\xi) := \mathbf{a}(\mathbf{x}-\mathbf{y}) \cdot (i\nabla_{\mathbf{x}}) (H_{R} - \xi)^{-1}(\mathbf{x},\mathbf{y}),\\
\label{WR2}
T_{R,2}(\mathbf{x},\mathbf{y};\xi) := \frac{1}{2} \mathbf{a}^{2}(\mathbf{x}-\mathbf{y})(H_{R} - \xi)^{-1}(\mathbf{x},\mathbf{y}).
\end{gather}
Remind that we have introduced the operators $T_{P,j}(\xi)$, $j=1,2$ via their kernel defined similarly to \eqref{WR1}-\eqref{WR2} but with $(H_{P}-\xi)^{-1}$ instead of $(H_{R}-\xi)^{-1}$, see \eqref{WP1}-\eqref{WP2}. Since $\vert \mathbf{a}(\mathbf{x}-\mathbf{y})\vert \leq \vert \mathbf{x}-\mathbf{y}\vert$, then under the conditions of Lemma \ref{lemhgfP} (below $\Xi:=R$ or $P$), one has on $\mathbb{R}^{6}\setminus D$:
\begin{equation}
\label{estWker}
\vert T_{\Xi,j}(\mathbf{x},\mathbf{y};\xi)\vert \leq p(\vert \xi\vert) \frac{\mathrm{e}^{- \vartheta_{\xi}  \vert \mathbf{x} - \mathbf{y}\vert}}{\vert \mathbf{x} - \mathbf{y}\vert}, \quad \vartheta_{\xi}:= \frac{\vartheta}{1 + \vert \xi\vert},\, j=1,2,
\end{equation}
for another constant $\vartheta>0$ and polynomial $p(\cdot\,)$. Due to \eqref{estWker} and \eqref{kertildR}, the operators $(H_{\Xi} - \xi)^{-1} T_{\Xi,1}(\xi) T_{\Xi,1}(\xi)$ and $(H_{\Xi} - \xi)^{-1} T_{\Xi,2}(\xi)$, $\Xi=R$ or $P$ are locally trace class on $L^{2}(\mathbb{R}^{3})$. Furthermore, both are integral operators with a jointly continuous integral kernel on $\mathbb{R}^{6}$, whose diagonal part is bounded above by some polynomial in $\vert \xi\vert$ uniformly in the spatial variable, see e.g. \cite[Lem. A.1]{BS}. This along with \eqref{expdeC} ensure that \eqref{ZFOSR} is well-defined.\\

Now, let us turn to the proof of \eqref{thmfond}. We point out that the main difficulty consists in isolating the main $R$-dependent contribution from \eqref{ZFOSR} in the tight-binding situation, while keeping a good control on the behavior of the 'remainder' term, even in the zero-temperature limit.\\

\indent The starting point is the approximation of the resolvent $(H_{R}-\xi)^{-1}$ derived in Sec. \ref{GPT}. By replacing in \eqref{ZFOSR} each resolvent $(H_{R}-\xi)^{-1}$ (including the ones appearing in \eqref{WR1}-\eqref{WR2}) with the r.h.s. of \eqref{complr}, and taking into account the features of the three last operators in the r.h.s. of \eqref{complr} we discussed in Sec. \ref{GPT}, we naturally expect the main $R$-dependent contribution from \eqref{ZFOSR} to be obtained  by replacing each $(H_{R}-\xi)^{-1}$ with $(H_{P}-\xi)^{-1}$. In this way, define $\forall \beta >0$, $\forall\mu \in \mathbb{R}$ and $\forall R>0$, the following quantities:
\begin{equation}
\label{ZFOSP}
\begin{split}
\widetilde{\mathcal{X}}_{R}(\beta,\mu,b=0) := &\left(\frac{q}{c}\right)^{2}\frac{1}{\beta} \frac{1}{\vert \Omega_{R}\vert} \frac{i}{\pi} \int_{\mathscr{C}_{\beta}^{(P)}} \mathrm{d}\xi\, \mathfrak{f}(\beta,\mu;\xi) \\ &\times \mathrm{Tr}_{L^{2}(\mathbb{R}^{3})}\left\{\chi_{\Omega_{R}}(H_{P} - \xi)^{-1}\left[T_{P,1}(\xi)T_{P,1}(\xi) - T_{P,2}(\xi)\right]\chi_{\Omega_{R}}\right\},
\end{split}
\end{equation}
\begin{equation}
\label{calGRbm}
\Delta_{R}(\beta,\mu) := \mathcal{X}_{R}(\beta,\mu,b=0) - \widetilde{\mathcal{X}}_{R}(\beta,\mu,b=0),
\end{equation}
where $\mathscr{C}_{\beta}^{(P)}$ appearing in \eqref{ZFOSP} denotes the counter-clockwise oriented simple contour around the interval $[\inf \sigma(H_{P}),\infty)$ defined by:
\begin{gather}
\label{cbetaP}
\mathscr{C}_{\beta}^{(P)} := \left\{ \Re \xi \in [\delta_{P},\infty), \Im\xi = \pm \frac{\pi}{2\beta}\right\} \cup \left\{ \Re \xi = \delta_{P}, \Im\xi \in \left[-\frac{\pi}{2\beta},\frac{\pi}{2\beta}\right]\right\} \\
\delta_{P} := \inf \sigma(H_{P}) - 1 = \lambda_{1} - 1. \nonumber
\end{gather}

\indent Now, we switch to the canonical conditions. Assume that the number of particles $n_{0} \in \mathbb{N}^{*}$ in the Wigner-Seitz cell  is fixed and obeys $n_{0} \leq \tau$, while the density is given by \eqref{fisdeni}. Let $\mu_{R}^{(0)}(\beta,\rho_{0}(R),b=0) \in \mathbb{R}$ be the unique solution of the equation $\rho_{R}(\beta,\mathrm{e}^{\beta \mu},b=0)=  \rho_{0}(R)$. Then, from \eqref{ZFOSrho} together with \eqref{ZFOSP}-\eqref{calGRbm}, one has $\forall \beta>0$ and $\forall R>0$:
\begin{equation}
\label{aprlimb}
\mathcal{X}_{R}\left(\beta,\rho_{0}(R),b=0\right) = \widetilde{\mathcal{X}}_{R}\left(\beta,\mu_{R}^{(0)}(\beta,\rho_{0}(R),b=0),b=0\right)
+ \Delta_{R}\left(\beta,\mu_{R}^{(0)}(\beta,\rho_{0}(R),b=0)\right).
\end{equation}
The next step of the proof consists in performing the zero-temperature limit in \eqref{aprlimb} in the tight-binding situation. Here, the crucial point is the insulating situation: for $R$ sufficiently large, the Fermi energy lies outside the spectrum of $H_{R}$ and is located in a neighborhood of the middle of the interval $(\lambda_{n_{0}},\lambda_{n_{0}+1})$ if $n_{0}<\tau$, $(\lambda_{\tau},0)$ otherwise; see Proposition \ref{locferm} along with Lemma \ref{lemRsuf}. Remind that the insulating situation results from the non-degeneracy assumption $(\mathscr{A}_{\mathrm{nd}})$ together with the condition \eqref{fisdeni}. To perform the zero-temperature limit in \eqref{aprlimb}, we need the two following propositions. Their proof are placed in Appendix, see Sec. \ref{append1}. Recall that $\{\lambda_{l}\}_{l=1}^{\tau}$, $\tau \in \mathbb{N}^{*}$ denotes the set of eigenvalues of $H_{P}$ in $(-\infty,0)$ counting in increasing order.

\begin{proposition}
\label{PRO1}
Let $I_{\varsigma}$, $\varsigma \in \{1,\ldots,\tau\}$ be an open interval s.t. $I_{\varsigma} \subsetneq (\lambda_{\varsigma}, \lambda_{\varsigma+1})$ and $\frac{\lambda_{\varsigma} + \lambda_{\varsigma+1}}{2} \in I_{\varsigma}$ if $\varsigma < \tau$; otherwise $I_{\tau} \subsetneq (\lambda_{\tau}, 0)$ and $\frac{\lambda_{\tau}}{2} \in I_{\tau}$.\\ Then, for any $R>0$ and for any compact subset $K \subset I_{\varsigma}$:
\begin{equation*}
\lim_{\beta \uparrow \infty} \widetilde{\mathcal{X}}_{R}(\beta,\mu,b=0) = \frac{1}{\vert \Omega_{R}\vert} \mathscr{X}_{R}(\mu,b=0),
\end{equation*}
uniformly in $\mu \in K$, with:
\begin{equation*}
\begin{split}
\mathscr{X}_{R}(\mu,b=0):= &\left(\frac{q}{c}\right)^{2}\frac{i}{\pi} \int_{\Gamma_{\varsigma}} \mathrm{d}\xi\, (\mu - \xi) \\
&\times \mathrm{Tr}_{L^{2}(\mathbb{R}^{3})}\left\{\chi_{\Omega_{R}}(H_{P} - \xi)^{-1}\left[T_{P,1}(\xi) T_{P,1}(\xi) - T_{P,2}(\xi)\right]\chi_{\Omega_{R}}\right\}.
\end{split}
\end{equation*}
$\Gamma_{\varsigma}$ is any positively oriented simple closed contour surrounding the $\varsigma$ smallest eigenvalues of $H_{P}$ in $(-\infty,0)$ while letting outside the rest of the spectrum.
\end{proposition}

\begin{proposition}
\label{PRO2}
Let $I_{\varsigma}$, $\varsigma \in \{1,\ldots,\tau\}$ be an open interval as in Proposition \ref{PRO1}. Then, for $R$ sufficiently large, $\lim_{\beta \uparrow \infty} \Delta_{R}(\beta,\mu)$ exists uniformly on compact subsets $K \subset I_{\varsigma}$. Denote it by $\Delta_{R}(\mu)$. Furthermore, $\forall 0< \alpha <1$ there exist two constants $c,C>0$ s.t. $\forall \mu \in I_{\varsigma}$ and for $R$ sufficiently large:
\begin{equation}
\label{supesmu}
\vert \Delta_{R}(\mu)\vert \leq C (1 + \vert \mu \vert) \mathrm{e}^{-c R^{\alpha}}.
\end{equation}
\end{proposition}

Let us emphasize that the exponentially decaying estimate appearing in \eqref{supesmu} arises from the fact that $u$ is compactly supported, see assumption ($\mathscr{A}_{\mathrm{r}}$). The proof of Proposition \ref{PRO2} is essentially based on the features of the three last operators in the r.h.s. of \eqref{complr} we mentioned in Sec. \ref{GPT}.\\

\indent Subsequently to Propositions \ref{PRO1} and \ref{PRO2}, we are in a position to isolate a first main $R$-dependent contribution from \eqref{aprlimb} in the tight-binding situation and in the zero-temperature regime. Under the conditions of \eqref{aprlimb}, let us show that $\forall 0< \alpha <1$, there exists a $R$-independent constant $c>0$ s.t.
\begin{equation}
\label{expandZFOS}
\lim_{\beta \uparrow \infty} \mathcal{X}_{R}(\beta,\rho_{0}(R),b=0) = \frac{1}{\vert \Omega_{R}\vert} \mathscr{X}_{R}\left(\mathscr{E}_{R}(\rho_{0}(R)),b=0\right) + \mathcal{O}\left(\mathrm{e}^{-c R^{\alpha}}\right),
\end{equation}
with:
\begin{equation}
\label{tildZFOSP}
\begin{split}
\mathscr{X}_{R}\left(\mathscr{E}_{R}(\rho_{0}(R)),b=0\right):= &\left(\frac{q}{c}\right)^{2} \frac{i}{\pi} \int_{\Gamma_{n_{0}}} \mathrm{d}\xi\, (\mathscr{E}_{R}(\rho_{0}(R)) - \xi) \\
\times &\mathrm{Tr}_{L^{2}(\mathbb{R}^{3})}\left\{\chi_{\Omega_{R}}(H_{P} - \xi)^{-1} \left[T_{P,1}(\xi)T_{P,1}(\xi)- T_{P,2}(\xi)\right]\chi_{\Omega_{R}}\right\}.
\end{split}
\end{equation}
To achieve that, let $I_{n_{0}}$ be an open interval s.t. $I_{n_{0}} \subsetneq (\lambda_{n_{0}}, \lambda_{n_{0}+1})$ and
$\frac{\lambda_{n_{0}} + \lambda_{n_{0}+1}}{2} \in I_{n_{0}}$ if $n_{0}<\tau$; otherwise
$I_{n_{0}} \subsetneq (\lambda_{\tau}, 0)$ and $\frac{\lambda_{\tau}}{2} \in I_{n_{0}}$. From Lemma
\ref{lemRsuf} and Proposition \ref{locferm}, for $R$ sufficiently large,
$\overline{I}_{n_{0}} \cap \sigma(H_{R}) = \emptyset$ and the Fermi energy
$\mathscr{E}_{R}(\rho_{0}(R)):= \lim_{\beta \uparrow \infty} \mu_{R}^{(0)}(\beta,\rho_{0}(R),b=0) \in I_{n_{0}}$ respectively. Then, \eqref{expandZFOS} follows from Propositions \ref{PRO1} and \ref{PRO2} together.

\begin{remark} In the case of $n_{0} < \tau$, due to the asymptotic expansion in Lemma \ref{locferm} $\mathrm{(ii)}$ along with \eqref{supesmu}, then one obtains instead of \eqref{tildZFOSP}:
\begin{equation*}
\lim_{\beta \uparrow \infty} \mathcal{X}_{R}(\beta,\rho_{0}(R),b=0) = \frac{1}{\vert \Omega_{R}\vert} \mathscr{X}_{R}\left(\mathscr{E}_{P}(n_{0}),b=0\right) + \mathcal{O}\left(\mathrm{e}^{-c R^{\alpha}}\right),
\end{equation*}
with $\mathscr{X}_{R}(\mathscr{E}_{P}(n_{0}),b=0)$ as in \eqref{tildZFOSP} but with $\mathscr{E}_{P}(n_{0})$ instead of $\mathscr{E}_{R}(\rho_{0}(R))$.
\end{remark}

\subsubsection{Isolating the main $R$-dependent contribution at zero-temperature - Continuation and end.}
\label{Fgh13}

The continuation of the proof of \eqref{thmfond} consists in removing the $R$-dependence arising from the presence of the indicator functions $\chi_{\Omega_{R}}$ inside the trace in \eqref{tildZFOSP}.\\
\indent The key-ingredient is Proposition \ref{PRO3} below whose proof lies in Appendix, see Sec. \ref{append2}. The exponential localization of the eigenfunctions associated with the eigenvalues of $H_{P}$ in $(-\infty,0)$ plays a crucial role in the proof.\\

Introduce the family of functions $\mathfrak{g}_{\theta,w} : \mathbb{C} \rightarrow \mathbb{C}$, $\theta \in \mathbb{C}$, $w=0,1$, defined as:
\begin{equation}
\label{frakg}
\mathfrak{g}_{\theta,1}(\xi) := \theta - \xi,\quad \mathfrak{g}_{\theta,0}(\xi) := \theta,\quad (\xi,\theta) \in \mathbb{C}^{2}.
\end{equation}

\begin{proposition}
\label{PRO3}
$\forall \theta \in \mathbb{C}$, $\forall w \in \{0,1\}$ and $\forall \varsigma \in\{1,\ldots,\tau\}$, there exist two constants $C= C(\theta,\varsigma)>0$ and $c>0$ s.t. $\forall \mathbf{x} \in \mathbb{R}^{3}$ and for any $j \in \{1,2\}$:
\begin{equation}
\label{vvvv2}
\begin{split}
&\max\left\{\left\vert \frac{i}{2\pi}  \int_{\Gamma_{\varsigma}} \mathrm{d}\xi\, \mathfrak{g}_{\theta,w}(\xi) \left\{(H_{P} - \xi)^{-1} T_{P,1}(\xi)T_{P,1}(\xi)\right\}(\mathbf{x},\mathbf{x})\right\vert \right. ,\\
& \left. \left\vert \frac{i}{2\pi}  \int_{\Gamma_{\varsigma}} \mathrm{d}\xi\, \mathfrak{g}_{\theta,w}(\xi) \left\{(H_{P} - \xi)^{-1} T_{P,j}(\xi)\right\}(\mathbf{x},\mathbf{x})\right\vert\right\}
\leq C \mathrm{e}^{-c \vert \mathbf{x}\vert}.
\end{split}
\end{equation}
$\Gamma_{\varsigma}$ is any positively oriented simple closed contour surrounding the $\varsigma$ smallest eigenvalues of $H_{P}$ in $(-\infty,0)$, while letting outside the rest of the spectrum.
\end{proposition}

As a result of Proposition \ref{PRO3}, the error made in getting the trace out the integration w.r.t. $\xi$ in \eqref{tildZFOSP} and extending the trace to the whole space behaves like $\mathcal{O}(\mathrm{e}^{-cR})$ for some $R$-independent constant $c>0$. Hence, under the conditions of \eqref{tildZFOSP}, there exists a $R$-independent $c>0$ s.t.:
\begin{equation}
\label{vvvv}
\begin{split}
\mathscr{X}_{R}(\mathscr{E}_{R}(\rho_{0}(R)),b=0) =  &\left(\frac{q}{c}\right)^{2}\frac{i}{\pi} \mathrm{Tr}_{L^{2}(\mathbb{R}^{3})}\biggl\{ \int_{\Gamma_{n_{0}}}  \mathrm{d}\xi\, \left(\mathscr{E}_{R}(\rho_{0}(R)) - \xi\right)  \\
&\times (H_{P} - \xi)^{-1} \left[T_{P,1}(\xi)T_{P,1}(\xi)- T_{P,2}(\xi)\right] \biggr\} + \mathcal{O}\left(\mathrm{e}^{-c R}\right).
\end{split}
\end{equation}

Gathering \eqref{expandZFOS}, \eqref{tildZFOSP} and \eqref{vvvv} together, then to complete the proof of \eqref{thmfond}-\eqref{defcalLp1}, it remains to show that the quantity containing the Fermi energy inside the trace of the leading term in the r.h.s. of \eqref{vvvv} plays no role (i.e. we can get rid of it without changing the value of the trace). This is contained in the below result, whose proof lies in Appendix, see Sec. \ref{append2}:

\begin{proposition}
\label{PRO4}
With the notation of Proposition \ref{PRO3}, one has:
\begin{equation*}
\frac{i}{2\pi} \mathrm{Tr}_{L^{2}(\mathbb{R}^{3})}\left\{\int_{\Gamma_{\varsigma}} \mathrm{d}\xi\, (H_{P} - \xi)^{-1} \left[T_{P,1}(\xi)T_{P,1}(\xi)- T_{P,2}(\xi)\right]\right\} = 0.
\end{equation*}
\end{proposition}


\subsection{Proof of $\mathrm{(ii)}$.}
\label{defcal}

Before beginning, let us recall and introduce some notation. Under the assumptions ($\mathscr{A}_{\mathrm{r}}$)-($\mathscr{A}_{\mathrm{m}}$)-($\mathscr{A}_{\mathrm{nd}}$), let $\{\lambda_{l}\}_{l=1}^{\tau}$, $\tau \in \mathbb{N}^{*}$ be the set of eigenvalues of $H_{P}$ in $(-\infty,0)$ counting in increasing order. Let $\gamma_{l}$, $l=1,\ldots,\tau$ be positively oriented simple closed contours assumed to be two by two disjoint, chosen in such a way that $\forall l \in \{1,\ldots,\tau\}$, $\gamma_{l}$ surrounds the eigenvalue $\lambda_{l}$, while letting outside the rest of the spectrum of $H_{P}$. Let $H_{P}(b)$, $b \in \mathbb{R}$ be the magnetic 'single atom' operator in \eqref{HamiWb}. From \cite[Thm. 1.1]{BC} (see also \cite{N1,HS}), there exists $\mathfrak{b}_{0}>0$ s.t. $\forall \vert b \vert \leq \mathfrak{b}_{0}$, $\cup_{l=1}^{\tau} \gamma_{l} \in \varrho(H_{P}(b))$ (the resolvent set). Moreover, by \cite[Thm. 6.1]{AHS}, the eigenvalues $\lambda_{l}$, $l=1,\ldots,\tau$ are stable under the perturbation $H_{P}(b)-H_{P}$ for
$b$ sufficiently small. Due to the assumption ($\mathscr{A}_{\mathrm{nd}}$), this means that there exists $\mathfrak{b}_{1}>0$ s.t. $\forall \vert b\vert \leq \mathfrak{b}_{1}$, $H_{P}(b)$ has exactly one and only one eigenvalue $\lambda_{l}(b)$ nearby $\lambda_{l}$, $l=1,\ldots,\tau$. Furthermore, by the asymptotic perturbation theory, $\forall l \in \{1,\ldots,\tau\}$ $\lambda_{l}(b)$ can be written in the first order in $b$ as $\lambda_{l}(b) = \lambda_{l} + b e_{l} + o(b)$, see e.g.
\cite[Sec. VIII, Thm. 2.6]{K}. Actually, the $\lambda_{l}(b)$'s can be written in terms of an asymptotic power series in $b$, see e.g. \cite[Thm. 1.2]{BC}. From the foregoing, then there exists
$0 < \mathfrak{b} \leq \min\{\mathfrak{b}_{0},\mathfrak{b}_{1}\}$ s.t. $\forall \vert b \vert \leq \mathfrak{b}$ and $\forall l \in \{1,\ldots,\tau\}$, $\lambda_{l}(b)$ lies inside the closed contour $\gamma_{l}$ surrounding $\lambda_{l}$. Under these conditions, we denote hereafter by $\Pi_{l}(b)$ the orthogonal projection onto the eigenvector corresponding to the eigenvalue $\lambda_{l}(b)$.\\

Let us turn to the proof of \eqref{defcalLp2}. Under the above conditions, the Riesz integral formula allows to write for any $l \in \{1,\ldots,\tau\}$:
\begin{equation*}
\forall \vert b \vert \leq \mathfrak{b},\quad \lambda_{l}(b) \Pi_{l}(b)  = H_{P}(b) \Pi_{l}(b) =  \frac{i}{2\pi} \int_{\gamma_{l}} \mathrm{d}\xi\, \xi (H_{P}(b) - \xi)^{-1}.
\end{equation*}
Since $\mathrm{dim}\, \mathrm{Ran}(\Pi_{l}(b))=1$ by stability of the $\lambda_{l}$'s, then for any $n_{0} \in \{1,\ldots,\tau\}$:
\begin{equation}
\label{avanpr}
\forall \vert b \vert \leq \mathfrak{b},\quad \sum_{l=1}^{n_{0}} \lambda_{l}(b) =  \frac{i}{2\pi} \mathrm{Tr}_{L^{2}(\mathbb{R}^{3})} \left\{\int_{\cup_{l=1}^{n_{0}}\gamma_{l}} \mathrm{d}\xi\, \xi (H_{P}(b) - \xi)^{-1} \right\}.
\end{equation}

Next, the proof of \eqref{defcalLp2} follows from the following result which is concerned with the quantity in the r.h.s. of \eqref{avanpr} (seen as a function of $b$):

\begin{proposition}
\label{dernPr}
Let $\{\mathfrak{g}_{\theta,w},\,\theta \in \mathbb{C}\}$, $w=0,1$ be the families of functions defined in \eqref{frakg}. Then, there exists a neighborhood $\mathcal{I}$ of $b=0$ s.t. $\forall \theta \in \mathbb{C}$ and $\forall w \in \{0,1\}$, the following map:
\begin{equation}
\label{scrF}
b \mapsto \mathscr{F}_{\theta,w}(b) := \frac{i}{2\pi} \mathrm{Tr}_{L^{2}(\mathbb{R}^{3})} \left\{\int_{\cup_{l=1}^{n_{0}}\gamma_{l}} \mathrm{d}\xi\, \mathfrak{g}_{\theta,w}(\xi) (H_{P}(b) - \xi)^{-1} \right\},
\end{equation}
is twice differentiable on $\mathcal{I}$. Moreover, its second derivative at $b=0$ reads as:
\begin{equation}
\label{derscrF}
\frac{\mathrm{d}^{2} \mathscr{F}_{\theta,w}}{\mathrm{d}b^{2}}(b=0) :=
\frac{i}{\pi} \mathrm{Tr}_{L^{2}(\mathbb{R}^{3})} \left\{ \int_{\cup_{l=1}^{n_{0}}\gamma_{l}} \mathrm{d}\xi\, \mathfrak{g}_{\theta,w}(\xi) (H_{P}-\xi)^{-1}\left[T_{P,1}(\xi) T_{P,1}(\xi) - T_{P,2}(\xi)\right]\right\}.
\end{equation}
\end{proposition}

Since the $\lambda_{l}(\cdot\,)$'s are twice differentiable w.r.t. $b$ in a neighborhood of $b=0$, then \eqref{defcalLp2} follows directly from \eqref{avanpr} along with Proposition \ref{dernPr}.\\

The rest of this section is devoted to the proof of Proposition \ref{dernPr}. It is essentially based on the so-called \textit{gauge invariant magnetic perturbation theory}, see
\cite{C0, CN0, C1, N2, BC, BCL, CN3, BCL2, CN1, C2, BCS2, BS, Sa2} and references therein.\\
\indent Before starting, let us introduce some notation.
Define $\forall (\mathbf{x},\mathbf{y}) \in \mathbb{R}^{6}$ the magnetic phase $\phi$ as:
\begin{equation}
\label{phase}
\phi(\mathbf{x},\mathbf{y}) := \frac{1}{2} \mathbf{e}_{3} \cdot (\mathbf{y} \times \mathbf{x}) = - \phi(\mathbf{y},\mathbf{x}), \quad \mathbf{e}_{3} := (0,0,1).
\end{equation}
By \cite[Thm. B.7.2]{Si1}, $\forall b \in \mathbb{R}$ and $\forall \xi \in \varrho(H_{P}(b))$, the resolvent $(H_{P}(b)-\xi)^{-1}$ is an integral operator with integral kernel $(H_{P}(b)-\xi)^{-1}(\cdot\,,\cdot\,)$ jointly continuous on $\mathbb{R}^{6}\setminus D$. Introduce on $L^{2}(\mathbb{R}^{3})$ the operators $T_{P,j}(b,\xi)$, $j=1,2$ via their kernel respectively defined on $\mathbb{R}^{6}\setminus D$ by:
\begin{gather*}
T_{P,1}(\mathbf{x},\mathbf{y};b,\xi) := \mathbf{a}(\mathbf{x} - \mathbf{y}) \cdot (i\nabla_{\mathbf{x}} + b \mathbf{a}(\mathbf{x})) (H_{P}(b)-\xi)^{-1}(\mathbf{x},\mathbf{y}),\\
T_{P,2}(\mathbf{x},\mathbf{y};b,\xi) := \frac{1}{2} \mathbf{a}^{2}(\mathbf{x} - \mathbf{y})  (H_{P}(b)-\xi)^{-1}(\mathbf{x},\mathbf{y}).
\end{gather*}
From \cite[Eq. (2.9)]{BS} together with \cite[Lem. 2.4]{BS}, $\forall \eta>0$ there exists a constant $\vartheta= \vartheta(\eta)>0$ and a polynomial $p(\cdot\,)$ s.t. $\forall b \in \mathbb{R}$, $\forall \xi \in \mathbb{C}$ satisfying $\mathrm{dist}(\xi, \sigma(H_{P}(b))) \geq \eta$ and $\forall(\mathbf{x},\mathbf{y})\in\mathbb{R}^{6}\setminus D$:
\begin{gather}
\label{reesti1}
\left\vert (H_{P}(b) - \xi)^{-1}(\mathbf{x},\mathbf{y})\right\vert \leq p(\vert \xi\vert) \frac{\mathrm{e}^{- \vartheta_{\xi} \vert \mathbf{x} - \mathbf{y}\vert}}{\vert \mathbf{x} - \mathbf{y}\vert},\quad \vartheta_{\xi} := \vartheta (1 + \vert \xi\vert)^{-1},\\
\label{reesti2}
\left\vert T_{P,j}(\mathbf{x},\mathbf{y};b,\xi)\right\vert \leq p(\vert \xi\vert)(1+\vert b \vert)^{3} \frac{\mathrm{e}^{- \vartheta_{\xi} \vert \mathbf{x} - \mathbf{y}\vert}}{\vert \mathbf{x} - \mathbf{y}\vert}, \quad j=1,2.
\end{gather}
By the Shur-Holmgren criterion, the $T_{P,j}(\xi)$'s are bounded on $L^{2}(\mathbb{R}^{3})$:
\begin{equation}
\label{L2bound}
\left\Vert (H_{P}(b) - \xi)^{-1} \right\Vert \leq p(\vert \xi\vert),\quad \left\Vert  T_{P,j}(b,\xi) \right\Vert \leq p(\vert \xi\vert)(1 + \vert b \vert)^{3},
\end{equation}
for another polynomial $p(\cdot\,)$. Let $j \in \mathbb{N}^{*}$ and $\bold{i}:=(i_{1},\ldots,i_{j}) \in \{1,2\}^{j}$. For any $k \in \mathbb{N}^{*}$ with $k \geq j \geq 1$, let $\chi_{j}^{k}(\bold{i})$ be the characteristic function given by:
\begin{equation*}
\chi_{j}^{k}(\mathbf{i}) := \left\{\begin{array}{ll}
1 &\textrm{if $i_{1}+\dotsb + i_{j} = k$}, \\
0 &\textrm{otherwise}
\end{array}\right., \quad 1 \leq j\leq k.
\end{equation*}
Restricting to $k \in \{1,2\}$, define $\forall m\in \{0,1\}$ and $\forall b \in \mathbb{R}$ the function on $\mathbb{R}^{6}$:
\begin{equation}
\label{calTPkm}
\begin{split}
\mathfrak{T}_{P,k}^{m}(\mathbf{x},\mathbf{y};b,\xi) := \sum_{j=1}^{k} (-1)^{j} &\sum_{\mathbf{i} \in \{1,2\}^{j}} \chi_{j}^{k}(\mathbf{i}) \int_{\mathbb{R}^{3}} \mathrm{d}\mathbf{z}_{1} \dotsb \int_{\mathbb{R}^{3}} \mathrm{d}\mathbf{z}_{j}
(i\phi(\mathbf{z}_{j},\mathbf{y}) - i \phi(\mathbf{z}_{j},\mathbf{x}))^{m} \\ &\times (H_{P}(b)-\xi)^{-1}(\mathbf{x},\mathbf{z}_{1})  T_{P,i_{1}}(\mathbf{z}_{1},\mathbf{z}_{2};b,\xi) \dotsb T_{P,i_{j}}(\mathbf{z}_{j},\mathbf{y};b,\xi),
\end{split}
\end{equation}
where by convention, we set $0^{0} = 1$. Note that, due to the antisymmetry of $\phi$, the terms in the r.h.s. of \eqref{calTPkm} containing the magnetic phases vanish when $\mathbf{x}=\mathbf{y}$. Moreover, $\forall \eta>0$, $\forall b \in \mathbb{R}$ and $\forall \xi \in \mathbb{C}$ satisfying $\mathrm{dist}(\xi, \sigma(H_{P}(b))) \geq \eta$, $\mathfrak{T}_{P,k}^{m}(\cdot\,,\cdot\,;b,\xi)$ is jointly continuous on $\mathbb{R}^{6}$. This follows by applying $j$-times \cite[Lem. A.1]{BS} together with \eqref{reesti1}-\eqref{reesti2}. Furthermore, from \eqref{phase}, \eqref{reesti1}-\eqref{reesti2} along with \cite[Lem. A.2 (ii)]{BS}, there exists a $b$-independent polynomial $p(\cdot\,)$ s.t. for any $k\in \{1,2\}$, $m \in \{0,1\}$ and $\forall (\mathbf{x},\mathbf{y})\in \mathbb{R}^{6}$:
\begin{equation}
\label{frakTes}
\vert \mathfrak{T}_{P,k}^{m}(\mathbf{x},\mathbf{y};b,\xi) \vert \leq p(\vert \xi\vert)(1 + \vert b \vert)^{6} \times \left\{\begin{array}{ll}
(\vert \mathbf{x}\vert^{m} + \vert \mathbf{y}\vert^{m})\,\,\,&\textrm{if $\mathbf{x} \neq \mathbf{y}$,} \\
1 &\textrm{if $\mathbf{x} = \mathbf{y}$,}\end{array}\right.
\end{equation}
where, in the case of $\mathbf{x}\neq \mathbf{y}$, we used the rough estimate:
\begin{equation*}
\forall (\mathbf{x},\mathbf{y}) \in \mathbb{R}^{6},\quad \vert \phi(\mathbf{x},\mathbf{y})\vert \leq \vert \mathbf{y}\vert \vert \mathbf{x}-\mathbf{y}\vert.
\end{equation*}

\begin{remark}
\label{explicit}
In view of \eqref{calTPkm}, one has on $\mathbb{R}^{3}$:
\begin{gather*}
\mathfrak{T}_{P,1}^{0}(\mathbf{x},\mathbf{x};b,\xi) = - \int_{\mathbb{R}^{3}} \mathrm{d}\mathbf{z}\, (H_{P}(b)-\xi)^{-1}(\mathbf{x},\mathbf{z}) T_{P,1}(\mathbf{z},\mathbf{x};b,\xi),\\
\mathfrak{T}_{P,1}^{1}(\mathbf{x},\mathbf{x};b,\xi) = 0,\\
\begin{split}
&\mathfrak{T}_{P,2}^{0}(\mathbf{x},\mathbf{x};b,\xi) = - \int_{\mathbb{R}^{3}} \mathrm{d}\mathbf{z}\, (H_{P}(b)-\xi)^{-1}(\mathbf{x},\mathbf{z}) T_{P,2}(\mathbf{z},\mathbf{x};b,\xi) \\
&+ \int_{\mathbb{R}^{3}} \mathrm{d}\mathbf{z}_{1} \int_{\mathbb{R}^{3}} \mathrm{d}\mathbf{z}_{2}\, (H_{P}(b)-\xi)^{-1}(\mathbf{x},\mathbf{z}_{1}) T_{P,1}(\mathbf{z}_{1},\mathbf{z}_{2};b,\xi) T_{P,1}(\mathbf{z}_{2},\mathbf{x};b,\xi).
\end{split}
\end{gather*}
\end{remark}

Now, let us turn to the proof of Proposition \ref{dernPr}. It requires the two following technical  results whose proofs lie in Appendix, see Sec. \ref{append3}.\\
\indent The first one deals with the regularity of the integral kernel of $(H_{P}(b)-\xi)^{-1}$, seen as a function of the $b$-variable, in a neighborhood of $b=0$:

\begin{lema}
\label{oubli}
Let $K \subset \left(\varrho(H_{P}) \cap \left\{\zeta \in \mathbb{C}: \Re \zeta < 0\right\}\right)$ be a compact subset. Let $\mathfrak{b}_{K}>0$ s.t. $\forall \vert b \vert \leq \mathfrak{b}_{K}$, $K \subset \left(\varrho(H_{P}(b)) \cap \left\{\zeta \in \mathbb{C}: \Re \zeta < 0\right\}\right)$.\\
Then, $\forall \xi \in K$ and $\forall(\mathbf{x},\mathbf{y}) \in \mathbb{R}^{6}\setminus D$, the map $b \mapsto (H_{P}(b)-\xi)^{-1}(\mathbf{x},\mathbf{y})$ is twice differentiable on the interval $(-\mathfrak{b}_{K},\mathfrak{b}_{K})$. Furthermore, its first two derivatives at $b_{0} \in (-\mathfrak{b}_{K},\mathfrak{b}_{K})$ read as:
\begin{equation*}
\left.\frac{\partial^{s}}{\partial b^{s}}(H_{P}(b) - \xi)^{-1}(\mathbf{x},\mathbf{y})\right\vert_{b=b_{0}} =
(i\phi(\mathbf{x},\mathbf{y}))^{s} (H_{P}(b_{0})-\xi)^{-1}(\mathbf{x},\mathbf{y}) \\
 + s \sum_{k=1}^{s} \mathfrak{T}_{P,k}^{s-k}(\mathbf{x},\mathbf{y};b_{0},\xi),\quad s=1,2,
\end{equation*}
where the functions $\mathfrak{T}_{P,k}^{m}(\cdot\,,\cdot\,;b_{0},\xi)$, $k=1,2$, $m=0,1$ are defined in \eqref{calTPkm}.
\end{lema}

The second one deals with the 'uniform' exponential localization of the eigenfunctions associated with the $\lambda_{l}(b)$'s, $l=1,\ldots,\tau$ for $b$ sufficiently small:

\begin{lema}
\label{stabb}
Let $\lambda_{l}$, $l \in \{1,\ldots,\tau\}$ be a simple eigenvalue of $H_{P}$ in $(-\infty,0)$. Let $b$ sufficiently small s.t. $H_{P}(b)$ has exactly one and only one eigenvalue $\lambda_{l}(b)$ near $\lambda_{l}$. Denote by $\Phi_{l}(\cdot\,;b)$ the associated (normalized) eigenfunction. Then, there exists $\mathfrak{b}>0$ and two constants $c,C>0$ s.t. for any $k \in \{1,2,3\}$:
\begin{equation}
\label{exp2}
\forall \vert b \vert\leq \mathfrak{b},\, \forall \mathbf{x} \in \mathbb{R}^{3},\quad \max\left\{\left\vert \Phi_{l}(\mathbf{x};b)\right\vert, \left\vert \partial_{x_{k}} \Phi_{l}(\mathbf{x};b)\right\vert\right\} \leq C \mathrm{e}^{-c \vert \mathbf{x}\vert}.
\end{equation}
\end{lema}

The exponential decay for the $\Phi_{l}(\cdot\,;b)$'s is a well-known result, see e.g. \cite[Thm. 4.4]{I} and also \cite[Thm. 1.10]{Ip}, \cite[Sec. 7.2]{FH}. For $b$ sufficiently small, we stress the point that all the constants can be chosen $b$-independent. As a result of Lemma \ref{stabb}, one directly gets as a corollary of Proposition \ref{PRO3}:

\begin{corollary}
\label{coro1}
Let $\{\lambda_{l}\}_{l=1}^{\tau}$ be the set of eigenvalues of $H_{P}$ in $(-\infty,0)$. Let $\mathfrak{b}>0$ s.t. $\forall \vert b\vert \leq \mathfrak{b}$ and $\forall l \in \{1,\ldots,\tau\}$:\\
$\mathrm{(i)}$. $H_{P}(b)$ has exactly one and only one eigenvalue $\lambda_{l}(b)$ located nearby $\lambda_{l}$.\\
$\mathrm{(ii)}$. The (normalized) eigenfunction associated with $\lambda_{l}(b)$ obeys estimate \eqref{exp2}.\\
Then $\forall \theta \in \mathbb{C}$, $\forall w \in \{0,1\}$ and $\forall \varsigma \in \{1,\ldots,\tau\}$, there exist two constants $C= C(\theta,\varsigma)>0$ and $c>0$ s.t. $\forall \vert b \vert \leq \mathfrak{b}$, $\forall \mathbf{x} \in \mathbb{R}^{3}$ and for any $j \in \{1,2\}$:
\begin{equation}
\label{vvvv3}
\begin{split}
&\max\left\{\left\vert \frac{i}{2\pi}  \int_{\Gamma_{\varsigma}} \mathrm{d}\xi\, \mathfrak{g}_{\theta,w}(\xi) \left\{(H_{P}(b) - \xi)^{-1} T_{P,j}(b,\xi)\right\}(\mathbf{x},\mathbf{x})\right\vert, \right.\\
& \left. \left\vert \frac{i}{2\pi}  \int_{\Gamma_{\varsigma}} \mathrm{d}\xi\, \mathfrak{g}_{\theta,w}(\xi) \left\{(H_{P}(b) - \xi)^{-1}  T_{P,1}(b,\xi)T_{P,1}(b,\xi) \right\} (\mathbf{x},\mathbf{x})\right\vert \right\} \leq C (1 + \vert b \vert)^{6} \mathrm{e}^{-c \vert \mathbf{x}\vert},
\end{split}
\end{equation}
where $\mathfrak{g}_{\theta,w}$ are the maps in \eqref{frakg} and $\Gamma_{\varsigma}$ a contour as in Proposition \ref{PRO3}.
\end{corollary}

Note that the presence of the factor $(1 + \vert b\vert)^{6}$ in the upper bound in \eqref{vvvv3} comes from the estimate \eqref{reesti2} (the estimate in \eqref{reesti1} is $b$-independent).\\

\indent We are now ready for:
\noindent \textit{\textbf{Proof of Proposition \ref{dernPr}}}. Let $\theta \in \mathbb{C}$ and $w \in \{0,1\}$. Let $\mathfrak{b}>0$ s.t. $\forall \vert b\vert \leq \mathfrak{b}$, \eqref{avanpr} holds. Let us first prove that:
\begin{equation*}
\label{opcdf}
\forall \vert b \vert \leq \mathfrak{b},\quad \mathcal{F}_{\theta,w}(b) := \frac{i}{2\pi} \int_{\cup_{l=1}^{n_{0}} \gamma_{l}} \mathrm{d}\xi\, \mathfrak{g}_{\theta,w}(\xi) (H_{P}(b) - \xi)^{-1},
\end{equation*}
has an integral kernel jointly continuous on $\mathbb{R}^{6}$. By the Dunford-Pettis theorem, $\mathcal{F}_{\theta,w}(b)$ is an integral operator since $(H_{P}(b)-\xi)^{-1}$ is bounded from $L^{2}(\mathbb{R}^{3})$ to $L^{\infty}(\mathbb{R}^{3})$ by some polynomial in $\vert \xi\vert$, see \eqref{reesti1}. Let $\xi_{0} < \inf \sigma(H_{P})$ and large enough s.t. $\xi_{0} < \min_{1\leq l\leq \tau} \{\gamma_{l} \cap \mathbb{R}\}$. By the first resolvent equation:
\begin{equation*}
\begin{split}
\mathcal{F}_{\theta,w}(b) = &\frac{i}{2\pi} \left(\int_{\cup_{l=1}^{n_{0}} \gamma_{l}} \mathrm{d}\xi\, \mathfrak{g}_{\theta,w}(\xi) \right) (H_{P}(b) - \xi_{0})^{-1} \\
&+ \frac{i}{2\pi} \int_{\cup_{l=1}^{n_{0}} \gamma_{l}} \mathrm{d}\xi\, \mathfrak{g}_{\theta,w}(\xi) (\xi-\xi_{0}) (H_{P}(b) - \xi)^{-1} (H_{P}(b) - \xi_{0})^{-1}.
\end{split}
\end{equation*}
By the Cauchy-Goursat theorem, the first term in the above r.h.s. is identically zero. The second term has a jointly continuous integral kernel on $\mathbb{R}^{6}$ due to \cite[Lem. A.1]{BS} along with \eqref{reesti1}. Denoting by $\mathcal{F}_{\theta,w}(\cdot\,,\cdot\,;b)$ the integral kernel of $\mathcal{F}_{\theta,w}(b)$, its diagonal part reads $\forall b \in (-\mathfrak{b},\mathfrak{b})$ as:
\begin{equation}
\label{diagke}
\forall \bold{x} \in \mathbb{R}^{3},\quad  \mathcal{F}_{\theta,w}(\mathbf{x},\mathbf{x};b) :=
\frac{i}{2\pi} \left. \left(\int_{\cup_{l=1}^{n_{0}} \gamma_{l}} \mathrm{d}\xi\, \mathfrak{g}_{\theta,w}(\xi) (H_{P}(b) - \xi)^{-1}(\mathbf{x},\mathbf{y})\right)\right\vert_{\mathbf{y}=\mathbf{x}}.
\end{equation}
Next, from Lemma \ref{oubli} together with \eqref{frakTes}, one can prove that for any $\bold{x} \in \mathbb{R}^{3}$, the map $b \mapsto \mathcal{F}_{\theta,w}(\mathbf{x},\mathbf{x};b)$ is twice differentiable on $(-\mathfrak{b},\mathfrak{b})$. Moreover, its first two derivatives at $b_{0} \in (-\mathfrak{b},\mathfrak{b})$ satisfy (below $s=1,2$):
\begin{equation*}
\forall \mathbf{x} \in \mathbb{R}^{3},\quad \frac{\partial^{s} \mathcal{F}_{\theta,w}}{\partial b^{s}}(\mathbf{x},\mathbf{x};b_{0})
:=  s \frac{i}{2\pi} \int_{\cup_{l=1}^{n_{0}} \gamma_{l}} \mathrm{d}\xi\, \mathfrak{g}_{\theta,w}(\xi) \sum_{k=1}^{s} \mathfrak{T}_{P,k}^{s-k}(\mathbf{x},\mathbf{x};b_{0},\xi).
\end{equation*}
Here, we used that $\phi(\mathbf{x},\mathbf{x})=0$. Finally, let $0<\hat{\mathfrak{b}}\leq \mathfrak{b}$ s.t. $\forall \vert b\vert \leq \hat{\mathfrak{b}}$ and for any $l\in\{1,\ldots,n_{0}\}$, the (normalized) eigenfunction associated with $\lambda_{l}(b)$ obeys an estimate of type \eqref{exp2}. From the explicit expressions in Remark \ref{explicit} together with the estimate \eqref{vvvv3}, then for any compact subset $K \subset (-\hat{\mathfrak{b}},\hat{\mathfrak{b}})$, there exist two constants $c>0$ and $C= C(n_{0},\theta,K)>0$ s.t.
\begin{equation*}
\forall \mathbf{x} \in \mathbb{R}^{3},\quad \sup_{b \in K} \left\vert \frac{\partial^{s} \mathcal{F}_{\theta,w}}{\partial b^{s}}(\mathbf{x},\mathbf{x};b)\right\vert \leq C \mathrm{e}^{- c \vert \mathbf{x}\vert}, \quad s=1,2.
\end{equation*}
Since the upper bound belongs to $L^{1}(\mathbb{R}^{3})$, the proposition follows. \qed

\subsection{Proof of $\mathrm{(iii)}$.}
\label{Fesh}

Let us recall some notation. Under the assumptions ($\mathscr{A}_{\mathrm{r}}$)-($\mathscr{A}_{\mathrm{m}}$)-($\mathscr{A}_{\mathrm{nd}}$), let $\{\lambda_{l}\}_{l=1}^{\tau}$ be the set of eigenvalues of $H_{P}$ in $(-\infty,0)$ counting in increasing order. For any $l \in \{1,\ldots,\tau\}$, denote by $\Phi_{l}$ the normalized eigenfunction associated with $\lambda_{l}$, and by $\Pi_{l}= \vert \Phi_{l}\rangle \langle \Phi_{l}\vert$ the orthogonal projection onto $\Phi_{l}$. We define $\Pi_{l}^{\perp}:= \mathbbm{1} - \Pi_{l}$. From
\cite[Thm. 6.1]{AHS}, there exists $\mathfrak{b}_{1}>0$ s.t. $\forall \vert b \vert \leq \mathfrak{b}_{1}$ and $\forall l \in \{1,\ldots,\tau\}$, $\lambda_{l}$ is stable under the perturbation $W(b):= H_{P}(b) - H_{P} = -b \mathbf{a}\cdot (-i\nabla) + \frac{1}{2} b^{2} \mathbf{a}^{2}$. This means that $\forall \vert b \vert \leq \mathfrak{b}_{1}$, $H_{P}(b)$ has exactly one and only one eigenvalue $\lambda_{l}(b)$ nearby $\lambda_{l}$ which reduces to $\lambda_{l}$ in the limit $b\rightarrow 0$. For such $b$'s and any $l \in \{1,\ldots,\tau\}$, denote by $\Phi_{l}(b)$ the normalized eigenfunction associated with $\lambda_{l}(b)$, and by $\Pi_{l}(b)=\vert \Phi_{l}(b) \rangle \langle \Phi_{l}(b)\vert$ the orthogonal projection onto the eigenvector $\Phi_{l}(b)$. We define $\Pi_{l}^{\perp}(b) := \mathbbm{1} - \Pi_{l}(b)$.\\

Let us turn to the proof of $\mathrm{(iii)}$. Let $l \in \{1,\ldots,\tau\}$ and $K$ be a compact neighborhood of $\lambda_{l}$. Let $0<\mathfrak{b}\leq\mathfrak{b}_{1}$ s.t. $\forall \vert b\vert \leq \mathfrak{b}$, $K \cap \big(\sigma (H_{P}(b)) \setminus \lambda_{l}(b)\big) = \emptyset$ and $\lambda_{l}(b) \in K$. From the Feshbach formula in \cite{Fesch} and under our conditions, $\forall \vert b \vert \leq \mathfrak{b}$ and small enough, $\lambda_{l}(b)$ is the unique number $\zeta$ near $\lambda_{l}$ for which:
\begin{equation}
\label{quant1}
(\lambda_{l} - \zeta)\Pi_{l} + \Pi_{l}W(b)\Pi_{l} - \Pi_{l}W(b)\left\{\Pi_{l}^{\perp}(H_{P}+W(b) - \zeta)\Pi_{l}^{\perp}\right\}^{-1}W(b)\Pi_{l},
\end{equation}
is not invertible. Note that $W(b)\Pi_{l}$ is bounded on $L^{2}(\mathbb{R}^{3})$, and its operator norm behaves like $\mathcal{O}(\vert b\vert)$. This follows from the exponential localization of $\Phi_{l}$ and $\nabla \Phi_{l}$ in \eqref{exp2}. We point out that $W(b)\Pi_{l}$ remains bounded even with an exponential weight. More precisely, there exists $\varepsilon_{0}>0$ and a constant $C>0$ s.t.
\begin{equation}
\label{weight}
\forall 0<\varepsilon \leq \varepsilon_{0}, \quad \left\Vert W(b)\Pi_{l} \mathrm{e}^{\varepsilon \langle \cdot\,\rangle}\right\Vert \leq C\vert b\vert, \quad \langle \cdot\,\rangle := \sqrt{1 + \vert \cdot\,\vert^{2}}.
\end{equation}
Now, let us justify that the operator $\Pi_{l}^{\perp}(H_{P}(b) - \xi)\Pi_{l}^{\perp}$ is invertible $\forall \vert b\vert \leq \mathfrak{b}$ and small enough, and $\forall \xi \in K$. To do so, introduce the Sz-Nagy transformation in \cite[Sec. I.4.6]{K} corresponding to the pair of projections $\Pi_{l}(b)$, $\Pi_{l}$:
\begin{equation}
\label{Nagy}
U(b) = \left(\mathbbm{1} - (\Pi_{l}(b)- \Pi_{l})^{2}\right)^{-\frac{1}{2}}\left\{\Pi_{l}(b)\Pi_{l} + \Pi_{l}^{\perp}(b)\Pi_{l}^{\perp}\right\}.
\end{equation}
Since $\Pi_{l}(b)$ converges to $\Pi_{l}$ in norm as $b \rightarrow 0$, see e.g.
\cite[Sec. VIII.2]{K}, the square-root in \eqref{Nagy} is well-defined by a binomial series. $U(b)$ is a unitary operator, and it intertwines both projections:
\begin{equation}
\label{intertw}
\Pi_{l}(b) = U(b) \Pi_{l} U^{*}(b).
\end{equation}
From \eqref{intertw}, one gets $\Pi_{l}^{\perp}(b) = U(b)\Pi_{l}^{\perp}U^{*}(b)$ what leads to $U^{*}(b) \Pi_{l}^{\perp}(b) = \Pi_{l}^{\perp}U^{*}(b)$. As a result, $\forall\vert b \vert \leq \mathfrak{b}$ and small enough, and for any $\xi \in K$:
\begin{equation}
\label{unite}
U(b) \Pi_{l}^{\perp}(H_{P}(b) - \xi)\Pi_{l}^{\perp}U^{*}(b) = \Pi_{l}^{\perp}(b) U(b) (H_{P}(b) - \xi) U^{*}(b) \Pi_{l}^{\perp}(b),
\end{equation}
i.e. $\Pi_{l}^{\perp}(H_{P}(b) - \xi)\Pi_{l}^{\perp}$ is unitarily equivalent with the operator in the r.h.s. of \eqref{unite}. Next, from the asymptotic perturbation theory in \cite[Sec. VIII.2.4]{K}:
\begin{gather*}
U(b) = \mathbbm{1} + \left\{\Pi_{l}W(b)(H_{P}-\xi)^{-1}(\mathbbm{1}-\Pi_{l}) - (\mathbbm{1}-\Pi_{l})(H_{P}-\xi)^{-1}W(b)\Pi_{l}\right\}+ \mathfrak{u}_{1}(b),\\
U^{*}(b) = \mathbbm{1} + \left\{(\mathbbm{1}-\Pi_{l})(H_{P}-\xi)^{-1}W(b)\Pi_{l} - \Pi_{l}W(b)(H_{P}-\xi)^{-1}(\mathbbm{1}-\Pi_{l})\right\}
+ \mathfrak{u}_{2}(b),
\end{gather*}
where the $\mathfrak{u}_{j}(b)$'s, $j=1,2$ are operators satisfying $b^{-1}\mathfrak{u}_{j}(b)  \rightarrow 0$ in the strong sense. Putting these asymptotic expansions into \eqref{unite}, then we obtain:
\begin{equation*}
U(b) \Pi_{l}^{\perp}(H_{P}(b) - \xi)\Pi_{l}^{\perp}U^{*}(b) = \Pi_{l}^{\perp}(b) (H_{P}(b) - \xi) \Pi_{l}^{\perp}(b)\left[\mathbbm{1} + \Upsilon(\xi,b)\right],
\end{equation*}
where $\Upsilon(\xi,b)$ is an operator s.t. $\Vert \Upsilon(\xi,b)\Vert =\mathcal{O}(\vert b\vert)$ uniformly in $\xi$. It follows that $\Pi_{l}^{\perp}(H_{P}(b) - \xi)\Pi_{l}^{\perp}$ is invertible $\forall \vert b\vert \leq \mathfrak{b}$ and small enough, and $\forall \xi \in K$.\\
\indent Let us get back to the quantity in \eqref{quant1}. By using the scalar product, $\forall l \in \{1,\ldots,\tau\}$ and $\forall\vert b \vert \leq \mathfrak{b}$ and small enough, $\lambda_{l}(b)$ has to obey  the equation:
\begin{equation}
\label{comre}
\lambda_{l}(b) = \lambda_{l} + \left\langle \Phi_{l}, W(b) \Phi_{l}\right\rangle
 - \left\langle W(b) \Phi_{l}, \left(\Pi_{l}^{\perp}\left(H_{P}(b) - \lambda_{l}(b)\right) \Pi_{l}^{\perp}\right)^{-1} W(b) \Phi_{l}\right\rangle.
\end{equation}
By iterating the identity in \eqref{comre}, one obtains under the same conditions:
\begin{equation}
\label{Fesch2}
\lambda_{l}(b) = \lambda_{l} + \left\langle \Phi_{l}, W(b) \Phi_{l}\right\rangle
- \left\langle W(b) \Phi_{l}, \left(\Pi_{l}^{\perp}\left(H_{P} - \lambda_{l}\right) \Pi_{l}^{\perp}\right)^{-1} W(b) \Phi_{l} \right\rangle + \mathcal{O}\left(\vert b\vert^{3}\right).
\end{equation}
To control the behavior in $b$ of the remainder term, we used that:
\begin{equation*}
\left\Vert \mathrm{e}^{-\varepsilon\langle \cdot\,\rangle} \left(\Pi_{l}^{\perp}\left(H_{P}(b) - \lambda_{l}(b)\right)\Pi_{l}^{\perp}\right)^{-1} - \left(\Pi_{l}^{\perp}\left(H_{P}  - \lambda_{l}\right)\Pi_{l}^{\perp}\right)^{-1} \mathrm{e}^{-\varepsilon\langle \cdot\,\rangle}\right\Vert = \mathcal{O}(\vert b\vert),
\end{equation*}
with $0<\varepsilon \leq \varepsilon_{0}$ as in \eqref{weight}. It can be proved by using an asymptotic expansion for the projection $\Pi_{l}(b)$ together with the application of the gauge invariant magnetic perturbation theory for the kernel of the unperturbed projector as we did for the kernel of the resolvent in Sec. \ref{defcal}. Note that the fourth term involved in the expansion \eqref{Fesch2} can be identified with:
\begin{equation*}
\begin{split}
&- \left\langle \Phi_{l},W(b) \Phi_{l} \right\rangle \left\langle W(b) \Phi_{l}, \left(\Pi_{l}^{\perp}\left(H_{P}- \lambda_{l}\right) \Pi_{l}^{\perp}\right)^{-2} W(b) \Phi_{l} \right\rangle \\
&+ \left\langle W(b) \Phi_{l}, \left(\Pi_{l}^{\perp} \left(H_{P}-\lambda_{l}\right)\Pi_{l}^{\perp}\right)^{-1} W(b) \left(\Pi_{l}^{\perp}\left(H_{P}-\lambda_{l}\right)\Pi_{l}^{\perp}\right)^{-1} W(b) \Phi_{l} \right\rangle.
\end{split}
\end{equation*}
Since $\lambda_{l}(\cdot\,)$ has an asymptotic series expansion for $b$ small enough, then one obtains from \eqref{Fesch2}:
\begin{equation}
\label{dert2}
\frac{\mathrm{d}^{2} \lambda_{l}}{\mathrm{d}b^{2}}(b=0) =  \left\langle \Phi_{l}, \mathbf{a}^{2} \Phi_{l}\right\rangle - 2 \left\langle \mathbf{a} \cdot (-i\nabla) \Phi_{l}, \left(\Pi_{l}^{\perp}\left(H_{P} - \lambda_{l}\right) \Pi_{l}^{\perp}\right)^{-1} \mathbf{a}\cdot (-i\nabla) \Phi_{l} \right\rangle.
\end{equation}
From \eqref{dert2}, the proof of \eqref{deride} directly follows from \eqref{defcalLp2}.

\section{Appendix.}
\label{append}

\subsection{Proof of Propositions \ref{PRO1} and \ref{PRO2}.}
\label{append1}

We start by introducing some notation. Recall that under the assumptions ($\mathscr{A}_{\mathrm{r}}$)-($\mathscr{A}_{\mathrm{m}}$)-($\mathscr{A}_{\mathrm{nd}}$), $\{\lambda_{l}\}_{l=1}^{\tau}$ with $\tau \in \mathbb{N}^{*}$ denotes the set of eigenvalues of $H_{P}$ in $(-\infty,0)$ counting in increasing order. For simplicity's sake, we set $\lambda_{\tau +1} := 0$. Let $I_{\varsigma}$, $\varsigma \in \{1,\ldots,\tau\}$ be an open interval s.t. $I_{\varsigma} \subsetneq (\lambda_{\varsigma}, \lambda_{\varsigma+1})$ and $\frac{\lambda_{\varsigma} + \lambda_{\varsigma+1}}{2} \in I_{\varsigma}$. Without loss of generality, we give an explicit form to $I_{\varsigma}$:
\begin{equation}
\label{expIs}
I_{\varsigma} := \left(\frac{2 \lambda_{\varsigma} + \lambda_{\varsigma + 1}}{3}, \frac{\lambda_{\varsigma} + 2 \lambda_{\varsigma + 1}}{3}\right),\quad \textrm{with the convention $\lambda_{\tau+1}:=0$}.
\end{equation}
\noindent Introduce the following decomposition of the contour $\mathscr{C}_{\beta}^{(P)}$  defined in \eqref{cbetaP}:
\begin{gather}
\label{decomCbeta}
\mathscr{C}_{\beta}^{(P)} =  \hat{\Gamma}_{\varsigma,\beta} \cup \gamma_{\varsigma,\beta}^{(2)} \cup \gamma_{\varsigma,\beta}^{(1)},\quad \varsigma \in \{1,\ldots,\tau\},\\
\gamma_{\varsigma,\beta}^{(1)} := \left\{ \Re \xi \in [\omega_{\varsigma}^{+},\infty),\, \Im\xi = \pm \frac{\pi}{2\beta}\right\} \cup \left\{ \Re \xi = \omega_{\varsigma}^{+},\,\, \Im\xi \in \left[-\frac{\pi}{2\beta},\frac{\pi}{2\beta}\right]\right\},\nonumber\\
\gamma_{\varsigma,\beta}^{(2)} := \left\{ \Re \xi \in [\omega_{\varsigma}^{-},\omega_{\varsigma}^{+}],\, \Im\xi = \pm \frac{\pi}{2\beta}\right\} \cup \left\{ \Re \xi = \omega_{\varsigma}^{\pm},\,\, \Im\xi \in \left[-\frac{\pi}{2\beta},\frac{\pi}{2\beta}\right]\right\}, \nonumber\\
\begin{split}
\hat{\Gamma}_{\varsigma,\beta} := \left\{ \Re \xi \in [\delta_{P},\omega_{\varsigma}^{-}],\, \Im\xi = \pm \frac{\pi}{2\beta}\right\} &\cup \left\{ \Re \xi = \delta_{P},\,\, \Im\xi \in \left[-\frac{\pi}{2\beta},\frac{\pi}{2\beta}\right]\right\} \\
&\cup \left\{ \Re \xi = \omega_{\varsigma}^{-},\,\, \Im\xi \in \left[-\frac{\pi}{2\beta},\frac{\pi}{2\beta}\right]\right\},\nonumber
\end{split}
\end{gather}
where $\omega_{\varsigma}^{-} := \frac{19 \lambda_{\varsigma} + 5\lambda_{\varsigma + 1}}{24}$ and $\omega_{\varsigma}^{+} := \frac{5 \lambda_{\varsigma} + 19\lambda_{\varsigma + 1}}{24}$, with the convention $\lambda_{\tau+1}=0$.\\

Let us turn to the proof of Proposition \ref{PRO1}:

\noindent \textit{\textbf{Proof of Proposition \ref{PRO1}}}. Let $\varsigma \in \{1,\ldots,\tau\}$. From \eqref{ZFOSP} and \eqref{decomCbeta}, $\forall \mu \in I_{\varsigma}$:
\begin{gather}
\label{integr}
\widetilde{\mathcal{X}}_{R}(\beta,\mu,b=0) = \left(\frac{q}{c}\right)^{2}\frac{1}{\beta} \frac{1}{\vert \Omega_{R}\vert} \frac{i}{\pi} \int_{\hat{\Gamma}_{\varsigma,\beta} \cup \gamma_{\varsigma,\beta}^{(2)} \cup \gamma_{\varsigma,\beta}^{(1)}} \mathrm{d}\xi\, \mathfrak{f}(\beta,\mu;\xi) \Theta_{R}(\xi),\\
\Theta_{R}(\xi):= \mathrm{Tr}_{L^{2}(\mathbb{R}^{3})} \left\{\chi_{\Omega_{R}} (H_{P} - \xi)^{-1}\left[T_{P,1}(\xi) T_{P,1}(\xi) - T_{P,2}(\xi)\right] \chi_{\Omega_{R}}\right\},\nonumber
\end{gather}
and \eqref{integr} holds $\forall R>0$. Since $[\omega_{\varsigma}^{-},\omega_{\varsigma}^{+}] \cap \sigma(H_{P}) = \emptyset$, then $\forall \mu \in I_{\varsigma}$ the closed subset surrounding by $\gamma_{\varsigma,\beta}^{(2)}$ is a strict subset of the holomorphic domain of the integrand in \eqref{integr}. Therefore, the Cauchy-Goursat theorem yields:
\begin{equation*}
\int_{\gamma_{\varsigma,\beta}^{(2)}} \mathrm{d}\xi\, \mathfrak{f}(\beta,\mu;\xi)\Theta_{R}(\xi) = 0.
\end{equation*}
Now, the contours $\hat{\Gamma}_{\varsigma,\beta}$ and $\gamma_{\varsigma,\beta}^{(1)}$ can be deformed respectively in $\hat{\Gamma}_{\varsigma,1}$ and $\gamma_{\varsigma,1}^{(1)}$ (set $\beta=1$ in their definition) due to the location of the interval $I_{\varsigma}$. In the wake of the deformation of the contours, then under the conditions of Lemma \ref{lemhgfP}, from \eqref{kertildR} and \eqref{estWker} with $\Xi = P$ and by setting $\eta=1$, there exists a polynomial $p(\cdot\,)$ independent of $\beta$ (and $R$) s.t.
\begin{equation}
\label{trac}
\vert \Theta_{R}(\xi) \vert \leq p(\vert \xi\vert) \vert \Omega_{R}\vert.
\end{equation}
It remains to use the Lebesgue's dominated convergence theorem which yields:
\begin{gather*}
\lim_{\beta \uparrow \infty} \frac{1}{\beta} \int_{\hat{\Gamma}_{\varsigma,1}} \mathrm{d}\xi\,  \mathfrak{f}(\beta,\mu;\xi)\Theta_{R}(\xi) = \int_{\hat{\Gamma}_{\varsigma,1}} \mathrm{d}\xi\,  (\mu-\xi) \Theta_{R}(\xi),\\
\lim_{\beta \uparrow \infty} \frac{1}{\beta}
\int_{\gamma_{\varsigma,1}^{(1)}} \mathrm{d}\xi\,  \mathfrak{f}(\beta,\mu;\xi)\Theta_{R}(\xi)  = 0,
\end{gather*}
where we used the pointwise convergence:
\begin{equation}
\label{pointwis}
\lim_{\beta \uparrow \infty} \frac{1}{\beta} \mathfrak{f}(\beta,\mu;\xi) = (\mu - \xi) \chi_{(-\infty,\mu)}(\Re \xi).
\end{equation}
The uniform convergence on compact subsets $K \subset I_{\varsigma}$ is straightforward. \qed

\begin{remark} We emphasize that the deformation of $\hat{\Gamma}_{\varsigma,\beta}$ and $\gamma_{\varsigma,\beta}^{(1)}$ in some $\beta$-independent contours is crucial to make the estimate in \eqref{trac} $\beta$-independent.
\end{remark}

Next, let us turn to the proof of Proposition \ref{PRO2}. To do that, we need beforehand to introduce some operators. Recall that $\mathcal{R}_{R}(\xi)$ and $\mathscr{R}_{R}(\xi)$, $R\geq R_{0}$ are respectively defined in \eqref{tildSR} and \eqref{rewtildSR2} ($R_{0}$ is defined through \eqref{defR0}). Introduce $\forall R \geq R_{0}$, the bounded operators $\mathcal{T}_{R,j}(\xi)$ and $\mathscr{T}_{R,j}(\xi)$, $j=1,2$ on $L^{2}(\mathbb{R}^{3})$ generated via their kernel respectively defined on $\mathbb{R}^{6}\setminus D$ by:
\begin{gather*}
\mathcal{T}_{R,1}(\mathbf{x},\mathbf{y};\xi) := \mathbf{a}(\mathbf{x} - \mathbf{y})\cdot (i \nabla_{\mathbf{x}}) (\mathcal{R}_{R}(\xi))(\mathbf{x},\mathbf{y}), \\
\mathcal{T}_{R,2}(\mathbf{x},\mathbf{y};\xi) := \frac{1}{2}\mathbf{a}^{2}(\mathbf{x} - \mathbf{y})(\mathcal{R}_{R}(\xi))(\mathbf{x},\mathbf{y}),\\
\mathscr{T}_{R,1}(\mathbf{x},\mathbf{y};\xi) := \mathbf{a}(\mathbf{x} - \mathbf{y}) \cdot (i \nabla_{\mathbf{x}})(\mathscr{R}_{R}(\xi))(\mathbf{x},\mathbf{y}),\\
\mathscr{T}_{R,2}(\mathbf{x},\mathbf{y};\xi):= \frac{1}{2} \mathbf{a}^{2}(\mathbf{x} - \mathbf{y}) (\mathscr{R}_{R}(\xi))(\mathbf{x},\mathbf{y}).
\end{gather*}

The proof relies on the following four lemmas whose proofs can be found in Appendix, Sec. \ref{append4}. Recall that $R_{0},R_{1} \geq 1$ are respectively defined through \eqref{defR0}-\eqref{defR1}. With the shorthand notation introduced in \eqref{shorthand}:

\begin{lema}
\label{lem1ma}
Let $0<\alpha < 1$. For every $\eta>0$, there exists a constant $\vartheta=\vartheta(\eta)>0$ and a polynomial $p(\cdot\,)$ s.t. $\forall R \geq R_{0}$ and $\forall \xi \in \mathbb{C}$ satisfying $\mathrm{dist}(\xi,\sigma(H_{R})\cap \sigma(H_{P})) \geq \eta$:
\begin{equation}
\label{firequi1}
\begin{split}
\frac{1}{\vert \Omega_{R} \vert} &\left\vert \mathrm{Tr}_{L^{2}(\mathbb{R}^{3})}\left\{\chi_{\Omega_{R}} (H_{R} - \xi)^{-1} \left[T_{R,1}(\xi)T_{R,1}(\xi) - T_{R,2}(\xi)\right] \chi_{\Omega_{R}}\right\} \right. \\
&\left. -  \mathrm{Tr}_{L^{2}(\mathbb{R}^{3})}\left\{\chi_{\Omega_{R}} \mathcal{R}_{R}(\xi) \left[\mathcal{T}_{R,1}(\xi) \mathcal{T}_{R,1}(\xi) - \mathcal{T}_{R,2}(\xi)\right]\chi_{\Omega_{R}}\right\} \right\vert
\leq p(\vert \xi \vert) \mathrm{e}^{-\vartheta_{\xi} R^{\alpha}}.
\end{split}
\end{equation}
\end{lema}

\begin{lema}
\label{lem2ma}
Let $0<\alpha < 1$. For every $\eta>0$, there exists a constant $\vartheta=\vartheta(\eta)>0$ and a polynomial $p(\cdot\,)$ s.t. $\forall R \geq \max\{R_{0},R_{1}\}$ and $\forall \xi \in \mathbb{C}$ satisfying $\mathrm{dist}(\xi,\sigma(H_{R})\cap \sigma(H_{P})) \geq \eta$:
\begin{equation}
\label{firequi2}
\frac{1}{\vert \Omega_{R} \vert} \left\vert \mathrm{Tr}_{L^{2}(\mathbb{R}^{3})}\left\{\chi_{\Omega_{R}} \mathcal{R}_{R}(\xi) \mathcal{T}_{R,2}(\xi) \chi_{\Omega_{R}}\right\}
- \mathrm{Tr}_{L^{2}(\mathbb{R}^{3})}\left\{\chi_{\Omega_{R}} \mathscr{R}_{R}(\xi) \mathscr{T}_{R,2}(\xi) \chi_{\Omega_{R}}\right\} \right\vert
\leq p(\vert \xi \vert) \mathrm{e}^{-\vartheta_{\xi} R^{\alpha}}.
\end{equation}
\end{lema}

\begin{lema}
\label{lem3ma}
Let $0<\alpha < 1$. For every $\eta>0$, there exists a constant $\vartheta=\vartheta(\eta)>0$ and a polynomial $p(\cdot\,)$ s.t. $\forall R \geq \max\{R_{0},R_{1}\}$ and $\forall \xi \in \mathbb{C}$ satisfying $\mathrm{dist}(\xi,\sigma(H_{R})\cap \sigma(H_{P})) \geq \eta$:
\begin{equation}
\label{firequi3}
\begin{split}
\frac{1}{\vert \Omega_{R} \vert} &\left\vert \mathrm{Tr}_{L^{2}(\mathbb{R}^{3})}\left\{\chi_{\Omega_{R}} \mathcal{R}_{R}(\xi) \mathcal{T}_{R,1}(\xi) \mathcal{T}_{R,1}(\xi) \chi_{\Omega_{R}}\right\}\right.\\ &\left. -  \mathrm{Tr}_{L^{2}(\mathbb{R}^{3})}\left\{\chi_{\Omega_{R}} \mathscr{R}_{R}(\xi) \mathscr{T}_{R,1}(\xi) \mathscr{T}_{R,1}(\xi) \chi_{\Omega_{R}}\right\} \right\vert
\leq p(\vert \xi \vert) \mathrm{e}^{-\vartheta_{\xi} R^{\alpha}}.
\end{split}
\end{equation}
\end{lema}

\begin{lema}
\label{lem4ma}
Let $0<\alpha < 1$. For every $\eta>0$, there exists a constant $\vartheta=\vartheta(\eta)>0$ and a polynomial $p(\cdot\,)$ s.t. $\forall R \geq R_{0}$ and $\forall \xi \in \mathbb{C}$ satisfying $\mathrm{dist}(\xi,\sigma(H_{R})\cap \sigma(H_{P})) \geq \eta$:
\begin{equation}
\label{firequi4}
\begin{split}
\frac{1}{\vert \Omega_{R} \vert} &\left\vert \mathrm{Tr}_{L^{2}(\mathbb{R}^{3})}\left\{\chi_{\Omega_{R}} \mathscr{R}_{R}(\xi) \left[\mathscr{T}_{R,1}(\xi) \mathscr{T}_{R,1}(\xi) - \mathscr{T}_{R,2}(\xi)\right] \chi_{\Omega_{R}}\right\}\right. \\
&\left. - \mathrm{Tr}_{L^{2}(\mathbb{R}^{3})}\left\{\chi_{\Omega_{R}} (H_{P} - \xi)^{-1}\left[T_{P,1}(\xi)T_{P,1}(\xi) - T_{P,2}(\xi)\right] \chi_{\Omega_{R}}\right\} \right\vert
\leq p(\vert \xi \vert) \mathrm{e}^{- \vartheta_{\xi} R^{\alpha}}.
\end{split}
\end{equation}
\end{lema}

\indent We are now ready for:\\
\textit{Proof of Proposition \ref{PRO2}.} In view of \eqref{ZFOSR} and \eqref{ZFOSP}, then $\forall \beta >0$ and $\forall \mu \in \mathbb{R}$ the quantity $\Delta_{R}(\beta,\mu)$ in \eqref{calGRbm} can be rewritten for $R$ sufficiently large as:
\begin{gather}
\label{introscrA}
\Delta_{R}(\beta,\mu) = \left(\frac{q}{c}\right)^{2} \frac{1}{\beta} \frac{1}{\vert \Omega_{R}\vert} \frac{i}{\pi} \int_{\mathscr{C}_{\beta}^{(P)}} \mathrm{d}\xi\, \mathfrak{f}(\beta,\mu;\xi) \mathscr{K}_{R}(\xi),\\
\label{calKR2}
\begin{split}
\mathscr{K}_{R}(\xi) := \mathrm{Tr}_{L^{2}(\mathbb{R}^{3})}&\left\{\chi_{\Omega_{R}}\left\{(H_{R} - \xi)^{-1}\left[T_{R,1}(\xi)T_{R,1}(\xi) - T_{R,2}(\xi)\right]\right.\right. \\
 &\left. \left. - (H_{P} - \xi)^{-1}\left[T_{P,1}(\xi)T_{P,1}(\xi) - T_{P,2}(\xi)\right]\right\} \chi_{\Omega_{R}}\right\}.
\end{split}
\end{gather}
By $R$ sufficiently large, we mean $R \geq R_{2} \geq 1$ chosen in such a way that $\forall R \geq R_{2}$, $\inf \sigma(H_{R}) \geq \lambda_{1} - \frac{1}{2}$. Such $R_{2}$ exists since $\inf \sigma(H_{R})$ has to coincide with $\inf \sigma(H_{P})$ when $R \uparrow \infty$, see Lemma \ref{lemRsuf}. Next, let us estimate the quantity in \eqref{calKR2}. In view of the definition of $\mathscr{C}_{\beta}^{(P)}$ in \eqref{cbetaP}, set $\eta:= \min\{\frac{1}{2},\frac{\pi}{2\beta}\}>0$. Pick $0<\alpha<1$ and set $R_{3}:= \max\{R_{0},R_{1},R_{2}\} \geq 1$. From Lemmas \ref{lem1ma}-\ref{lem4ma}, there exists a constant $\vartheta=\vartheta(\eta)>0$ and a polynomial $p(\cdot\,)$ s.t. $\forall R \geq R_{3}$:
\begin{equation}
\label{estbetadep}
\forall \xi \in \mathscr{C}_{\beta}^{(P)}, \quad \vert \mathscr{K}_{R}(\xi) \vert \leq p(\vert \xi \vert) \vert \Omega_{R}\vert \mathrm{e}^{-\vartheta_{\xi} R^{\alpha}}, \quad \vartheta_{\xi} := \vartheta (1 + \vert \xi\vert)^{-1}.
\end{equation}
We point out that $\vartheta$ and $p(\cdot\,)$ in \eqref{estbetadep} are $\beta$-dependent (at least for large $\beta$).\\
Let $\varsigma \in \{1,\ldots,\tau\}$. From now on, we limit the domain of $\mu$ to the interval $I_{\varsigma}$. Without loss of generality, we give the explicit form \eqref{expIs} to $I_{\varsigma}$. Justified by Lemma \ref{lemRsuf}, we assume that $R \geq R_{3}$ and large enough so that $[\frac{5\lambda_{\varsigma} + \lambda_{\varsigma+1}}{6}, \frac{\lambda_{\varsigma} + 5\lambda_{\varsigma+1}}{6}] \cap \sigma(H_{R})  = \emptyset$ (we set $\lambda_{\tau+1} := 0$). Let $\omega_{\varsigma}^{\pm}$ be the reals defined below \eqref{decomCbeta}. Since $I_{\varsigma} \subset [\omega_{\varsigma}^{-},\omega_{\varsigma}^{+}] \subset [\frac{5\lambda_{\varsigma} + \lambda_{\varsigma+1}}{6}, \frac{\lambda_{\varsigma} + 5\lambda_{\varsigma+1}}{6}]$, then $\forall R \geq R_{3}$ and large enough, $\overline{I}_{\varsigma} \cap \sigma(H_{R}) = \emptyset$ and $\hat{\Gamma}_{\varsigma,\beta}, \gamma_{\varsigma,\beta}^{(1)} \cap \sigma(H_{R}) = \emptyset$. Here $\hat{\Gamma}_{\varsigma,\beta}$ and $\gamma_{\varsigma,\beta}^{(1)}$ are the contours coming from the decomposition of  $\mathscr{C}_{\beta}^{(P)}$ in \eqref{decomCbeta}. From the foregoing, it remains to mimic the proof of Proposition \ref{PRO1} to perform the zero-temperature limit. One arrives, for any compact subset $K \subset I_{\varsigma}$, at:
\begin{equation*}
\Delta_{R}(\mu) := \lim_{\beta \uparrow \infty} \Delta_{R}(\beta,\mu) = \left(\frac{q}{c}\right)^{2} \frac{1}{\vert \Omega_{R}\vert} \frac{i}{\pi} \int_{\hat{\Gamma}_{\varsigma,1}} \mathrm{d}\xi\, (\mu - \xi)  \mathscr{K}_{R}(\xi),
\end{equation*}
uniformly in $\mu \in K$. We emphasize that, as in the proof of Proposition \ref{PRO1}, the deformation of $\hat{\Gamma}_{\varsigma,\beta}$ and $\gamma_{\varsigma,\beta}^{(1)}$ in some $\beta$-independent contours makes the estimate in \eqref{estbetadep} $\beta$-independent on these contours, see above the definition of the $\eta$. Finally, it remains to use \eqref{firequi1}-\eqref{firequi4} along with \eqref{estbetadep} which lead to the existence of two constants $c,C>0$ s.t. $\forall R\geq R_{3}$ and large enough:
\begin{equation*}
\forall \mu \in I_{\varsigma},\quad \vert \Delta_{R}(\mu) \vert \leq C(1+ \vert \mu\vert) \mathrm{e}^{-c R^{\alpha}}. \tag*{\qed}
\end{equation*}

\subsection{Proof of Propositions \ref{PRO3} and \ref{PRO4}.}
\label{append2}

\noindent \textit{\textbf{Proof of Proposition \ref{PRO3}}}. Let $\{\lambda_{l}\}_{l=1}^{\tau}$, $\tau \in \mathbb{N}^{*}$ be the set of eigenvalues of $H_{P}$ in $(-\infty,0)$ counting in increasing order. Let $\varsigma \in \mathbb{N}^{*}$ with $1\leq \varsigma \leq \tau$. For any $l \in \{1,\ldots,\varsigma\}$, denote by $\Pi_{l}$ the orthogonal projection onto the eigenvector corresponding to the eigenvalue $\lambda_{l}$. Recall that $\Pi_{l}$ is given by a Riesz integral:
\begin{equation*}
\Pi_{l} = \frac{i}{2\pi} \int_{\gamma_{l}} \mathrm{d}\xi\, (H_{P} - \xi)^{-1},
\end{equation*}
where $\gamma_{l}$ is any positively oriented closed simple contour surrounding the eigenvalue $\lambda_{l}$, while letting outside the rest of the spectrum of $H_{P}$. Denote by $\Pi_{\aleph} := \sum_{l=1}^{\varsigma} \Pi_{l}$. Introduce the following decomposition of the resolvent:
\begin{gather}
\label{decompproj}
\forall \xi \in \varrho(H_{P}),\quad (H_{P} - \xi)^{-1} = (H_{P} - \xi)_{\aleph}^{-1} + (H_{P} - \xi)_{\perp}^{-1},\\
\label{resprojn0}
(H_{P}-\xi)_{\aleph}^{-1} := \Pi_{\aleph}(H_{P} - \xi)^{-1} = \sum_{l=1}^{\varsigma} \frac{1}{\lambda_{l} - \xi} \Pi_{l},\\
\label{resprojperp}
(H_{P} - \xi)_{\perp}^{-1} := (\mathbbm{1} - \Pi_{\aleph})(H_{P} - \xi)^{-1} = -\frac{i}{2\pi} \int_{\cup_{l=1}^{\varsigma}\gamma_{l}} \frac{\mathrm{d}\zeta }{\zeta - \xi} (H_{P} -\zeta)^{-1}.
\end{gather}
Due to \eqref{resprojn0}, we have in the kernels sense on $\mathbb{R}^{6}$:
\begin{equation}
\label{kerProj}
(H_{P}-\xi)_{\aleph}^{-1}(\mathbf{x},\mathbf{y}) = \sum_{l=1}^{\varsigma} \frac{1}{\lambda_{l} - \xi} \Phi_{l}(\mathbf{x}) \overline{\Phi_{l}(\mathbf{y})},
\end{equation}
where $\Phi_{l}$ denotes the normalized eigenfunction associated with the eigenvalue $\lambda_{l}$. Note that under our conditions, the $\Phi_{l}$'s  with $=1,\ldots,\varsigma$ satisfy the following, see e.g.
\cite[Eq. (5.8)]{Dau}. There exists a constant $C>0$ s.t.
\begin{equation}
\label{fpexpo}
\forall \mathbf{x} \in \mathbb{R}^{3},\quad \max\left\{\vert \Phi_{l}(\mathbf{x}) \vert, \vert \partial_{x_{k}} \Phi_{l}(\mathbf{x}) \vert\right\}  \leq C \mathrm{e}^{-\sqrt{\vert \lambda_{l}\vert} \vert \mathbf{x}\vert},\quad k \in \{1,2,3\}.
\end{equation}
Moreover, denote by $T_{P,j}^{\wp}(\xi)$, $j=1,2$ and $\wp =\aleph,\perp$, the operators on $L^{2}(\mathbb{R}^{3})$ defined via their kernel as in \eqref{WP1}-\eqref{WP2} but with $(H_{P} - \xi)_{\wp}^{-1}$ instead of $(H_{P}-\xi)^{-1}$. Due to \eqref{resprojperp}, the kernel of $T_{P,j}^{\perp}(\xi)$, $j=1,2$ still obey \eqref{estWker}.\\
\indent Let $\theta \in \mathbb{C}$, $w \in \{0,1\}$ and $j\in \{1,2\}$. Due to the decomposition in \eqref{decompproj}, introduce for any $(\wp_{1},\wp_{2}) \in \{\aleph,\perp\}^{2}$ the bounded operator on $L^{2}(\mathbb{R}^{3})$:
\begin{equation}
\label{M2diez}
\mathscr{M}_{\theta,w}^{(j),\wp_{1},\wp_{2}}:= \frac{i}{2\pi} \int_{\Gamma_{\varsigma}} \mathrm{d}\xi\, \mathfrak{g}_{\theta,w}(\xi) (H_{P} - \xi)_{\wp_{1}}^{-1} T_{P,j}^{\wp_{2}}(\xi).
\end{equation}
It is clearly an integral operator with a jointly continuous integral kernel on $\mathbb{R}^{6}$. We denote it by $\mathscr{M}_{\theta,w}^{(j),\wp_{1},\wp_{2}}(\cdot\,,\cdot\,)$. The first part of the proof consists in showing the existence of two constants $c>0$ and $C=C(\theta,\varsigma)>0$ s.t.
\begin{equation*}
\forall \mathbf{x} \in \mathbb{R}^{3},\quad \sum_{(\wp_{1}, \wp_{2}) \in \{\aleph,\perp\}^{2}} \left \vert \mathscr{M}_{\theta,w}^{(j),\wp_{1},\wp_{2}}(\mathbf{x},\mathbf{x}) \right\vert \leq C  \mathrm{e}^{-c \vert \mathbf{x}\vert}.
\end{equation*}
At first, due to the location of $\Gamma_{\varsigma}$, $\mathscr{M}_{\theta,w}^{(j),\perp,\perp}(\mathbf{x},\mathbf{x})= 0$ on $\mathbb{R}^{3}$ by the Cauchy-Goursat theorem. Secondly, from \eqref{kerProj} followed by the residue theorem:
\begin{equation*}
\label{hhhf}
\forall \mathbf{x} \in \mathbb{R}^{3},\quad \mathscr{M}_{\theta,w}^{(j),\aleph,\perp}(\mathbf{x},\mathbf{x}) =   \sum_{l=1}^{\varsigma} \mathfrak{g}_{\theta,w}(\lambda_{l}) \Phi_{l}(\mathbf{x}) \int_{\mathbb{R}^{3}} \mathrm{d}\mathbf{z}\, \overline{\Phi_{l}(\mathbf{z})} T_{P,j}^{\perp}(\mathbf{z},\mathbf{x};\lambda_{l}).
\end{equation*}
Due to \eqref{fpexpo} and \eqref{estWker}, use next that there exists a constant $C>0$ s.t.
\begin{equation*}
\forall 1\leq l \leq \varsigma,\quad \sup_{\mathbf{x} \in \mathbb{R}^{3}} \int_{\mathbb{R}^{3}} \mathrm{d}\mathbf{z}\, \left\vert \overline{\Phi_{l}(\mathbf{z})} T_{P,j}^{\perp}(\mathbf{z},\mathbf{x};\lambda_{l}) \right\vert \leq C.
\end{equation*}
Then, it follows that $\vert \mathscr{M}_{\theta,w}^{(j),\aleph,\perp}(\mathbf{x},\mathbf{x})\vert \leq C \mathrm{e}^{-c \vert \mathbf{x}\vert}$ $\forall \mathbf{x} \in \mathbb{R}^{3}$, for another $c>0$ and $C=C(\theta,\varsigma)>0$. A similar estimate still holds true for $\vert \mathscr{M}_{\theta,w}^{(j),\perp,\aleph}(\mathbf{x},\mathbf{x})\vert$. This comes from the previous arguments, the upper bound \eqref{fpexpo} and the rough estimate $\vert \mathbf{a}(\mathbf{x}-\mathbf{y})\vert \leq (\vert \mathbf{x}\vert + \vert\mathbf{y}\vert)$, combined with:
\begin{equation}
\label{ingr}
\forall \nu\geq 0,\,\forall \mu>0,\quad t^{\nu} \mathrm{e}^{- \mu t} \leq C \mathrm{e}^{- \frac{\mu}{2} t},\quad t\geq 0,
\end{equation}
for another $C=C(\nu,\mu)>0$. The last term we have to treat reads as:
\begin{gather*}
\mathscr{M}_{\theta,1}^{(1),\aleph,\aleph}(\mathbf{x},\mathbf{x}):=  \sum_{l_{1},l_{2}=1}^{\varsigma} \Phi_{l_{1}}(\mathbf{x})  \int_{\mathbb{R}^{3}} \mathrm{d}\mathbf{z}\, \overline{\Phi_{l_{1}}(\mathbf{z})} \mathbf{a}(\mathbf{z}-\mathbf{x})\cdot(i \nabla_{\mathbf{z}} \Phi_{l_{2}})(\mathbf{z}) \overline{\Phi_{l_{2}}(\mathbf{x})},\\
\mathscr{M}_{\theta,1}^{(2),\aleph,\aleph}(\mathbf{x},\mathbf{x}) :=  \sum_{l_{1},l_{2}=1}^{\varsigma} \Phi_{l_{1}}(\mathbf{x})  \int_{\mathbb{R}^{3}} \mathrm{d}\mathbf{z}\, \overline{\Phi_{l_{1}}(\mathbf{z})} \frac{1}{2} \mathbf{a}^{2}(\mathbf{z}-\mathbf{x}) \Phi_{l_{2}}(\mathbf{z}) \overline{\Phi_{l_{2}}(\mathbf{x})},
\end{gather*}
if $w=1$ and $j=1$ or $j=2$ respectively; and if $w=0$ regardless of $j=1,2$:
\begin{equation*}
\mathscr{M}_{\theta,0}^{(j),\aleph,\aleph}(\mathbf{x},\mathbf{x}) = 0;
\end{equation*}
where we used the following identity provided by the residue theorem:
\begin{equation*}
\sum_{l_{1},l_{2}=1}^{\varsigma} \int_{\Gamma_{\varsigma}} \mathrm{d}\xi\, \frac{\mathfrak{g}_{\theta,w}(\xi)}{(\lambda_{l_{1}} - \xi)(\lambda_{l_{2}} - \xi)} = \displaystyle{\left\{\begin{array}{ll}
-2i\pi \sum_{l_{1},l_{2}=1}^{\varsigma}\,\,\,&\textrm{if $w=1$}, \\
0 &\textrm{if $w=0$}
\end{array}\right.}.
\end{equation*}
In the case of $w=1$, it is enough to use \eqref{fpexpo}, the rough estimate $\vert \mathbf{a}(\mathbf{x}-\mathbf{y})\vert \leq (\vert \mathbf{x}\vert + \vert\mathbf{y}\vert)$ combined with
\eqref{ingr}. This leads to $\vert \mathscr{M}_{\theta,1}^{(j),\aleph,\aleph}(\mathbf{x},\mathbf{x}) \vert \leq C \mathrm{e}^{-c \vert \mathbf{x}\vert}$ $\forall \mathbf{x} \in \mathbb{R}^{3}$ and $j=1,2$, for another constant $C=C(\theta,\varsigma)>0$ and $c>0$.\\
\indent Afterwards, we turn to the second part of the proof. Under the conditions of \eqref{M2diez}, introduce $\forall (\wp_{1},\wp_{2},\wp_{3}) \in \{\aleph,\perp\}^{3}$ the operator on $L^{2}(\mathbb{R}^{3})$:
\begin{equation*}
\mathscr{N}_{\theta,w}^{\wp_{1},\wp_{2},\wp_{3}}:= \frac{i}{2\pi} \int_{\Gamma_{\varsigma}} \mathrm{d}\xi\, \mathfrak{g}_{\theta,w}(\xi) (H_{P} - \xi)_{\wp_{1}}^{-1} T_{P,1}^{\wp_{2}}(\xi)T_{P,1}^{\wp_{3}}(\xi).
\end{equation*}
It is an integral operator with a jointly continuous integral kernel on $\mathbb{R}^{6}$. We denote it by $\mathscr{N}_{\theta,w}^{\wp_{1},\wp_{2},\wp_{3}}(\cdot\,,\cdot\,)$. Let us show that there exist $c,C >0$ s.t.
\begin{equation*}
\forall \bold{x} \in \mathbb{R}^{3},\quad \sum_{(\wp_{1}, \wp_{2},\wp_{3}) \in \{\aleph,\perp\}^{3}} \left \vert \mathscr{N}_{\theta,w}^{\wp_{1},\wp_{2},\wp_{3}}(\mathbf{x},\mathbf{x}) \right\vert \leq C  \mathrm{e}^{-c \vert \mathbf{x}\vert}.
\end{equation*}
Firstly, $\mathscr{N}_{\theta,w}^{\perp,\perp,\perp}(\bold{x},\bold{x}) = 0$ on $\mathbb{R}^{3}$ by the Cauchy-Goursat theorem. Also, by a direct calculation, $\mathscr{N}_{\theta,w}^{\aleph,\aleph,\aleph}(\bold{x},\bold{x}) = 0$ since the residue theorem leads to:
\begin{equation*}
\sum_{l_{1},l_{2},l_{3}=1}^{\varsigma} \int_{\Gamma_{\varsigma}} \mathrm{d}\xi\, \frac{\mathfrak{g}_{\theta,w}(\xi)}{(\lambda_{l_{1}} -\xi) (\lambda_{l_{2}} -\xi) (\lambda_{l_{3}} -\xi)} = 0.
\end{equation*}
It remains to treat six terms. Let us treat the trickiest one (we make it clear hereafter) which is $\mathscr{N}_{\theta,w}^{\perp,\aleph,\perp}(\bold{x},\bold{x})$. From the residue theorem:
\begin{equation*}
\begin{split}
\mathscr{N}_{\theta,w}^{\perp,\aleph,\perp}(\mathbf{x},\mathbf{x}) =
\sum_{l=1}^{\varsigma} \mathfrak{g}_{\theta,w}(\lambda_{l}) &\int_{\mathbb{R}^{3}} \mathrm{d}\mathbf{z}_{1}   \int_{\mathbb{R}^{3}} \mathrm{d}\mathbf{z}_{2}\, (H_{P}-\lambda_{l})_{\perp}^{-1}(\mathbf{x},\mathbf{z}_{1}) \\ &\times\mathbf{a}(\mathbf{z}_{1}-\mathbf{z}_{2}) \cdot (i\nabla_{\mathbf{z}_{1}} \Phi_{l})(\mathbf{z}_{1}) \overline{\Phi_{l}(\mathbf{z}_{2})} T_{P,1}^{\perp}(\mathbf{z}_{2},\mathbf{x};\lambda_{l}).
\end{split}
\end{equation*}
Define for any $1 \leq l \leq \varsigma$, $k \in \{1,2\}$ and $\forall (\mathbf{x},\mathbf{z}_{1},\mathbf{z}_{2}) \in \mathbb{R}^{9}$ with $\mathbf{x} \neq \mathbf{z}_{1} \neq \mathbf{z}_{2}$:
\begin{equation*}
\mathcal{J}_{l}^{(k)}(\mathbf{x},\mathbf{z}_{1},\mathbf{z}_{2}) := \left\vert (H_{P}-\lambda_{l})_{\perp}^{-1}(\mathbf{x},\mathbf{z}_{1}) (i\nabla_{\mathbf{z}_{1}} \Phi_{l})(\mathbf{z}_{1})\right\vert \left\vert \mathbf{z}_{k}\right\vert \left\vert \overline{\Phi_{l}(\mathbf{z}_{2})} T_{P,1}^{\perp}(\mathbf{z}_{2},\mathbf{x};\lambda_{l}) \right\vert.
\end{equation*}
From \eqref{kertildR}, \eqref{estWker} and \eqref{fpexpo}, there exist two constants $C,c>0$ s.t.
\begin{equation*}
\mathcal{J}_{l}^{(1)}(\mathbf{x},\mathbf{z}_{1},\mathbf{z}_{2})\leq C \frac{\mathrm{e}^{-c \vert \mathbf{x} - \mathbf{z}_{1}\vert}}{\vert \mathbf{x} - \mathbf{z}_{1}\vert} \vert \mathbf{z}_{1}\vert \mathrm{e}^{-\sqrt{\vert \lambda_{l} \vert} \vert \mathbf{z}_{1} \vert} \mathrm{e}^{- \sqrt{\vert \lambda_{l} \vert} \vert \mathbf{z}_{2} \vert} \mathrm{e}^{-\frac{c}{2} \vert \mathbf{z}_{2} - \mathbf{x}\vert} \frac{\mathrm{e}^{-\frac{c}{2} \vert \mathbf{z}_{2} - \mathbf{x}\vert}}{\vert \mathbf{z}_{2} - \mathbf{x}\vert}.
\end{equation*}
Using that $\mathrm{e}^{- \min\{\frac{c}{2}, \sqrt{\vert \lambda_{l} \vert}\}( \vert \mathbf{z}_{2} \vert + \vert \mathbf{z}_{2} - \mathbf{x}\vert)} \leq \mathrm{e}^{- \min\{\frac{c}{2}, \sqrt{\vert \lambda_{\varsigma} \vert}\} \vert \mathbf{x} \vert}$ $\forall \mathbf{x} \in \mathbb{R}^{3}$, along with \eqref{ingr} leading to the estimate $\vert \mathbf{z}_{1}\vert \mathrm{e}^{- \sqrt{\vert \lambda_{\varsigma} \vert} \vert\mathbf{z}_{1}\vert} \leq \vert \lambda_{\varsigma} \vert^{-\frac{1}{2}}$ $\forall \mathbf{z}_{1} \in \mathbb{R}^{3}$, then by \cite[Lem. A.2 $(\mathrm{ii})$]{BS}, there exist another constant $C>0$ s.t. for any $1\leq l \leq \varsigma$:
\begin{equation}
\label{majJ1}
\forall\mathbf{x} \in \mathbb{R}^{3},\quad \int_{\mathbb{R}^{3}} \mathrm{d}\mathbf{z}_{1} \int_{\mathbb{R}^{3}} \mathrm{d}\mathbf{z}_{2}\, \mathcal{J}_{l}^{(1)}(\mathbf{x},\mathbf{z}_{1},\mathbf{z}_{2}) \leq C \mathrm{e}^{- \min\{\frac{c}{2}, \sqrt{\vert \lambda_{\varsigma} \vert}\} \vert \mathbf{x} \vert}.
\end{equation}
Clearly, the upper bound in \eqref{majJ1} still holds true when replacing $\mathcal{J}_{l}^{(1)}$ with $\mathcal{J}_{l}^{(2)}$. Hence, one arrives at $\vert \mathscr{N}_{\theta,w}^{\perp,\aleph,\perp}(\mathbf{x},\mathbf{x})\vert \leq C \mathrm{e}^{-c \vert \mathbf{x}\vert}$ $\forall \mathbf{x} \in \mathbb{R}^{3}$, for another constant $C=C(\theta,\varsigma)>0$ and $c>0$. We do not treat the other terms which are simpler. Note that they have all the form $\mathscr{N}_{\theta,w}^{\aleph,\wp_{2},\wp_{3}}(\bold{x},\bold{x})$ with $(\wp_{2},\wp_{3}) \in \{\aleph,\perp\}^{2}$ and $\wp_{2},\wp_{3}\neq \aleph$, or $\mathscr{N}_{\theta,w}^{\wp_{1},\wp_{2},\aleph}(\bold{x},\bold{x})$ with $(\wp_{1},\wp_{2}) \in \{\aleph,\perp\}^{2}$ and $\wp_{1},\wp_{2}\neq \aleph$. Therefore, they can always be written on $\mathbb{R}^{3}$, via the residue theorem, as $\sum_{l=1}^{\varsigma} \Phi_{l}(\mathbf{x}) \mathcal{A}_{l}(\mathbf{x})$ with $\sup_{\mathbf{x} \in \mathbb{R}^{3}} \vert \mathcal{A}_{l}(\mathbf{x}) \vert \leq cste$.  Thus, the announced exponential decay in $\vert \mathbf{x}\vert$ results only from \eqref{fpexpo}. \qed \\

\noindent \textit{\textbf{Proof of Proposition \ref{PRO4}}}. The proof we give requires the notation introduced in Sec. \ref{defcal}. We do not recall them, and we refer to the beginning of Sec. \ref{defcal}. Let $\varsigma \in \{1,\ldots,\tau\}$. From the Riesz integral formula (for orthogonal projections) together with the definition \eqref{scrF}, one has:
\begin{equation*}
\forall \vert b\vert \leq \mathfrak{b},\quad \sum_{l=1}^{\varsigma} \mathrm{Tr}_{L^{2}(\mathbb{R}^{3})}\left\{\Pi_{l}(b)\right\} = \frac{i}{2\pi} \mathrm{Tr}_{L^{2}(\mathbb{R}^{3})}\left\{\int_{\cup_{l=1}^{\varsigma} \gamma_{l}} \mathrm{d}\xi\, (H_{P}(b)-\xi)^{-1}\right\}
=: \mathscr{F}_{1,0}(b).
\end{equation*}
Here, the $\mathfrak{b}$ and the $\gamma_{l}$'s are the same as the ones appearing in \eqref{avanpr}. Since the map $b \mapsto \mathscr{F}_{1,0}(b)$ is twice differentiable in a neighborhood of $b=0$ (see Proposition \ref{dernPr}), then by using the expression for the second derivative at $b=0$ given in \eqref{derscrF} (with $\theta=1$ and $w=0$), one gets the identity:
\begin{equation*}
\label{idzero}
\frac{\mathrm{d}^{2}}{\mathrm{d} b^{2}} \left. \left(\sum_{l=1}^{\varsigma} \mathrm{Tr}_{L^{2}(\mathbb{R}^{3})}\{\Pi_{l}(b)\}\right)\right\vert_{b=0}
= \frac{i}{\pi} \mathrm{Tr}_{L^{2}(\mathbb{R}^{3})} \left\{ \int_{\cup_{l=1}^{\varsigma}\gamma_{l}} \mathrm{d}\xi\, (H_{P}-\xi)^{-1}\left[T_{P,1}(\xi) T_{P,1}(\xi) - T_{P,2}(\xi)\right]\right\}.
\end{equation*}
But the quantity in the above l.h.s. is zero. Indeed, by stability of the eigenvalues of $H_{P}$ in $(-\infty,0)$, one has $\mathrm{dim}\,\mathrm{Ran} \Pi_{l}(b) = 1$ and thus the sum is a $b$-independent quantity (equals to $\varsigma$).\qed

\subsection{Proof of Lemma \ref{oubli}.}
\label{append3}

Introduce $\forall (b,b_{0}) \in \mathbb{R}^{2}$ and $\forall \xi \in \varrho(H_{P}(b_{0}))$, the operators $\tilde{R}_{P}(b,b_{0},\xi)$ and $\tilde{T}_{P,j}(b,b_{0},\xi)$, $j=1,2$ on $L^{2}(\mathbb{R}^{3})$ via their kernel respectively defined on $\mathbb{R}^{6}\setminus D$ by:
\begin{gather}
\label{tildR}
\tilde{R}_{P}(\mathbf{x},\mathbf{y};b,b_{0},\xi) := \mathrm{e}^{i \delta b \phi(\mathbf{x},\mathbf{y})} (H_{P}(b_{0}) - \xi)^{-1}(\mathbf{x},\mathbf{y}),\quad \delta b := b - b_{0},\\
\label{tildaWPj}
\tilde{T}_{P,j}(\mathbf{x},\mathbf{y};b,b_{0},\xi) := \mathrm{e}^{i \delta b \phi(\mathbf{x},\mathbf{y})} T_{P,j}(\mathbf{x},\mathbf{y};b_{0},\xi),
\end{gather}
where $\phi$ is the magnetic phase defined in \eqref{phase}. Also, set:
\begin{equation}
\label{tildaWj}
\tilde{T}_{P}(b,b_{0},\xi) := \delta b \tilde{T}_{P,1}(b,b_{0},\xi) + (\delta b)^{2} \tilde{T}_{P,2}(b,b_{0},\xi).
\end{equation}
Except for a gauge phase factor, the kernel of $\tilde{R}_{P}(b,b_{0},\xi)$ and $\tilde{T}_{P,j}(b,b_{0},\xi)$, $j=1,2$ is the same as the one of $(H_{P}(b_{0})-\xi)^{-1}$ and $T_{P,j}(b_{0},\xi)$ respectively. Therefore, $\forall \eta>0$ and $\forall \xi \in \mathbb{C}$ satisfying $\mathrm{dist}(\xi, \sigma(H_{P}(b_{0}))) \geq \eta$, then $\forall b \in \mathbb{R}$ $\tilde{R}_{P}(b,b_{0},\xi)$ and $\tilde{T}_{P,j}(b,b_{0},\xi)$ are bounded with operator norm obeying \eqref{L2bound} (with $b_{0}$ instead of $b$). Under the same conditions, introduce on $L^{2}(\mathbb{R}^{3})$:
\begin{gather}
\label{scrW1}
\tilde{\mathfrak{T}}_{P}^{(1)}(b,b_{0},\xi) := - \tilde{R}_{P}(b,b_{0},\xi) \tilde{T}_{P,1}(b,b_{0},\xi),\\
\label{scrW2}
\tilde{\mathfrak{T}}_{P}^{(2)}(b,b_{0},\xi) := \tilde{R}_{P}(b,b_{0},\xi)\left \{\left(\tilde{T}_{P,1}(b,b_{0},\xi)\right)^{2} - \tilde{T}_{P,2}(b,b_{0},\xi)\right\},
\end{gather}
as well as, with the additional condition $\xi \in \varrho(H_{P}(b))\cap \varrho(H_{P}(b_{0}))$:
\begin{equation}
\label{scrW3}
\begin{split}
\tilde{\mathfrak{T}}_{P}^{(3)}(b,b_{0},\xi) :=  &- (H_{P}(b) - \xi)^{-1}\left(\tilde{T}_{P}(b,b_{0},\xi)\right)^{3} \\
&+ (\delta b)^{3} \sum_{k=0}^{1} (\delta b)^{k} \sum_{\mathbf{i} \in \{1,2\}^{2}} \chi_{2}^{3+k}(\mathbf{i}) \tilde{R}_{P}(b,b_{0},\xi) \prod_{r = 1}^{2} \tilde{T}_{P,i_{r}}(b,b_{0},\xi).
\end{split}
\end{equation}
Now, we turn to the actual proof. Let $K \subset \left(\varrho(H_{P}) \cap \{\zeta \in \mathbb{C}: \Re \zeta < 0\}\right)$ be a compact subset. From \cite[Thm. 1.1]{BC}, then there exists $\mathfrak{b}_{K}>0$ s.t. $\forall \vert b \vert \leq \mathfrak{b}_{K}$, $K \subset \left(\varrho(H_{P}(b)) \cap \{\zeta \in \mathbb{C}: \Re \zeta < 0\}\right)$. From now on, let $b_{0} \in (-\mathfrak{b}_{K},\mathfrak{b}_{K})$ be fixed. The starting point of the gauge invariant magnetic perturbation theory is the following identity which holds $\forall \xi \in K$ in the bounded operators sense on $L^{2}(\mathbb{R}^{3})$, see \cite[Proof of Prop. 3.2]{CN1} and also
\cite[Lem. 3.2]{BS}:
\begin{equation}
\label{startequi}
\forall \vert b \vert \leq \mathfrak{b}_{K},\quad (H_{P}(b)-\xi)^{-1} = \tilde{R}_{P}(b,b_{0},\xi) - (H_{P}(b)-\xi)^{-1} \tilde{T}_{P}(b,b_{0},\xi).
\end{equation}
It results from \eqref{startequi} that, for $b$ sufficiently close to $b_{0}$, the operator norm of $(H_{P}(b)-\xi)^{-1} - \tilde{R}_{P}(b,b_{0},\xi)$ behaves like $\mathcal{O}(\vert \delta b \vert)$. This comes from \eqref{tildaWj}, the definitions \eqref{tildR}-\eqref{tildaWPj} and the estimates \eqref{reesti1}-\eqref{reesti2} which yield:
\begin{equation*}
\forall \vert b\vert \leq \mathfrak{b}_{K},\quad \max \left\{ \sup_{\xi \in K} \left\Vert \tilde{R}_{P}(b,b_{0},\xi)\right\Vert, \sup_{\xi \in K} \left\Vert \tilde{T}_{P,j}(b,b_{0},\xi) \right\Vert \right\} \leq C,\quad j=1,2,
\end{equation*}
for some constant $C=C(\vert b_{0} \vert,K)>0$. Next, by iterating twice \eqref{startequi} and in view of \eqref{scrW1}-\eqref{scrW3}, one arrives $\forall \vert b \vert \leq \mathfrak{b}_{K}$ and $\forall \xi \in K$ on $L^{2}(\mathbb{R}^{3})$ at:
\begin{equation}
\label{iteratw}
(H_{P}(b)-\xi)^{-1} = \tilde{R}_{P}(b,b_{0},\xi)
+ \sum_{k=1}^{2} (\delta b)^{k} \tilde{\mathfrak{T}}_{P}^{(k)}(b,b_{0},\xi)  + \tilde{\mathfrak{T}}_{P}^{(3)}(b,b_{0},\xi).
\end{equation}
In terms of corresponding integral kernels, one has on $\mathbb{R}^{6}\setminus D$:
\begin{equation}
\label{idkeraz}
(H_{P}(b)-\xi)^{-1}(\mathbf{x},\mathbf{y}) = \tilde{R}_{P}(\mathbf{x},\mathbf{y};b,b_{0},\xi) + \sum_{k=1}^{2} (\delta b)^{k} \tilde{\mathfrak{T}}_{P}^{(k)}(\mathbf{x},\mathbf{y};b,b_{0},\xi) + \tilde{\mathfrak{T}}_{P}^{(3)}(\mathbf{x},\mathbf{y};b,b_{0},\xi),
\end{equation}
where, for any $k \in \{1,2\}$, $\forall(\mathbf{x},\mathbf{y}) \in \mathbb{R}^{6}$ and $\forall\vert b \vert \leq \mathfrak{b}_{K}$:
\begin{equation}
\label{frakTk}
\begin{split}
\tilde{\mathfrak{T}}_{P}^{(k)}(\mathbf{x},\mathbf{y};b,b_{0},\xi)
:= &\sum_{j=1}^{k} (-1)^{j} \sum_{\mathbf{i} \in \{1,2\}^{j}} \chi_{j}^{k}(\mathbf{i}) \int_{\mathbb{R}^{3}} \mathrm{d}\mathbf{z}_{1} \dotsb \int_{\mathbb{R}^{3}} \mathrm{d}\mathbf{z}_{j}\, \mathrm{e}^{i \delta b (\phi(\mathbf{x},\mathbf{z}_{1})+ \dotsb + \phi(\mathbf{z}_{j},\mathbf{y}))} \\
&\times (H_{P}(b_{0}) - \xi)^{-1}(\mathbf{x},\mathbf{z}_{1}) T_{P,i_{1}}(\mathbf{z}_{1},\mathbf{z}_{2};b_{0},\xi) \dotsb T_{P,i_{j}}(\mathbf{z}_{j},\mathbf{y};b_{0},\xi),
\end{split}
\end{equation}
and $\tilde{\mathfrak{T}}_{P}^{(3)}(\cdot\,,\cdot\,;b,b_{0},\xi)$ in \eqref{idkeraz} stands for the kernel of $\tilde{\mathfrak{T}}_{P}^{(3)}(b,b_{0},\xi)$. Let us note that, in view of \eqref{scrW3} along with \eqref{tildR}-\eqref{tildaWPj} and the estimates \eqref{reesti1}-\eqref{reesti2}, the kernel $\tilde{\mathfrak{T}}_{P}^{(3)}(\cdot\,,\cdot\,;b,b_{0},\xi)$ behaves like $\mathcal{O}(\vert \delta b \vert^{3})$ uniformly in $\xi \in K$.\\
Afterwards, we remove the $b$-dependence in the first two terms of the r.h.s. of \eqref{idkeraz}. To achieve that, we expand in Taylor power series the exponential phase factor appearing in \eqref{tildR} and \eqref{frakTk} up to the second order in $ \delta b$. In this way, one obtains $\forall \vert b \vert \leq \mathfrak{b}_{K}$ and $\forall(\mathbf{x},\mathbf{y}) \in \mathbb{R}^{6}\setminus D$:
\begin{multline}
\label{mainexp}
\tilde{R}_{P}(\mathbf{x},\mathbf{y};b,b_{0},\xi) + \sum_{k=1}^{2} (\delta b)^{k} \tilde{\mathfrak{T}}_{P}^{(k)}(\mathbf{x},\mathbf{y};b,b_{0},\xi) = \\
 \sum_{k=0}^{2} (\delta b)^{k} \frac{\left(i\phi(\mathbf{x},\mathbf{y})\right)^{k}}{k!} (H_{P}(b_{0}) - \xi)^{-1}(\mathbf{x},\mathbf{y})
 + \sum_{k=1}^{2} (\delta b)^{k} \sum_{m=1}^{k} \mathfrak{T}_{P,m}^{k-m}(\mathbf{x},\mathbf{y};b_{0},\xi) + \mathfrak{T}_{P}^{(4)}(\mathbf{x},\mathbf{y};b,b_{0},\xi),
\end{multline}
where the function $\mathfrak{T}_{P,m}^{k-m}(\cdot\,,\cdot\,;b_{0},\xi)$, $2 \geq k \geq m \geq 1$  is defined in \eqref{calTPkm}, and the last term stands for the remainder term. We mention that we have used the explicit expressions in Remark \ref{explicit} to rewrite the second term in the r.h.s. of \eqref{mainexp} coming from \eqref{frakTk}. Note also that, by construction, $\mathfrak{T}_{P}^{(4)}(\cdot\,,\cdot\,;b,b_{0},\xi)$ satisfies the property that its first two derivatives at $b_{0}$ are identically zero.\\
From \eqref{mainexp}, it holds on $\mathbb{R}^{6} \setminus D$ for $b \in [-\mathfrak{b}_{K},\mathfrak{b}_{K}]$ sufficiently close to $b_{0}$:
\begin{multline*}
(H_{P}(b) - \xi)^{-1}(\mathbf{x},\mathbf{y}) - (H_{P}(b_{0}) - \xi)^{-1}(\mathbf{x},\mathbf{y})= \\
\delta b \left\{ i \phi(\mathbf{x},\mathbf{y}) (H_{P}(b_{0}) - \xi)^{-1}(\mathbf{x},\mathbf{y}) - \left((H_{P}(b_{0})-\xi)^{-1}T_{P,1}(b_{0},\xi)\right)(\mathbf{x},\mathbf{y})\right\} + o\left(\delta b\right).
\end{multline*}
It follows that the map $b \mapsto (H_{P}(b)-\xi)^{-1}(\mathbf{x},\mathbf{y})$ is differentiable at $b_{0}$, and:
\begin{equation*}
\left. \frac{\partial}{\partial b} (H_{P}(b)-\xi)^{-1}(\mathbf{x},\mathbf{y})\right\vert_{b=b_{0}} := i \phi(\mathbf{x},\mathbf{y}) (H_{P}(b_{0}) - \xi)^{-1}(\mathbf{x},\mathbf{y})
- \left((H_{P}(b_{0})-\xi)^{-1}T_{P,1}(b_{0},\xi)\right)(\mathbf{x},\mathbf{y}).
\end{equation*}
This can be extended to $(-\mathfrak{b}_{K},\mathfrak{b}_{K})$. The lemma follows by induction. \qed

\subsection{Proof of Lemmas \ref{lem1ma}-\ref{lem4ma}.}
\label{append4}

Throughout this section, $(\mathfrak{I}_{2}(L^{2}(\mathbb{R}^{3})),\Vert \cdot\,\Vert_{\mathfrak{I}_{2}})$ and $(\mathfrak{I}_{1}(L^{2}(\mathbb{R}^{3})),\Vert \cdot\,\Vert_{\mathfrak{I}_{1}})$ denote the Banach space of Hilbert-Schmidt (H-S) and trace class operators on $L^{2}(\mathbb{R}^{3})$ respectively.\\

\noindent \textit{\textbf{Proof of Lemma \ref{lem1ma}}}. Let us denote:
\begin{gather*}
\mathcal{Y}_{R,1}(\xi) := \vert \Omega_{R}\vert^{-1} \mathrm{Tr}_{L^{2}(\mathbb{R}^{3})}\left\{\chi_{\Omega_{R}}(H_{R} - \xi)^{-1} T_{R,1}(\xi) T_{R,1}(\xi)\chi_{\Omega_{R}}\right\},\\
\mathcal{Y}_{R,2}(\xi) := \vert \Omega_{R}\vert^{-1} \mathrm{Tr}_{L^{2}(\mathbb{R}^{3})}\left\{\chi_{\Omega_{R}}(H_{R} - \xi)^{-1} T_{R,2}(\xi)\chi_{\Omega_{R}}\right\}.
\end{gather*}
By replacing $(H_{R} - \xi)^{-1}$ with the r.h.s. of \eqref{eqsecresQ} in $\mathcal{Y}_{R,j}(\xi)$, $j=1,2$ then:
\begin{gather*}
\mathcal{Y}_{R,1}(\xi) = \vert \Omega_{R}\vert^{-1} \mathrm{Tr}_{L^{2}(\mathbb{R}^{3})}\left\{\chi_{\Omega_{R}}\mathcal{R}_{R}(\xi) \mathcal{T}_{R,1}(\xi) \mathcal{T}_{R,1}(\xi)\chi_{\Omega_{R}}\right\} + \mathcal{Q}_{R,1}(\xi), \\
\mathcal{Y}_{R,2}(\xi) = \vert \Omega_{R}\vert^{-1} \mathrm{Tr}_{L^{2}(\mathbb{R}^{3})}\left\{\chi_{\Omega_{R}}\mathcal{R}_{R}(\xi) \mathcal{T}_{R,2}(\xi)\chi_{\Omega_{R}}\right\}  + \mathcal{Q}_{R,2}(\xi),
\end{gather*}
where $\mathcal{Q}_{R,1}(\xi)$ and $\mathcal{Q}_{R,2}(\xi)$ consist of seven and three terms respectively. Let $0<\alpha<1$ and $\eta>0$ be fixed. Let us show that, under the conditions of Lemma \ref{lem1ma}, the quantities $\vert \mathcal{Q}_{R,j}(\xi)\vert$, $j=1,2$ obey an estimate of type \eqref{firequi1}. To do that, take one generical term from $\mathcal{Q}_{R,1}(\xi)$ and one from $\mathcal{Q}_{R,2}(\xi)$:
\begin{gather*}
\begin{split}
q_{R,1}(\xi) :=  \frac{1}{\vert \Omega_{R}\vert} &\int_{\Omega_{R}} \mathrm{d}\mathbf{x} \int_{\mathbb{R}^{3}} \mathrm{d}\mathbf{z}_{1} \int_{\mathbb{R}^{3}} \mathrm{d}\mathbf{z}_{2}\, (\mathcal{R}_{R}(\xi))(\mathbf{x},\mathbf{z}_{1})
 \\
& \times  \mathbf{a}(\mathbf{z}_{1} - \mathbf{z}_{2}) \cdot \nabla_{\mathbf{z}_{1}} (\mathcal{R}_{R}(\xi))(\mathbf{z}_{1},\mathbf{z}_{2}) \mathbf{a}(\mathbf{z}_{2} - \mathbf{x}) \cdot \nabla_{\mathbf{z}_{2}}\left\{(H_{R} - \xi)^{-1} \mathcal{W}_{R}(\xi)\right\}(\mathbf{z}_{2},\mathbf{x}),
\end{split} \\
q_{R,2}(\xi) := - \frac{1}{\vert \Omega_{R}\vert} \int_{\Omega_{R}} \mathrm{d}\mathbf{x} \int_{\mathbb{R}^{3}} \mathrm{d}\mathbf{z}\, (\mathcal{R}_{R}(\xi))(\mathbf{x},\mathbf{z}) \frac{1}{2}\mathbf{a}^{2}(\mathbf{z} - \mathbf{x}) \left\{(H_{R} - \xi)^{-1} \mathcal{W}_{R}(\xi)\right\}(\mathbf{z},\mathbf{x}).
\end{gather*}
 We need the following estimates. From \eqref{kertildR}-\eqref{kertildRder}, \eqref{kertildT} combined with \cite[Lem. A.2]{BS}, then under the conditions of Lemma \ref{lemhgf} $\mathrm{(i)}$, one has $\forall R \geq R_{0}$:
\begin{gather}
\label{oops1}
\left\vert \int_{\mathbb{R}^{3}} \mathrm{d}\mathbf{z}\, (H_{R} - \xi)^{-1}(\mathbf{x},\mathbf{z})(\mathcal{W}_{R}(\xi))(\mathbf{z},\mathbf{y})\right\vert \leq p(\vert \xi \vert) \mathrm{e}^{-\vartheta_{\xi} R^{\alpha}} \mathrm{e}^{- \vartheta_{\xi} \vert \mathbf{x} - \mathbf{y}\vert}, \\
\label{oops2}
\left\vert \int_{\mathbb{R}^{3}} \mathrm{d}\mathbf{z}\, \nabla_{\mathbf{x}} (H_{R} - \xi)^{-1}(\mathbf{x},\mathbf{z})(\mathcal{W}_{R}(\xi))(\mathbf{z},\mathbf{y})\right\vert \leq p(\vert \xi \vert) \mathrm{e}^{-\vartheta_{\xi} R^{\alpha}} \frac{\mathrm{e}^{- \vartheta_{\xi} \vert \mathbf{x} - \mathbf{y}\vert}}{\vert \mathbf{x} - \mathbf{y}\vert},
\end{gather}
for another $\vartheta>0$ and polynomial $p(\cdot\,)$ both $R$-independent. From \eqref{kertildS}-\eqref{kertildSder}, \eqref{oops1}-\eqref{oops2} combined  with
\cite[Lem. A.2 (ii)]{BS}, $\vert q_{R,j}(\xi) \vert \leq p(\vert \xi\vert) \mathrm{e}^{-\vartheta_{\xi} R^{\alpha}}$ for another $\vartheta >0$ and polynomial $p(\cdot\,)$ both $R$-independent. The other terms coming from $\mathcal{Q}_{R,j}(\xi)$, $j=1,2$ can be treated by using similar arguments. \qed \\

\noindent \textit{\textbf{Proof of Lemma \ref{lem2ma}}}. Let us denote:
\begin{equation*}
\mathscr{Y}_{R,2}(\xi) := \vert \Omega_{R} \vert^{-1} \mathrm{Tr}_{L^{2}(\mathbb{R}^{3})}\left\{\chi_{\Omega_{R}} \mathcal{R}_{R}(\xi) \mathcal{T}_{R,2}(\xi) \chi_{\Omega_{R}}\right\}.
\end{equation*}
By replacing $\mathcal{R}_{R}(\xi)$ with the r.h.s. of \eqref{rewtildSR} in $\mathscr{Y}_{R,2}(\xi)$, then we have:
\begin{equation*}
\mathscr{Y}_{R,2}(\xi) = \vert \Omega_{R} \vert^{-1} \mathrm{Tr}_{L^{2}(\mathbb{R}^{3})}\left\{\chi_{\Omega_{R}} \mathscr{R}_{R}(\xi) \mathscr{T}_{R,2}(\xi) \chi_{\Omega_{R}}\right\} + \mathscr{Q}_{R,2}(\xi),
\end{equation*}
where $\mathscr{Q}_{R,2}(\xi)$ consists of three terms. Let $0<\alpha<1$ and $\eta>0$ be fixed. Let us show that, under the conditions of Lemma \ref{lem2ma}, the quantity $\vert \mathscr{Q}_{R,2}(\xi)\vert$ obeys an estimate of type \eqref{firequi2}. To do so, take a generical term from $\mathscr{Q}_{R,2}(\xi)$:
\begin{equation*}
\mathfrak{q}_{R,2}(\xi) := - \frac{1}{\vert \Omega_{R} \vert} \int_{\Omega_{R}} \mathrm{d}\mathbf{x} \int_{\mathbb{R}^{3}} \mathrm{d}\mathbf{z}\,  (\mathscr{R}_{R}(\xi))(\mathbf{x},\mathbf{z}) \frac{1}{2} \mathbf{a}^{2}(\mathbf{z} - \mathbf{x}) (\mathscr{W}_{R}(\xi))(\mathbf{z},\mathbf{x}).
\end{equation*}
Introduce on $L^{2}(\mathbb{R}^{3})$ the operators $\mathscr{Z}_{R}(\xi)$ and $T_{\Xi}^{(1)}(\xi)$, with $\Xi:=R$ or $P$ and $R \geq R_{0}$, generated via their kernel respectively defined by:
\begin{gather}
\forall(\mathbf{x},\mathbf{y}) \in \mathbb{R}^{6},\quad \mathscr{Z}_{R}(\mathbf{x},\mathbf{y};\xi) := \frac{1}{2} \mathbf{a}^{2}(\mathbf{x} - \mathbf{y}) (\mathscr{W}_{R}(\xi))(\mathbf{x},\mathbf{y}), \nonumber\\
\label{TZ1}
\forall(\mathbf{x},\mathbf{y}) \in \mathbb{R}^{6}\setminus D,\quad T_{\Xi}^{(1)}(\mathbf{x},\mathbf{y};\xi) :=  \mathbf{a}(\mathbf{x} - \mathbf{y}) (H_{\Xi} - \xi)^{-1}(\mathbf{x},\mathbf{y}).
\end{gather}
Due to the estimates \eqref{kerhatT} and \eqref{kertildR}, the operators $\mathscr{Z}_{R}(\xi)$ and $T_{\Xi}^{(1)}(\xi)$ are bounded on $L^{2}(\mathbb{R}^{3})$. We stress the point that $\mathscr{Z}_{R}(\xi)$ can be rewritten as:
\begin{equation*}
\mathscr{Z}_{R}(\xi) = \hat{\hat{g}}_{R} \left\{T_{R,2}(\xi) \breve{V}_{R} (H_{P} -\xi)^{-1}
+ (H_{R} - \xi)^{-1} \breve{V}_{R} T_{P,2}(\xi) + T_{R}^{(1)}(\xi)\breve{V}_{R} T_{P}^{(1)}(\xi)\right\}(1 - g_{R}),
\end{equation*}
where we used that $\mathbf{a}^{2}(\mathbf{x}-\mathbf{y}) = \{\mathbf{a}(\mathbf{x} - \mathbf{z})+\mathbf{a}(\mathbf{z}-\mathbf{y})\}^{2}$ for any $\mathbf{x},\mathbf{y},\mathbf{z} \in \mathbb{R}^{3}$. Now, it comes down to prove that there exists a $\vartheta>0$ and a polynomial $p(\cdot\,)$ s.t. $\forall R \geq \max\{R_{0},R_{1}\}$ and $\forall \xi \in \mathbb{C}$ obeying $\mathrm{dist}(\xi,\sigma(H_{R})\cap \sigma(H_{P})) \geq \eta$:
\begin{equation}
\label{frakI1}
\vert \Omega_{R}\vert^{-1} \Vert \chi_{\Omega_{R}}\mathscr{R}_{R}(\xi) \mathscr{Z}_{R}(\xi)\chi_{\Omega_{R}}\Vert_{\mathfrak{I}_{1}} \leq p(\vert \xi \vert) \mathrm{e}^{- \vartheta_{\xi} R^{\alpha}}.
\end{equation}
To do so, it is enough to use these estimates which hold $\forall R \geq \max\{R_{0},R_{1}\}$:
\begin{equation}
\label{HSes1}
\max\left\{\Vert \chi_{\Omega_{R}}\hat{g}_{R} (H_{\Xi} - \xi)^{-1} \Vert_{\mathfrak{I}_{2}}, \Vert \chi_{\Omega_{R}} \hat{\hat{g}}_{R} (H_{\Xi} - \xi)^{-1} \Vert_{\mathfrak{I}_{2}}\right\} \leq p(\vert \xi\vert) \sqrt{\vert \Omega_{R}\vert},
\end{equation}
\begin{multline}
\label{HSes2}
\max \left\{\Vert \breve{V}_{R} (H_{\Xi} - \xi)^{-1} (1-g_{R}) \chi_{\Omega_{R}}\Vert_{\mathfrak{I}_{2}},
\Vert \breve{V}_{R} T_{\Xi}^{(1)}(\xi) (1 - g_{R}) \chi_{\Omega_{R}} \Vert_{\mathfrak{I}_{2}},\Vert \breve{V}_{R} T_{\Xi,2}(\xi) (1 - g_{R}) \chi_{\Omega_{R}} \Vert_{\mathfrak{I}_{2}}\right\} \\
\leq p(\vert \xi\vert) \sqrt{\vert \Omega_{R}\vert} \mathrm{e}^{-\vartheta_{\xi} R^{\alpha}},
\end{multline}
with $\Xi=R$ or $P$, and for another $R$-independent polynomial $p(\cdot\,)$. Here, we used the definitions \eqref{TZ1}, \eqref{WR2}-\eqref{WP2} with \eqref{kertildR} and Lemma \ref{lemhgf} $\mathrm{(iii)}$. The two other terms of $\mathscr{Q}_{R,2}(\xi)$ can be treated by using similar arguments.\qed \\

\noindent \textit{\textbf{Proof of Lemma \ref{lem3ma}}}. Let us denote:
\begin{equation*}
\mathscr{Y}_{R,1}(\xi) := \vert \Omega_{R} \vert^{-1} \mathrm{Tr}_{L^{2}(\mathbb{R}^{3})}\left\{\chi_{\Omega_{R}} \mathcal{R}_{R}(\xi) \mathcal{T}_{R,1}(\xi) \mathcal{T}_{R,1}(\xi) \chi_{\Omega_{R}}\right\}.
\end{equation*}
By replacing $\mathcal{R}_{R}(\xi)$ with the r.h.s. of \eqref{rewtildSR} in $\mathscr{Y}_{R,1}(\xi)$, then we have:
\begin{equation*}
\mathscr{Y}_{R,1}(\xi) = \vert \Omega_{R} \vert^{-1} \mathrm{Tr}_{L^{2}(\mathbb{R}^{3})}\left\{\chi_{\Omega_{R}} \mathscr{R}_{R}(\xi) \mathscr{T}_{R,1}(\xi) \mathscr{T}_{R,1}(\xi) \chi_{\Omega_{R}}\right\} + \mathscr{Q}_{R,1}(\xi),
\end{equation*}
where $\mathscr{Q}_{R,1}(\xi)$ consists of seven terms.  Let $0<\alpha<1$ and $\eta>0$ be fixed. Let us show that, under the conditions of Lemma \ref{lem3ma}, the quantity $\vert \mathscr{Q}_{R,1}(\xi)\vert$ obeys an estimate of type \eqref{firequi3}. To do so, take a generical term from $\mathscr{Q}_{R,1}(\xi)$:
\begin{equation*}
\begin{split}
\mathfrak{q}_{R,1}(\xi) := \frac{1}{\vert \Omega_{R} \vert} \int_{\Omega_{R}} \mathrm{d}\mathbf{x} &\int_{\mathbb{R}^{3}} \mathrm{d}\mathbf{z}_{1} \int_{\mathbb{R}^{3}} \mathrm{d}\mathbf{z}_{2}\, (\mathscr{R}_{R}(\xi))(\mathbf{x},\mathbf{z}_{1}) \\
&\times \mathbf{a}(\mathbf{z}_{1}-\mathbf{z}_{2}) \cdot \nabla_{\mathbf{z}_{1}} (\mathscr{R}_{R}(\xi))(\mathbf{z}_{1},\mathbf{z}_{2}) \mathbf{a}(\mathbf{z}_{2} - \mathbf{x}) \cdot \nabla_{\mathbf{z}_{2}} (\mathscr{W}_{R}(\xi))(\mathbf{z}_{2},\mathbf{x}).
\end{split}
\end{equation*}
Let us note that from \eqref{tild2SR} and \eqref{tild2TR}:
\begin{gather*}
\begin{split}
\nabla \mathscr{R}_{R}(\xi) = &\left[\left(\nabla \hat{g}_{R}\right)(H_{P}-\xi)^{-1} + \hat{g}_{R} \nabla (H_{P}-\xi)^{-1}\right] g_{R} \\
 &+ \left[\left(\nabla \hat{\hat{g}}_{R}\right) (H_{P}-\xi)^{-1}  + \hat{\hat{g}}_{R} \nabla (H_{P}-\xi)^{-1}\right](1-g_{R}),
 \end{split}\\
\nabla \mathscr{W}_{R}(\xi) = \left[\left(\nabla \hat{\hat{g}}_{R}\right) (H_{R}-\xi)^{-1} \breve{V}_{R} (H_{P} - \xi)^{-1} + \hat{\hat{g}}_{R} \nabla (H_{R}-\xi)^{-1} \breve{V}_{R} (H_{P} - \xi)^{-1}\right] (1-g_{R}).
\end{gather*}
By using the definitions \eqref{TZ1}, the quantity $\mathfrak{q}_{R,1}(\xi)$ can be rewritten as:
\begin{equation*}
\mathfrak{q}_{R,1}(\xi) = \mathfrak{q}_{R,1}^{(1)}(\xi) + \mathfrak{q}_{R,1}^{(2)}(\xi),\quad \textrm{where:}
\end{equation*}
\begin{equation}
\begin{split}
\mathfrak{q}_{R,1}^{(1)}(\xi) := \frac{1}{\vert \Omega_{R} \vert} &\int_{\Omega_{R}} \mathrm{d}\mathbf{x} \int_{\mathbb{R}^{3}} \mathrm{d}\mathbf{z}_{1} \int_{\mathbb{R}^{3}} \mathrm{d}\mathbf{z}_{2}\, (\mathscr{R}_{R}(\xi))(\mathbf{x},\mathbf{z}_{1})
 \left\{(\nabla \hat{g}_{R})(\mathbf{z}_{1}) T_{P}^{(1)}(\mathbf{z}_{1},\mathbf{z}_{2};\xi) g_{R}(\mathbf{z}_{2}) \right. \\
&+ \left. \left(\nabla \hat{\hat{g}}_{R}\right)(\mathbf{z}_{1}) T_{P}^{(1)}(\mathbf{z}_{1},\mathbf{z}_{2};\xi)(1-g_{R})(\mathbf{z}_{2})\right\}\mathbf{a}(\mathbf{z}_{2}- \mathbf{x}) \cdot \nabla_{\mathbf{z}_{2}} (\mathscr{W}_{R}(\xi))(\mathbf{z}_{2},\mathbf{x}), \nonumber
\end{split}
\end{equation}
\begin{equation*}
\begin{split}
\mathfrak{q}_{R,1}^{(2)}(\xi) := - \frac{i}{\vert \Omega_{R} \vert} &\mathrm{Tr}_{L^{2}(\mathbb{R}^{3})}\left\{\chi_{\Omega_{R}} \mathscr{R}_{R}(\xi)\left[\hat{g}_{R} T_{P,1}(\xi) g_{R} + \hat{\hat{g}}_{R} T_{P,1}(\xi)(1-g_{R})\right]\right. \\
&\left. \times \left\{\left(\nabla\hat{\hat{g}}_{R}\right) \left[T_{R}^{(1)}(\xi) \breve{V}_{R} (H_{P} - \xi)^{-1} + (H_{R} - \xi)^{-1}\breve{V}_{R}T_{P}^{(1)}(\xi)\right] \right. \right.\\ &\left. \left.+ \hat{\hat{g}}_{R}\left[-i T_{R,1}(\xi)\breve{V}_{R}(H_{P}-\xi)^{-1} +  \nabla (H_{R}-\xi)^{-1} \breve{V}_{R} T_{P}^{(1)}(\xi)\right]\right\} (1-g_{R})\chi_{\Omega_{R}}\right\}.
\end{split}
\end{equation*}
Let us first estimate $\vert \mathfrak{q}_{R,1}^{(1)}(\xi)\vert$. By mimicking the proof of \eqref{autilis}, then under the same conditions, there exists another $\vartheta>0$ and polynomial $p(\cdot\,)$ s.t. $\forall R \geq R_{0}$:
\begin{equation*}
\max\left\{\left\vert \left(\nabla \hat{g}_{R}\right)(\mathbf{x}) T_{P}^{(1)}(\mathbf{x},\mathbf{y};\xi) g_{R}(\mathbf{y})\right\vert, \left\vert \left(\nabla \hat{\hat{g}}_{R}\right)(\mathbf{x}) T_{P}^{(1)}(\mathbf{x},\mathbf{y};\xi)(1-g_{R})(\mathbf{y})\right\vert \right\} \leq p(\vert \xi\vert) \mathrm{e}^{-\vartheta_{\xi} R^{\alpha}} \mathrm{e}^{-\vartheta_{\xi} \vert \mathbf{x} - \mathbf{y}\vert}.
\end{equation*}
From \eqref{kertildS}, \eqref{ledernier} and the above estimate, one concludes by \cite[Lem. A.2]{BS} that $\vert \mathfrak{q}_{R,1}^{(1)}(\xi) \vert \leq p(\vert \xi\vert) \mathrm{e}^{-\vartheta_{\xi} R^{\alpha}}$ $\forall R \geq R_{0}$, for another constant $\vartheta>0$ and polynomial $p(\cdot\,)$ both $R$-independent. Next, let us estimate $\vert \mathfrak{q}_{R,1}^{(2)}(\xi)\vert$. By mimicking the arguments leading to \eqref{frakI1}, the trace norm of the operator inside the trace in the expression of $\mathfrak{q}_{R,1}^{(2)}(\xi)$ is bounded above $\forall R \geq \max\{R_{0},R_{1}\}$ by $polynomial \times \vert \Omega_{R}\vert \mathrm{e}^{-\vartheta_{\xi} R^{\alpha}}$  due to the H-S norms in \eqref{HSes1}-\eqref{HSes2} and the operator norms in \eqref{estgeN}-\eqref{estgeN2}. Then $\forall R \geq \max\{R_{0},R_{1}\}$, $\vert \mathfrak{q}_{R,1}^{(2)}(\xi) \vert \leq p(\vert \xi\vert) \mathrm{e}^{-\vartheta_{\xi} R^{\alpha}}$ for another constant $\vartheta>0$ and polynomial $p(\cdot\,)$ both $R$-independent. The other terms coming from $\mathscr{Q}_{R,1}(\xi)$ can be treated by similar arguments. \qed \\

\noindent \textit{\textbf{Proof of Lemma \ref{lem4ma}}}. Let us denote:
\begin{gather*}
\mathfrak{Y}_{R,1}(\xi) := \vert \Omega_{R} \vert^{-1} \mathrm{Tr}_{L^{2}(\mathbb{R}^{3})}\left\{ \chi_{\Omega_{R}} \mathscr{R}_{R}(\xi) \mathscr{T}_{R,1}(\xi) \mathscr{T}_{R,1}(\xi)\chi_{\Omega_{R}}\right\},\\
\mathfrak{Y}_{R,2}(\xi) := \vert \Omega_{R} \vert^{-1} \mathrm{Tr}_{L^{2}(\mathbb{R}^{3})}\left\{ \chi_{\Omega_{R}} \mathscr{R}_{R}(\xi) \mathscr{T}_{R,2}(\xi) \chi_{\Omega_{R}}\right\}.
\end{gather*}
By replacing $\mathscr{R}_{R}(\xi)$ with the r.h.s. of \eqref{rewtildSR2} in $\mathfrak{Y}_{R,j}(\xi)$, $j=1,2$ then:
\begin{gather*}
\mathfrak{Y}_{R,1}(\xi) = \vert \Omega_{R} \vert^{-1} \mathrm{Tr}_{L^{2}(\mathbb{R}^{3})}\left\{ \chi_{\Omega_{R}} (H_{P}-\xi)^{-1} T_{P,1}(\xi) T_{P,1}(\xi)\chi_{\Omega_{R}}\right\} + \mathfrak{Q}_{R,1}(\xi),\\
\mathfrak{Y}_{R,2}(\xi) = \vert \Omega_{R} \vert^{-1} \mathrm{Tr}_{L^{2}(\mathbb{R}^{3})}\left\{ \chi_{\Omega_{R}} (H_{P}-\xi)^{-1} T_{P,2}(\xi)\chi_{\Omega_{R}}\right\} + \mathfrak{Q}_{R,2}(\xi),
\end{gather*}
where $\mathfrak{Q}_{R,1}(\xi)$ and $\mathfrak{Q}_{R,2}(\xi)$ consist of seven and three terms respectively.  Let $0<\alpha<1$ and $\eta>0$ be fixed. Let us show that, under the conditions of Lemma \ref{lem4ma}, the quantities $\vert \mathfrak{Q}_{R,j}(\xi)\vert$, $j=1,2$ obey an estimate of type \eqref{firequi4}. To do so, take one generical term from $\mathfrak{Q}_{R,1}(\xi)$ and one from $\mathfrak{Q}_{R,2}(\xi)$:
\begin{gather*}
\label{hatr11Res}
\textswab{q}_{R,2}(\xi) := \frac{1}{\vert \Omega_{R} \vert} \int_{\Omega_{R}} \mathrm{d}\mathbf{x} \int_{\mathbb{R}^{3}} \mathrm{d}\mathbf{z}\, (H_{P}-\xi)^{-1}(\mathbf{x},\mathbf{z}) \frac{1}{2} \mathbf{a}^{2}(\mathbf{z} - \mathbf{x})(\mathfrak{W}_{R}(\xi))(\mathbf{z},\mathbf{x}),\\
\begin{split}
\textswab{q}_{R,1}(\xi) := - \frac{1}{\vert \Omega_{R} \vert} \int_{\Omega_{R}} \mathrm{d}\mathbf{x} &\int_{\mathbb{R}^{3}} \mathrm{d}\mathbf{z}_{1} \int_{\mathbb{R}^{3}} \mathrm{d}\mathbf{z}_{2}\, (H_{P}-\xi)^{-1}(\mathbf{x},\mathbf{z}_{1}) \mathbf{a}(\mathbf{z}_{1} - \mathbf{z}_{2}) \\
&\times \nabla_{\mathbf{z}_{1}} (H_{P}-\xi)^{-1}(\mathbf{z}_{1},\mathbf{z}_{2}) \mathbf{a}(\mathbf{z}_{2} - \mathbf{x}) \cdot \nabla_{\mathbf{z}_{2}} (\mathfrak{W}_{R}(\xi))(\mathbf{z}_{2},\mathbf{x}).
\end{split}
\end{gather*}
From \eqref{kertildR}-\eqref{kertildRder}, \eqref{kertildT} combined with \cite[Lem. A.2 (ii)]{BS}, then one obtains $\forall R\geq R_{0}$, $\vert \textswab{q}_{R,j}(\xi) \vert \leq p(\vert \xi\vert) \mathrm{e}^{- \vartheta_{\xi} R^{\alpha}}$ $j=1,2$, for another constant $\vartheta>0$ and polynomial $p(\cdot\,)$ both $R$-independent. The other terms coming from $\mathfrak{Q}_{R,j}(\xi)$, $j=1,2$ can be treated by using similar arguments. \qed

\section{Acknowledgments.}

The author was partially supported by the Lundbeck Foundation, and the European Research Council under the European Community's Seventh Framework Program (FP7/2007--2013)/ERC grant agreement 202859.
B.S. warmly thanks Horia D. Cornean and S\o ren Fournais for many fruitful and stimulating discussions. As well, the author thanks the referee for a number of relevant comments and helpful suggestions that helped improve the present manuscript.

\end{document}